\documentclass{emulateapj}
\usepackage{apjfonts}
\usepackage{lscape,graphicx}

\def\lumin{\,{\rm erg \ s^{-1}}}

\def\fluxu{\,{\rm erg} \ {\rm cm}^{-2} \ {\rm s}^{-1}}

\def\lsim{\mathrel{\hbox{\rlap{\lower.55ex \hbox {$\sim$}}\kern-.0em 
\raise.4ex \hbox{$<$}}}}  
\def\gsim{\mathrel{\hbox{\rlap{\lower.55ex \hbox {$\sim$}}\kern-.0em 
\raise.4ex \hbox{$>$}}}}

\def\arcmin{\hbox{$^\prime$}}
\def\arcsec{\hbox{$^{\prime\prime}$}}
\def\degs{$^\circ$}
\def\hardrange{2 -- 10 keV}
\def\softrange{0.5 -- 2 keV}
\def\hardfluxrange{$f_{\rm 2 - 10~keV}$}

\def\lognlogs{$\log N - \log S$}
\def\nh{$N_H$}
\def\lognh{$\log{N_H}$}
\def\lxlo{L_{\rm x}/L_{\rm o}}
\def\fxfo{\log(f_{\rm x}/f_{\rm o})}
\def\spose#1{\hbox to 0pt{#1\hss}}
\def\simlt{\mathrel{\spose{\lower 3pt\hbox{$\mathchar"218$}}
     \raise 2.0pt\hbox{$\mathchar"13C$}}}
\def\simgt{\mathrel{\spose{\lower 3pt\hbox{$\mathchar"218$}}
     \raise 2.0pt\hbox{$\mathchar"13E$}}}
\def\kms{km s$^{-1}$}

\def\htwo{\ion{H}{2}}

\def\lya{Ly$\alpha$}

\def\civ{\ion{C}{4} $\lambda$1549}

\def\ciii{\ion{C}{3}] $\lambda$1909}

\def\mgii{\ion{Mg}{2} $\lambda$2800}
\def\nev{[\ion{Ne}{5}] $\lambda \lambda$3346, 3426}
\def\nevsingle{[\ion{Ne}{5}] $\lambda$3426}
\def\oii{[\ion{O}{2}] $\lambda$3727}

\def\hten{H10 $\lambda$ 3798}
\def\hnine{H9 $\lambda$ 3835}
\def\neiii{[\ion{Ne}{3}] $\lambda$3869}
\def\hzeta{H$\zeta$ $\lambda$ 3889}
\def\hbeta{H$\beta$}
\def\oiii{[\ion{O}{3}] $\lambda$5007}
\def\oiiipair{[\ion{O}{3}] $\lambda \lambda$4959,5007}
\def\halpha{H$\alpha$}
\def\nii{[\ion{N}{2}] $\lambda$6583}
\def\cahk{CaHK $\lambda \lambda$3934, 3968}
\def\siiplus{\ion{S}{2} ($\lambda$6716 + $\lambda$6731)}

\begin{document} 

\title{The Serendipitous Extragalactic X-Ray Source \\ Identification
(SEXSI) Program. III. Optical Spectroscopy\altaffilmark{1}}

\author{Megan E. Eckart\altaffilmark{2}, 
Daniel Stern\altaffilmark{3}, 
David J. Helfand\altaffilmark{4}, \\
Fiona A. Harrison\altaffilmark{2},
Peter H. Mao\altaffilmark{2}$^,$\altaffilmark{5}, 
and Sarah A. Yost\altaffilmark{2}$^,$\altaffilmark{6}}

\begin{abstract}

We present the catalog of 477 spectra from the {\em Serendipitous Extragalactic
X-ray Source Identification (SEXSI)} program, a survey designed to 
probe the dominant contributors to the \hardrange~cosmic X-ray background. Our 
survey covers 1 deg$^2$ of sky to \hardrange~fluxes of $1 \times 10^{-14} 
\fluxu$, and 2 deg$^2$ for fluxes of $3 \times 10^{-14} \fluxu$.
Our spectra reach to $R$-band magnitudes of $\simlt 24$ and have produced 
identifications and redshifts for 438 hard X-ray sources.  Typical completeness 
levels in the 27 {\em Chandra} fields studied are 40 -- 70\%. The vast majority 
of the \hardrange-selected sample are active galactic nuclei (AGN) with 
redshifts between 0.1 and 3; our highest-redshift source lies at $z = 4.33$.
We find that few sources at $z < 1$ have high X-ray luminosities, reflecting a
dearth of high-mass, high-accretion-rate sources at low redshift, a result
consistent with other recent wide-area surveys.  We find that half of 
our sources show significant obscuration, with \nh$> 10^{22}$~cm$^{-2}$, 
independent of unobscured luminosity.  We classify 168 sources as emission-line
galaxies; all are X-ray luminous ($L_{\rm x} > 10^{41} \lumin$) objects with 
optical spectra lacking both high-ionization lines and evidence of a non-stellar
continuum. The redshift distribution of these emission-line galaxies peaks at a
significantly lower redshift than does that of the sources we spectroscopically
identify as AGN.
We conclude that few of these sources, even at the low-luminosity end, can be
powered by starburst activity. Stacking spectra for a subset of these sources in
a similar redshift range, we detect \nevsingle~emission, a clear signature of
AGN activity, confirming that the majority of these objects are Seyfert~2 
galaxies in which the high-ionization lines are diluted by stellar emission. We
find a total of 33 objects lacking broad lines in their optical spectra which
have quasar X-ray luminosities ($L_x>10^{44}$ erg s$^{-1}$), the largest sample
of such objects identified to date. In addition, we explore seventeen AGN 
associated with galaxy clusters and find that the cluster-member AGN sample has
a lower fraction of broad-line AGN than does the background sample.
\end{abstract}


\altaffiltext{1}{The majority of data presented herein
were obtained at the W.M. Keck Observatory, which is operated as a
scientific partnership among the California Institute of Technology,
the University of California and NASA. The Observatory was made possible
by the generous financial support of the W.M. Keck Foundation.}
\altaffiltext{2}{Space Radiation Laboratory, 220-47, California Institute
of Technology, Pasadena, CA 91125} 
\altaffiltext{3}{Jet Propulsion Laboratory, California Institute of
Technology, Mail Stop 169-506, Pasadena, CA 91109} 
\altaffiltext{4}{Columbia University, Department of Astronomy, 550 West
120th Street, New York, NY 10027} 
\altaffiltext{5}{University of California at Los Angeles, Department
of Earth and Space Sciences, 595 Charles Young Dr. East, 
Los Angeles, CA 90095}
\altaffiltext{6}{University of Michigan, Department of Physics, 
2464 Randall Laboratory, 450 Church St., Ann Arbor, MI 48109}

\section{Introduction}
\label{sec:intro}

A primary goal of extragalactic X-ray surveys is to determine the
nature and evolution of accretion power in the Universe. Accreting massive
black holes are observed over more than five orders of magnitude in
luminosity, and exhibit a broad range of intrinsic X-ray absorption
(from negligible levels to Compton-thick obscuration with
\nh $\simgt 10^{24}$~cm$^{-2}$). Additionally, cosmic X-ray sources
undergo significant evolution between
the current epoch and redshifts of $z \sim 3$.  Measuring this
enormous phase space requires broadband X-ray surveys extending from
essentially the whole sky (to constrain the bright end) to the deepest
surveys carried out with the most sensitive telescopes available over
sky regions comparable to the telescope field of view.

Enormous progress has been made at the faintest end over the last five
years with Mega-second surveys performed by {\em Chandra} and {\em XMM} 
\citep[see review by][]{Brandt:05}.  Together, these
surveys have covered more than a thousand square arcminutes to depths
of \hardfluxrange~$\simlt 10^{-15} \fluxu$. These projects have
resolved a significant fraction of the diffuse extragalactic X-ray
background 
\citep[at least in the lower half of the accessible energy band --][]{Worsley:05}. 
Spectroscopic optical followup has been
successful in classifying and measuring redshifts for a large fraction
(over half) of the resolved sources.

Also very important in covering the interesting phase space are
surveys with depths \hardfluxrange~$\simlt 10^{-14} \fluxu$. The slope
of the extragalactic X-ray \lognlogs~relation breaks at \hardfluxrange~$= 1 - 2
\times 10^{-14} \fluxu$  \citep{Cowie:02,Harrison:03}, so 
that sources in this flux range dominate the integrated light from
accretion.  In this brightness range, source densities on
the sky are a few hundred per square degree, requiring surveys covering
on the order of a square degree or more to obtain statistically useful
samples for the study of source properties and the evolution of the population.

A number of programs are surveying regions
of this size and depth, accompanied by significant optical
followup efforts. The CLASXS survey \citep{Yang:04}~obtained data
in a 0.4 deg$^2$ contiguous region in the Lockman Hole; optical
spectroscopy has identified about half of the sample of 525 objects \citep{Steffen:04}.  
The ChaMP \citep{Kim:03}~survey utilizes extragalactic {\em Chandra}
pointings largely from the guest observer (GO) program to identify
sources that are not associated with the primary
target.  ChaMP, which
ultimately aims to cover several square degrees over a range of depths, is
also accompanied by an optical source identification effort \citep{Green:04, Silverman:05}.  
The HELLAS2XMM survey \citep[e.g,][]{Baldi:02, Fiore:03, Perola:04}~is taking a
similar approach with fields from {\em XMM-Newton}.

The subject of this paper is the {\em Serendipitous Extragalactic X-ray Source Identification (SEXSI)} program, 
a survey using {\em Chandra} GO
and GTO fields specifically selected to obtain a significant sample of
identified objects in the flux range from a few times $10^{-13}$ to $10^{-15}
\fluxu$.  To accomplish this, SEXSI covers more than 2 deg$^{2}$ of
sky.  \citet[][hereafter Paper I]{Harrison:03}~describes the
X-ray source sample, \citet[][hereafter Paper II]{Eckart:05}
describes the optical imaging followup, and this paper presents
results of the optical spectroscopy. We have 477 spectra, of 
which 438 are of sufficient quality to provide redshifts
and optical classifications. The $L_{\rm x} - z$ phase
space covered by our survey is shown in Figure \ref{fig:lum_z}.

In our sample of 438 spectroscopically-identified sources (which
have counterpart magnitudes $R \simlt 24$) we confirm with high
significance a number of results found in other surveys.  We find that
few AGN at $z < 1$ have high rest-frame X-ray luminosities,
reflecting a dearth of high-mass, high-accretion-rate sources at low
redshift. In addition, our sample of broad-lined AGN peaks at a
significantly higher redshift ($z > 1$) than do sources 
identified as emission-line galaxies.  We find that 50\% of our sources
show significant obscuration, with $N_H > 10^{22}$~cm$^{-2}$,
independent of intrinsic luminosity.  We have identified nine narrow-lined AGN
at $z > 2$ having quasar luminosities ($L_{\rm x} > 10^{44} \lumin$).
This is consistent with predictions based
on unified AGN models.

We investigate in some detail the nature of the large sample of
168 sources classified as emission-line galaxies.  These X-ray
luminous ($10^{41} - 10^{44} \lumin$) galaxies have optical
spectra lacking both high-ionization lines and evidence for
a non-stellar continuum.  We conclude that few of these galaxies, even at
the low-luminosity end, can be powered by starburst activity.  By
stacking 21 spectra for sources in a similar redshift range in order
to increase the signal to noise, we detect \nevsingle~emission,
an unambiguous signature of AGN activity.  This suggests that the
majority of these sources are Seyfert~2 galaxies, where the
high-ionization lines are diluted by stellar emission and reduced in intensity
by nuclear extinction.

We organize the paper as follows: \S~\ref{sec:surveydesign} discusses
the overall design of the spectroscopic followup program; \S~\ref{sec:data} describes the
data collection and reduction; 
\S~\ref{sec:redshiftandclass} details how we determine
redshifts and source classifications; \S~\ref{sec:catalog} presents the catalog;
\S~\ref{sec:pop_stats} discusses the population statistics of the sample; 
\S~\ref{sec:sourceclasses} provides details on the characteristics of each
source class as well as the line-free spectra;
\S~\ref{sec:selection} discusses the sample completeness and selection effects; 
\S~\ref{sec:global_comp} presents the global characteristics of the sample and
provides a comparison to other surveys;
\S~\ref{sec:nature_of_elg} explores the nature of emission-line galaxies; 
\S~\ref{sec:clusters} provides a discussion 
of spectroscopically-identified AGN associated with
galaxy clusters; and \S~\ref{sec:summary} provides a summary. 
We adopt the standard cosmology throughout: 
$\Omega_{\rm m} = 0.3, ~\Omega_{\lambda}=0.7$, and $H_0 = 65$ km s$^{-1}$ Mpc$^{-1}$. 
Unless otherwise mentioned, error estimates and error bars refer to 
1 $\sigma$ errors calculated with Poissonian counting statistics.

\vspace{1cm}

\section{Survey Design}
\label{sec:surveydesign}

The SEXSI survey is designed to obtain optical identifications for a
large sample of hard (\hardrange) X-ray sources detected in extragalactic {\em
Chandra} fields in the flux range $10^{-13} - 10^{-15} \fluxu$. 
This range contains sources which are the dominant
contributors to the \hardrange~extragalactic background, filling the gap between
wide-area, shallow surveys ({\em e.g.,} HELLAS --
\citealt{LaFranca:02}; ASCA Large Sky Survey -- \citealt{Akiyama:00};
ASCA Medium Sensitivity Survey -- \citealt{Akiyama:03}) and the deep,
pencil-beam surveys ({\em e.g.,} CDF-N -- \citealt{Alexander:03,
Barger:03}; CDF-S -- \citealt{Rosati:02, Szokoly:04}).

Covering this phase space requires surveying 1 -- 2 deg$^2$ with
$\sim$ 50~ksec exposures.  SEXSI selected 27 archival, high Galactic
latitude fields ($|b| >20$\degs), covering a total survey area of
more than 2 deg$^2$ at $f_{\rm 2-10~{\rm keV}} \geq 3 \times
10^{-14} \fluxu$ and more than 1~deg$^2$ for $f_{2 - 10~{\rm keV}}
\geq 1 \times 10^{-14} \fluxu$.  To maximize sensitivity in the
hard band, we selected archival observations taken taken with the
Advanced Camera for Imaging Spectroscopy (ACIS I- and S-modes;
\citealt{Bautz:98}).  The exposure times range from 18 to 186 ks, with
three quarters of the fields having good-time integrations of between
40 and 100 ks.

Paper~I provides details of the X-ray source extraction and analysis;
we provide a brief summary here.  In each field we initially used
{\tt wavdetect} to identify sources in soft (0.3 -- 2.1 keV) and hard
(2.1 -- 7 keV) band images. In a subsequent step, we tested the
significance of each source and eliminated sources with nominal chance
occurrence probability $P> 10^{-6}$, which led to an average expected 
rate of $\leq 1$
false detections per field. We extracted photons from each
source, and used energy-weighted exposure maps to convert
background-subtracted source counts to fluxes in the standard soft
(\softrange) and hard (\hardrange) bands, adopting a power-law
spectral model with photon index $\Gamma=1.5$.  In addition, we
corrected the source fluxes for Galactic absorption.  We eliminated
all {\em Chandra} target objects from the sample, with the exception
of possible galaxy cluster members which we include in the catalog but
flag accordingly. The X-ray catalog contains 1034 hard-band-selected
sources. An additional catalog of 879 sources which
have soft-band detections but which lack hard-band detections is
presented in the Appendix of Paper I.

The SEXSI optical followup program is designed to maximize the
fraction of spectroscopically identified sources in the survey area.
We primarily used the MDM 2.4 m and the Palomar 60-inch and 200-inch telescopes 
for imaging, and
the Keck telescopes for spectroscopy. We image each field in the
$R$-band to minimum limiting magnitudes $R_{\rm limit} \sim 23$, a
depth chosen to match the typical limit where classifiable optical
spectra can be obtained in 1-hour integrations with Keck.  Since the
majority of sources in our X-ray flux range have optical
counterparts at this limit, this is a good tradeoff between areal
coverage and depth.

Paper~II describes the optical imaging and counterpart identification
in detail.  We iteratively matched the optical images to the X-ray
catalog, utilizing optical astrometry to correct the {\em Chandra}
pointing error for each field (typically these corrections are $\simlt
1$\arcsec).  For the 262 sources with imaging depths $22 < R_{\rm
limit} < 23$, 160 (61\%) have identified counterparts, while for the 434
sources with $23 < R_{\rm limit} < 24$, 291 (67\%) have identified
counterparts, and for the 167 sources with $R_{\rm limit}$ $>$ 24, 124
(74\%) have identified counterparts.  Our total sample of 947 sources
with unambiguous photometry (e.g., no contamination from nearby bright
stars, etc.) identifies 603 counterparts (64\%).  

Optical spectra of X-ray counterparts were primarily obtained using multi-slit
spectrographs at the W.M. Keck Observatory: the Low Resolution Imaging
Spectrometer \citep[LRIS;][]{Oke:95}~on Keck I and the Deep Extragalactic
Imaging Multi-Object Spectrograph \citep[DEIMOS;][]{Faber:03} on Keck II.  Our
basic slit mask design strategy was to place slits on all identified
\hardrange\ SEXSI sources with counterpart magnitudes $R \simlt 23$
(or, occasionally, from imaging in other bands when $R$-band images
were not yet available). For the majority of the masks made for LRIS,
these sources received the highest priority; we then filled any extra
space on the mask with sources from our soft-band-only catalog, and
then with fainter optical counterparts. The soft-band-only spectra are not
included in this paper. For DEIMOS masks we followed basically the same procedure.
However, DEIMOS's large field-of-view affords space to place ``blind''
slits at the X-ray positions of hard-band sources which lack optical
identifications.

Table~\ref{tbl:fields}~provides a summary of the SEXSI fields and our
spectroscopic completeness for each. Note that the optical photometric
identification completeness should be taken into account when gauging
spectroscopic completeness -- most sources for which we have found
either very faint optical counterparts or only a limiting
magnitude were not pursued spectroscopically.

\section{Optical Spectroscopy: Data Collection and Reduction}
\label{sec:data}

Although the majority of the SEXSI spectroscopy was obtained
using LRIS on Keck I and DEIMOS on Keck II, 
a small fraction ($\sim 2$\%) of the spectra
were collected with Doublespec \citep{Oke:82} at the Palomar 200 inch (5 m)
telescope. Below we describe the data collection and 
reduction techniques for each of the three instruments.
A small subset (19 sources) of the spectra were previously published in 
\citet{Stern:02a,Stern:02b}~and \citet{Stern:03}.
Section \ref{sec:selection}~addresses
the composition of the sample obtained from each instrument and
possible sample biases that might occur as a consequence of the differing
capabilities of the spectrographs; we show any such effects are
small in our final sample.

\subsection{LRIS Data}
The 293 LRIS spectra included in our catalog
were collected between September 2000 and June 2002.
LRIS has a $5.5\arcmin \times 8.0\arcmin$ field of view that we 
typically filled with 5 -- 20 slitlets. 
Our aim was followup of $R\simlt 23$ SEXSI sources.  Exposures of 1 -- 2 hours
provided sufficient signal to determine redshifts and perform classifications
for most such objects (see \S~\ref{sec:redshiftandclass}). The SEXSI source 
density varies with {\em Chandra} exposure time and
off-axis angle, leading to a large range in slitlets per mask. 
The masks were machined with 1.4\arcsec\ wide slitlets.

LRIS is a dual-beam spectrograph, with simultaneous blue (LRIS-B) and
red (LRIS-R) arms. 
LRIS-R has a $2048 \times 2048$ detector with 0.212\arcsec\ pixel$^{-1}$. 
From September 2000 to early June 2002, LRIS-B had a $2048 \times 2048$ pixel
engineering-grade CCD with a similar platescale to the red-side. 
In June 2002, prior to our final LRIS observing run, 
the CCD was replaced by a science-grade mosaic of two $2048 \times 4096$
CCDs with 0.135\arcsec\ pixel$^{-1}$. The new CCDs were 
selected to have high near-UV and blue quantum efficiency. 
\citet{Steidel:04}~provides a more detailed description of the new LRIS-B. 

We used the 300 lines mm$^{-1}$ ($\lambda_{\rm blaze} = 5000$ \AA) grism 
for blue-side observations providing a dispersion of 2.64 \AA\ pixel$^{-1}$ 
pre-upgrade and 1.43 \AA\ pixel$^{-1}$ post-upgrade. 
For red-side observations we employed either the 
150 lines mm$^{-1}$ ($\lambda_{\rm blaze} = 7500$ \AA) 
grating providing a dispersion of 4.8 \AA\ pixel$^{-1}$
or the 400 lines mm$^{-1}$ ($\lambda_{\rm blaze} =8500$) grating 
providing 1.86 \AA\ pixel$^{-1}$. 
In cases where only LRIS-R was available, 
we used the 150 lines mm$^{-1}$ grating. 
The 400 lines mm$^{-1}$ 
grating was only employed when we were using both arms of the spectrograph. 
We typically split the red and blue channels at 5600 \AA\, though 
occasionally the 6800 \AA\ dichroic was used.
These spectrometer configurations provide wavelength coverage across most of
the optical window. The wavelength window for 
each individual spectrum is included in the catalog (\S~\ref{sec:catalog}), 
since coverage depends on the source position on the slitmask and the particular
setup parameters.

The majority of the LRIS observations (227 sources) 
used both arms of the spectrograph with the 400 lines mm$^{-1}$ grating, while 
51 of the earliest LRIS spectra used only LRIS-R.  A
final 19 spectra used a dichroic, but have only blue-side (4 sources)
or red-side (15 sources) coverage due to technical problems during the observations.

Most of our LRIS masks were observed for a total integration time of 1 -- 1.5
hr, usually consisting of three consecutive exposures.
Between exposures we dithered $\sim 3$\arcsec~along the slit
in order to facilitate
removal of fringing at long wavelengths ($\lambda \simgt 7200$ \AA).
The LRIS data reductions were performed using 
IRAF\footnote{IRAF is distributed by the National Optical Astronomy Observatory, 
which is operated by the Association of the Universities for Research in 
Astronomy, Inc., under cooperative agreement with the National Science 
Foundation.}
and followed standard
slit-spectroscopy procedures.  Some aspects of treating the slit mask
data were facilitated by a home-grown software package, 
BOGUS\footnote{BOGUS is available online at \\
http://zwolfkinder.jpl.nasa.gov/$\sim$stern/homepage/bogus.html}, 
created by D. Stern, A.J. Bunker, and S.A. Stanford.
We calculated the pixel-to-wavelength
transformation using Hg, Ne, Ar, and Kr arc lamps and employed telluric
emission lines to adjust the wavelength zero point.  The spectra on
photometric nights were flux-calibrated using long-slit observations of standard
stars from \citet{Massey:90} taken with the same configuration as the 
multislit observations.

\subsection{DEIMOS Data}
Our 163 DEIMOS spectra were collected over three nights in August 
2003\footnote{A serendipitous galaxy at $z = 6.545$ in SEXSI field 
MS 1621 was also
identified during the DEIMOS observing run.  This was the third most
distant object known at the time of the discovery, and had interesting
implications for the ionization history of the 
Universe \citep{Stern:05}.}.
The DEIMOS field of view is $4\arcmin \times 16\arcmin$,  
approximately four times that of LRIS, allowing more slitlets per 
mask. Observations used the 600 lines mm$^{-1}$ grating blazed at 7500 \AA\ with
the GG455 order-blocking filter, eliminating flux below 4550 \AA.
With this setup, the spectrograph afforded spectral coverage
from roughly 4600 \AA\ to 1 $\mu$m, covering most of the optical 
window, though the blue-side sensitivity does not extend as far 
into the near-UV as does LRIS-B.
The observations of each mask were broken into three exposures
of 1200 s to allow rejection of cosmic rays; 
no dithering was performed between exposures to allow for
easy adoption of the pipeline reduction software (see below).

Calibration data consisting of three internal quartz flats 
and an arc lamp spectrum (Xe, Hg, Ne, Cd, and Zn) were obtained
for each mask during the afternoon. The DEIMOS flexure 
compensation system ensures that the calibration images match 
the science data to better than $\pm 0.25$ pixels.

The DEIMOS data reduction was performed using the 
automated pipeline developed by the DEEP2 Redshift Survey
Team \citep{Newman:05}. 
Minor adjustments to the code were needed to process data from slit masks
with too few slitlets or a slitlet that was too long for the original code.
These changes were performed by both the authors and
DEEP2 team members
M. Cooper and J. Newman.
The pipeline follows standard slit spectroscopy reduction procedures, 
performing all steps up to and including extraction and wavelength 
calibration.

\subsection{Doublespec Data}

While it is impractical to use the Palomar 200 inch (5 m) telescope for spectroscopy of sources fainter 
than $R=21$, Doublespec was used in long slit mode 
for brighter sources that did not fit well onto Keck slit
masks. Doublespec is a 
dual-beam spectrograph; 
we used the 600 lines mm$^{-1}$ ($\lambda_{\rm blaze} = 3780$ \AA) grating 
for blue-side observations (1.07 \AA\ pixel$^{-1}$), the 158 lines mm$^{-1}$ ($\lambda_{\rm blaze} = 7560$ \AA) grating for 
red-side observations (4.8 \AA\ pixel$^{-1}$), and the
5200 \AA~dichroic, which provided coverage of most of the optical window.
Most of our Doublspec observations were performed for a total integration time of 30 min, usually consisting of three consecutive exposures.
Between exposures we dithered along the slit
in order to facilitate
removal of fringing at long wavelengths.
 Our small
number of Doublespec spectra (ten) were reduced
using standard IRAF slit-spectroscopy procedures.

\section{Redshift Determination and Source Classification}
\label{sec:redshiftandclass}

From the 477 spectra collected, we have obtained spectroscopic redshifts for 438
of the 1034 \hardrange\ sources from Paper I.  We do not include
spectroscopic followup of any of the soft-only sources 
presented in the Appendix of Paper I, as the goal of our program has always been
to focus our telescope and analysis time on the hard-band populations.

To obtain source redshifts, we measure the observed line centers and average the
corresponding redshifts. When possible, we avoid using broad lines in 
determining source redshifts; in particular, we exclude lines such as \civ~that
are known to be systematically blueshifted from the object's systemic redshift
\citep{VandenBerk:01}.  When possible we measure the narrow oxygen
lines, \oii~or \oiii, although determining source redshifts to $<1$\% is not
essential for our scientific goals. When our emission or absorption line
identification is tentative, we flag the source in the catalog. This occurs
in $\sim 5$\% of the cases -- typically faint sources which lack bright, high-ionization
lines.

When a source has a reasonable signal yet lacks identifiable spectral features,
we include it in the source catalog and document the wavelength range
observed. These sources mainly show faint, power-law-like continua, 
although in a few cases the signal-to-noise is quite high.  Sources so faint that the 
continuum is not clearly detected are excluded from the catalog.

In addition to determining redshifts, our spectroscopic data 
allows us to group the sources into broad classes based on their
spectral features. This classification is independent of the sources'
X-ray properties. The broad goal of this classification
is to separate sources that {\em appear} to have normal galaxy spectra
from those that exhibit features characteristic of an active
nucleus -- high-ionization lines that are either broad or narrow.
In detail, our spectral classification is as follows:

\begin{itemize}
\item{\bf Broad-Lined AGN (BLAGN):}   
We classify sources as BLAGN if they have 
broad (FWHM $\simgt 2000$ \kms) emission lines such 
as \lya, \civ, \ciii, \mgii, \nev, \hbeta, \halpha.
These sources include Type 1 Seyferts and QSOs, which
in the unified theory~\citep{Antonucci:93}~are objects viewed with the obscuring
torus face on and the central nuclear region unobscured. An example of a 
typical BLAGN optical spectrum is shown in Figure~\ref{fig:blagn}.

\item{\bf Narrow-Lined AGN (NLAGN):}   We classify sources as NLAGN if they have
high-ionization emission lines similar to those seen in BLAGN, but 
with FWHM $\simlt 2000$ \kms. 
Typical high-ionization lines indicating the presence of an AGN are \civ, \ciii, and \nevsingle. 
Low-ionization lines such as \lya, \mgii, \hbeta, \halpha, etc., will usually also be present 
given appropriate wavelength coverage, but are not alone sufficient 
to classify a source as a NLAGN. Figure~\ref{fig:nlagn}~provides three examples.
These sources are the obscured AGN in the unified model~\citep{Antonucci:93}, 
viewed edge-on with an obscured view of the nucleus. 
In earlier studies of the lower-$z$ universe, line ratios such as 
\oiii/\hbeta, \nii/\halpha, etc., have been used to differentiate
spectra that show narrow lines due to ionization by hot stars 
from spectra that show narrow lines due to an active nucleus 
\citep[e.g.,][]{Veilleux:87}.
We do not measure such line ratios or apply them in our classification. 
Our sources span a large range in redshift and most are faint in 
the optical. Thus the emitted-frame spectral coverage varies greatly
from source to source, and our spectral and spatial resolutions are too low to
deblend and measure ratios accurately. There may be a handful of
sources classifed as ELG (see below) that could be re-classifed 
as AGN-dominated based solely on their line ratios in our data, but this number
of sources is expected to be small ($<10$).

\item{\bf Emission-Line Galaxies (ELG):} Extragalactic sources
with narrow emission lines, but with no obvious AGN features in their optical spectra (e.g., high-ionization and/or broad lines) are classified as
ELG. The emission lines in these spectra indicate that the ionization 
mechanism dominating the {\em optical} light we receive is from hot 
stars, not from a hard, power-law source.
This classification does not rule out the presence of an underlying 
active nucleus; indeed, we believe the X-ray emission from the vast majority of this
subsample does arise from AGN activity.
Figure~\ref{fig:elg} shows two example
ELG spectra. These objects typically exhibit narrow lines such as \oii, \hbeta,
and \oiiipair, and often have narrow \neiii~emission, \cahk~absorption, and the continuum
break at 4000 \AA\ (D4000). Narrow \nevsingle~and other high 
ionization lines are not detected in our ELG spectra -- sources with such
lines are classified as NLAGN.

\item{\bf Absorption-Line Galaxies (ALG):} We distinguish between 
galaxies showing emission lines (ELG) and
early-type galaxies (ALG), where the latter have continua marked only by
absorption features, notably the D4000 continuum break
and the \cahk~absorption lines. Figure~\ref{fig:alg}~shows an
example spectrum.

\item{\bf Stars:} SEXSI fields are selected to be at high Galactic 
latitude to avoid contaminating our extragalactic sample with 
 \hardrange~emitting stars, but we do identify a small
number of Galactic stars as optical counterparts (at $z=0$). Seven of the 
optically-bright SEXSI sources are identified in the literature as 
stars (Paper II); the other sources so identified are from our 
spectroscopy of fainter sources.
Section~\ref{sec:stars} discusses the bright 
stars in more detail and assesses the possibility that the optically-fainter
objects are chance coincidences.

\end{itemize}

As mentioned above, these classifications depend only on the optical
spectroscopic appearance, not on X-ray properties such as luminosity or
intrinsic obscuring column density.
With the exception of only a few sources, the identified extragalactic SEXSI 
sources have X-ray luminosities which suggest the presence of an accreting supermassive black hole.
The emission-line galaxies, which are prevalent in our 
sample, do not show any optical indication of emission lines
from atoms ionized by an X-ray source with copious hard X-ray emission. 
Instead, the ELG lines are typical of normal galaxies with 
lines from atoms excited by at best moderately energetic photons  
that can be produced thermally by the hottest
stars. 
This apparent discrepancy in the optical and X-ray source properties 
is discussed further in \S~\ref{sec:nature_of_elg}.

\section{The Catalog}
\label{sec:catalog}

In Table~\ref{tbl:catalog} we present the catalog of 477 
hard-band SEXSI
sources with optical spectroscopic data; the catalog is also available
in machine-readable format in the online version of the {\em
Astrophysical Journal}.  Complete X-ray data and optical photometry
for these sources are presented in Papers I and II, respectively. 
The remaining $\sim550$ unidentified sources 
from the complete sample of 1034 hard-band SEXSI sources do not have
optical spectroscopic data and are omitted from this catalog. The 
first six columns present the X-ray source data, while the following 
columns present optical counterpart information -- photometric data followed 
by spectroscopic data. The final columns describe the X-ray 
luminosity and the column density -- quantities determined by 
combining the X-ray data with the redshift.

Column 1 presents source names, designated by ``CXOSEXSI'' followed by 
standard truncated source coordinates. X-ray source positions, 
$\alpha_{\rm x}$ and $\delta_{\rm x}$, 
corrected for the mean X-ray-to-optical offsets to eliminate 
{\em Chandra} pointing errors, are shown in columns 2 -- 3.  Column 4 
lists the off-axis angle (OAA, the angular distance in 
arcmin of the source position from the telescope aim point). 
The~\hardrange\ flux (in units of $10^{-15} \fluxu$), converted from counts 
assuming $\Gamma=1.5$ and corrected for Galactic absorption, is presented in 
column 5, while column 6 gives the source hardness ratio, 
$HR=(H-S)/(H+S)$, where $H$ and $S$ are photons cm$^{-2}$ s$^{-1}$ in 
the \hardrange\ and \softrange~bands, respectively. Here, as in Paper II, we 
quote hardness ratios derived from the net soft X-ray counts 
recorded at the hard-band source position when there was not a 
significant soft-band source detected (as distinct from Paper I, in 
which these cases were reported as $HR=1$). In addition, for a subset of 
these cases, when the soft-band counts recorded at the hard-band 
position were less than twice the soft-band background counts, the $HR$ 
is considered a lower-limit, is flagged as such in the catalog, and is set to 
$HR=(H-S_{\rm limit})/(H+S_{\rm limit})$, where $S_{\rm limit} = 2 
~\times$ soft-band background counts corrected by the exposure map.

The data describing the optical counterparts begin in column 7,
with the photometric optical counterpart flag (optflag), a code essential for
interpreting the optical photometric data (see Table~\ref{tbl:catalog}~footnote). 
Note that in this spectroscopic catalog the majority of entries in this
column are optflag = 1, indicating a solid optical identification, since
a successful spectroscopic identification depends on having a  
counterpart bright enough that emission lines or absorption features are 
detectable. In a handful of cases the optical flag is a 2, indicating a
limiting magnitude. These cases occur either when we 
placed a slitlet on an optical counterpart identified in a band other than
$R$, or when a slitlet was placed `blindly' at the X-ray source
position even though no optical counterpart was present to the depth of our imaging data.

Column 8 is the $R$-band magnitude of the optical counterpart; in the
next column we show the limiting $R$ magnitude for the image from which
the optical counterpart, or a limit thereto, was derived. Next we
present the logarithmic X-ray-to-optical flux ratio (column 10), given
by the relation \begin{equation} \log{(f_{\rm x}/f_{\rm o})} =
\log{f_{2-10~{\rm keV}}} + (R/2.5) + 5.50, \end{equation} derived using the
Kron-Cousins $R$-band filter transmission function (see Paper II and
references therein).  Note that special attention must be paid to the 
optical flag when interpreting the data of columns 8 -- 10. For 
example, if the code is a 2 (optical counterpart not detected), then 
columns 8 and 10 describe limits on each quantity.

The optical spectroscopic data begins with the redshift in column 11,
followed by the classification in column 12 (see
\S~\ref{sec:redshiftandclass}~ for details of the redshift and classification
determinations). Column 13 provides notes on individual sources where 
necessary. 

The rest-frame wavelength range in angstroms for each optical spectrum  
is recorded in column 14; when a source spectrum shows 
continuum only and no redshift has been 
determined, we present the observed-frame wavelength range. This column is
essential for determining which spectral features are accessible for a given 
source. It is important to know, for example, whether the absence
of high-ionization lines (typical of AGN and necessary for an AGN
classification) is a consequence of non-detection in the measured
spectrum or as the result of inadequate spectral coverage. In most cases both
the blue and 
red sides of the spectrographs were employed; if this is the case and there is 
a significant spectral coverage gap ($>100$ \AA) between the red and blue
sides, a note is added in column 13.

Column 15 presents the log of the absorbed rest-frame \hardrange~X-ray 
luminosity in
erg s$^{-1}$, calculated from the hard-band flux (column 5) and the redshift
assuming an X-ray photon index of $\Gamma=1.5$. 

The final three columns (16--18) present the log of the neutral hydrogen
column density ($N_H$) in cm$^{-2}$ and the associated 1 $\sigma$ lower- 
and upper-limit to the value. These values are determined by X-ray spectral
fitting of each source using XSPEC\footnote{Available at 
http://heasarc.gsfc.nasa.gov/webspec/webspec.html}, a spectral
fitting program.  
Since many SEXSI sources have a low number of counts in the X-ray
we did not allow a many-free-parameter fit; instead, we performed the
fits assuming an intrinsic power-law spectrum with 
photon index $\Gamma=1.9$ typical of AGN continua.
We fixed $\Gamma$ and the Galactic $N_H$ at $z=0$ and allowed only the intrinsic 
column density at the source redshift to vary. 
The Galactic $N_H$ value used for each field is given in Paper I, Table 2. 
The photo-electric absorption was determined using 
Wisconsin cross-sections \citep{Morrison:83}.
The fitting was performed using C-statistic minimization instead of chi-squared minimization
since the observed data bins have few counts 
\citep[][and the XSPEC Manual, Appendix B]{Cash:79}. The fits use data
from 0.3 -- 7.0 keV, to match the energy range we used to extract
counts in Paper I.
The X-ray spectral data analysis was aided greatly by 
{\tt acis\_extract}\footnote{Available at http://www.astro.psu.edu/xray/docs/TARA/ae\_users\_guide.html.} Version 3.91 \citep{Broos:02}, 
software written in IDL
that assists in performing the many CIAO and XSPEC tasks involved in 
analyzing the spectra of large numbers of sources observed with ACIS.
For the spectral analysis presented here we use CIAO Version 3.2 and
CALDB Version 3.1.

Spectra were extracted for each source with a spectroscopic redshift
(excluding stellar sources at $z=0$). 
We choose 1.4967 keV as the primary
PSF energy at which the PSF fraction is to be computed, and a PSF fraction
of 0.8. Individual auxiliary response files (ARF's) and 
redistribution matrix files (RMF's) were calculated for each source.

We extracted a background spectrum for each source from
a local circular background region that includes at least 100 counts 
and an exposure ratio between background and source region of at least four, 
taking care to mask out all X-ray sources (SEXSI sources, including 
soft-only sources, target point sources, and extended cluster emission).
The background spectra were scaled based on the ratio of total exposure in 
the source extraction region to that of the background region.

For ACIS-I observations, all sources on all chips were fit; for 
ACIS-S observations, only sources on chips 6-8 were fit since PSF libraries
do not exist for chips 2-3 given an ACIS-S pointing. Fourteen SEXSI sources
fall in this category and are marked in catalog;  
the $N_H$ values reported for them are derived instead using
hardness ratios:
we determine what column density is necessary, given the source
redshift, to produce the measured hardness ratio assuming an
underlying intrinsic power-law spectrum with photon index $\Gamma=1.9$.
We use WebPIMMS\footnote{Available at 
http://heasarc.gsfc.nasa.gov/Tools/w3pimms.html} for this calculation. 

Throughout the paper, the luminosities we refer to are the obscured
(observed) luminosities presented in column 15, unless we specifically 
indicate that we are using intrinsic, unobscured
luminosities, corrected for absorption by the obscuring column density
at the source redshift. To calculate
the unobscured luminosity we multiply $L_x$ in column 15 by an approximate
correction factor calculated using WebPIMMS, assuming an 
intrinsic $\Gamma=1.9$, the best-fit \nh~value, and the source
redshift.

We now discuss the 23 sources flagged 
with an `A': sources whose optical spectra are from potential counterparts
which lie just outside the formal matching area. In the photometry catalog 
from Paper II, these
sources were listed as having limiting $R$-magnitudes (optflag=2)
since there were no optical counterparts in the search area. 
We include the spectroscopic information for these sources
in the main catalog here (Table~\ref{tbl:catalog}) and present 
the new photometric and astrometric information in 
Table~{\ref{tbl:newrmags}. This Table lists the X-ray 
source position as well as the new $R$-magnitude and its 
error and the offset between the X-ray and 
optical position (in
R.A. ($\Delta \alpha = \alpha_{x} - \alpha_{o}$), 
Dec. ($\Delta \delta = \delta_{x} - \delta_{o}$), 
and total position error 
($\Delta r = \sqrt{\Delta \alpha^2 + \Delta \delta^2}$)).
Column (9) gives the original X-ray search radius
(which depends on {\em Chandra} OAA), and the final 
column presents~$\fxfo$.  

Of these 23 sources, 5 are BLAGN, 1 is a NLAGN, 12 are ELGs,  
2 are ALGs, and 3 are stars. Given the low surface density of AGN 
on the sky, we assume that all of the six sources with
with spectra indicative of AGN activity are true counterparts. 
Comparing the BLAGN and NLAGN to the ELGs, we see that $6/212 = 3$\% of
the active galaxies are outside the search radius (as expected for radii
defined as approximately $2\sigma$ error radii), while $12/168=7$\% of the ELG 
are outside. The surface density of ELGs is also much higher than that of AGN,
and thus it is likely that some of the 12 
ELG are chance coincidences. The numbers for stars ($3/19=16$\%) and ALG 
($2/8=25$\%) are even higher; these objects are even more likely to 
be chance coincidences (see also \S~\ref{sec:sourceclasses}).

The number of false matches due to an optical source randomly overlapping
the matching search area depends on the depth of both the X-ray and
optical images as well as the size of the match area, which is determined
by the X-ray source off-axis angle. In Paper II, Section 3 we estimated the 
number of false matches considering the probability that an optical source would
overlap the X-ray match area; now that we have our set of identifications, we 
can estimate the number of false matches on a source-by source basis, taking into
account the particular match areas and optical magnitudes of the
identified sources in each class as well as the optical source density of
that class.

The optical sources that dominate in our high-Galactic latitude fields are
normal galaxies, sources that would be classified spectroscopically as ELG
or ALG. The false match rate will be lower than that estimated in Paper II
for several reasons, most importantly that the $R$-magnitude of many of
the spectroscopically identified sources are considerably brighter than the 
limiting magnitudes of the optical images. In addition, since {\em Chandra} 
is more sensitive closer to the aim point, the source density is highest for
low-OAA sources with the smaller match areas, and spectroscopic observing
efficiency favors these high-space-density sources.  Taking into account
each ELG and ALG match area and $R$-magnitude, we calculate the
false match probability for each source. Considering the 176 ELG and ALG
we estimate a total of $\simlt 4$ false ELG or ALG matches. The surface
density of (optically-identifiable) active galaxies is $\sim 100 \times$
lower than the normal-galaxy surface density for $R \simlt 23$ in optical
surveys \citep[e.g.,][]{Wolf:03}, thus we estimate that there will be $<1$
false NLAGN or BLAGN match. These estimates support our assertion that all
of the BLAGN and NLAGN from Table \ref{tbl:newrmags}~are true
counterparts.

\section{Hard X-ray Source Population Statistics}
\label{sec:pop_stats}

\subsection{Redshift distribution}
\label{sec:z_dist}

Figure~\ref{fig:z_hist}~shows the redshift distribution for the 419
spectroscopically-identified extragalactic sources (the 19
stars are excluded). 
The top panel (a) presents the entire 
distribution with the optical spectroscopic classifications
indicated in different shades. Panels (b) -- (e) present the 
same $z$-distribution with a different source class shaded in 
each panel to highlight their very different redshift distributions.

These plots show that the~\hardrange\ sources 
are dominated by two classes: BLAGN which show the typical 
broad, high-ionization-line signature indicative of gas near a 
source emitting copious hard X-rays, and ELGs,  a class of sources 
that show only emission lines typical of normal galaxies. 

The BLAGN population (shaded in panel (b)) exhibits a broad redshift
distribution; it includes 50\% of the identified sources and has a mean 
redshift $\langle z \rangle_{\rm BLAGN} = 1.46 \pm 0.75$, with objects ranging 
from $z=0.06$ to our highest redshift source at $z=4.33$. 

The next panel (c) shows the emission-line galaxies, which 
comprise 40\% of the identified sources. This redshift 
distribution is distinct from that of the BLAGN, with a much 
lower average redshift as well as a much tighter distribution 
about the average: $\langle z \rangle_{\rm ELG} = 0.75 \pm 0.36$. Note that
there are no ELGs found above $z=1.56$, and only three are above $z=1.4$ (two of
which are flagged as tentative identifications). 
There are two essential facts to keep in mind when discussing
the ELG redshift distribution. First, although the optical 
spectra of these sources are identified as a result of emission from the 
host galaxy, the~\hardrange~X-ray luminosities are, with the exception of 
eleven sources, above $10^{42} \lumin$, which is too high to be produced purely
from stars and stellar remnants in a star-forming galaxy (see \S~\ref{sec:starformation}). 
Second, the decline in sources above $z \sim 0.8$ does 
not necessarily represent the underlying distribution of hard X-ray emitting
ELG in the universe, but is more likely a result of observational biases. 
There are several strong selection effects that dominate at higher
redshift, most importantly the fact that as a typical galaxy is seen at
greater distances, it gets too faint for us to identify optically
(see \S~\ref{sec:selection}); the 
redshift desert -- the absence of strong emission lines within
the typical wavelength regime covered -- also militates 
against finding
ELGs above $z \approx 1.4$. Figure \ref{fig:r_z}~presents a scatter plot of 
$R$-magnitude versus redshift, showing the rapid rise in ELG (triangles) 
$R$-magnitude as $z$ increases as compared to, for example, the broad spread
in $R$ over a large $z$ range exhibited by the BLAGN (black crosses).

Panel (d) of Figure~\ref{fig:z_hist}~displays the redshift range for the 32 NLAGN, sources that show 
narrow high-ionization lines indicative of an obscured active nucleus. 
These sources comprise only $\sim 8$\% of the total identified 
sources. The redshift distribution is broad, with
 $\langle z \rangle_{\rm NLAGN} = 1.41 \pm 1.01$, similar to the 
BLAGN distribution. 

The final category of sources are the absorption line galaxies (ALG)
that make up only a very small fraction of our sample. These, like
the ELG, are found primarily at $z<1$ ($\langle z \rangle_{\rm ALG} = 0.64 \pm 0.23$), 
with none at $z>1.2$. Selection effects similar to those
that affect the identification of ELGs also constrain this distribution. 

\subsection{X-ray flux and X-ray-to-optical flux ratio distributions}

Figure \ref{fig:hflux_hist}~shows the \hardrange\ flux distribution 
for spectroscopically-identified extragalactic sources with $R<22$ (top panel) 
and $R>22$ (bottom panel); source classes are indicated by shading. 
Not surprisingly, the distribution of optically-fainter sources has a lower 
average \hardfluxrange~than does the distribution of 
$R<22$ sources. This effect has been seen previously in 
many of the {\em Chandra} and {\em XMM-Newton} surveys 
\citep[e.g.,][]{Hornschemeier:01}. At 
all but the faintest fluxes the optically-brighter distribution 
is dominated by BLAGN (black), while the optically-fainter distribution 
is dominated by ELG (dark gray). The relationship between 
$R$-magnitude distribution, redshift, and source class 
will be explored in detail in \S~\ref{sec:non_stellar_emission}. 

A related distribution is the $\fxfo$ distribution for identified 
extragalactic SEXSI sources presented in 
Figure \ref{fig:fxfo_hist}.  Again, source classes
are indicated by shading. The large majority of 
SEXSI sources are found with $-1 < \fxfo < 1$, 
typical values for AGN. The higher values of $\fxfo$
generally indicate more obscuration -- 
the \hardrange~light is relatively unaffected by the obscuring material
while the AGN optical light is absorbed. The highest $\fxfo$ sources
in the identified sample are dominated by ELG.

\subsection{X-ray luminosity distribution}
\label{sec:luminosity}

The majority of the spectroscopically-identified SEXSI sources have 
\hardrange~luminosities between $10^{43}$ and $10^{45} \lumin$; 
all but twelve sources (eleven ELGs, one NLAGN) have $L_{\rm x} > 10^{42} \lumin$. 
Figure \ref{fig:lum_hist}~shows the (a) \hardrange~and
(b) \softrange~luminosity distributions. Since SEXSI is a purely
hard-band-selected sample, the top panel includes all identified
sources, while the bottom panel displays the subset of these
sources having significant soft-band {\em Chandra} detections as well as
a dashed histogram showing upper limits to sources
undetected below 2~keV. In addition, 
Figure \ref{fig:lum_hr}~shows a scatter plot of $HR$ versus 
\hardrange~luminosity. 

The BLAGN tend to be softer sources -- they mainly fall below $HR\sim 0.2$ in
Figure \ref{fig:lum_hr} -- and thus most have significant soft detections and 
are included in black in both panels of Figure \ref{fig:lum_hist}. The BLAGN 
dominate the higher-luminosity end 
of the \hardrange~distribution, while they comprise almost the entirety of the 
high-luminosity end of the \softrange~distribution. 
The other contribution to the high-luminosity hard-band sources is from
the NLAGN. These tend to be harder (probably obscured) sources; thus, their 
corresponding \softrange~luminosities are lower (four have
no soft-band detections). 

The ELG dominate the \hardrange~luminosity distribution 
below $\sim 10^{43.5} \lumin$ where the BLAGN distribution
falls away and the ELG numbers rise. The ELGs have a broad range of 
$HRs$; thus,
there are many ELG (26) without significant soft detections. 
The limit to these detections are shown in the dashed histogram. 

Although they are few in number, the ALG we do detect all have \hardrange~
luminosities between $10^{42.5}$ and $10^{44} \lumin$, indicating all
likely host hidden AGN. A subset of our 
ELG plus all of our ALG represent the population of
sources dubbed X-ray Bright, Optically Normal Galaxies 
\citep[XBONGs; see][and references within]{Comastri:02}, sources 
that show X-ray luminosities indicative of AGN activity but which 
lack optical spectroscopic indication of the underlying AGN.\footnote{We
refrain from using the term ``XBONG'' further in this paper because
it lacks a clear definition; e.g., does XBONG refer to all X-ray luminous
sources that lack optical AGN lines or {\em only} such sources
that have optical spectra dominated by 
stellar absorption features?}

\subsection{Absorbing column density distribution}

Figure~\ref{fig:nhhist}~presents the \nh~distribution for 
BLAGN, ELG, and NLAGN. ALG are omitted from this plot owing to 
their small number (see \S~\ref{sec:alg}). The BLAGN (top panel) are distributed
broadly in \nh, with many more unobscured ($N_H < 10^{22}$ cm$^{-2}$) 
sources than for either 
the ELG or NLAGN populations. A total of 60 (29\%) of the 
BLAGN have \nh $>10^{22}$ cm$^{-2}$, while 149 (71\%) have
\nh $<10^{22}$ cm$^{-2}$. 
Note that inclusion in our
sample requires a significant hard-band X-ray detection --
we expect that many of the 879 soft-only sources presented
in Table 6 of Paper I will be unobscured BLAGN, 
and thus the population statistics and \nh~distributions will be
very different for a sample selected in the full-band (0.5 -- 8 keV). 
The middle panel of Figure~\ref{fig:nhhist}~presents 
the ELG. The majority of these sources have \nh $>10^{22}$ cm$^{-2}$
 -- only 27\% (43 sources) have \nh $<10^{22}$ cm$^{-2}$.
The NLAGN sources appear even more obscured than the ELG. Of
the 32 NLAGN only six (19\%) have \nh $<10^{22}$ cm$^{-2}$,
with more than half the sample having \nh $>10^{23}$ cm$^{-2}$.

\section{Source Classes}
\label{sec:sourceclasses}

\subsection{Broad-lined AGN}
\label{sec:blagn}

The BLAGN are the most common source type in our spectroscopically-identified   
sample, comprising over half of all sources; they are by far the dominant source
type above $z \simgt 1.4$. 
At the highest luminosities, these are (type 1) quasars; the lower
luminosity sources are Seyfert 1 galaxies. 
They are also the easiest type of source to identify 
over a broad redshift range; their optical counterparts tend to be bright 
and they have easily-detectable broad emission lines. Table 
\ref{tbl:lines}~gives a list of commonly detected 
lines and the percentage of time each 
is detected. Figure~\ref{fig:hr_hist} shows the hardness ratio distribution
split by $R$-band magnitude, with the top panel including sources with $R<22$ 
and the bottom panel showing the optically-fainter sources. At $R<22$ the BLAGN,
shown in black, dominate the distribution, with a sharp peak at $HR\sim -0.5$. 
In Paper II we speculated that this $HR$ peak near $-0.5$ was due to unobscured 
Seyferts and quasars, since that HR corresponds to a
power-law photon index ($\Gamma$) between 1.8 and 1.9, a 
typical value for unobscured broad-line AGNs; the spectroscopy presented here
confirms that conjecture.

Figure~\ref{fig:nhhist} shows the \nh~distribution of the BLAGN in the top
panel. In the unified AGN model, the existence of broad lines indicates a 
relatively unobstructed view to the central regions of the nucleus with low
obscuring column densities (\nh $<10^{22}$ cm$^{-2}$). While our BLAGN do have
the lowest mean $HR$ among our four source classes, we still find that 29\%
$\pm$ 4\% have \nh$>10^{22}$ cm$^{-2}$, indicating significant X-ray absorption.
However, a significant fraction of these sources are at a high enough redshift
that their \nh~values are less-well constrained due to the absorption edges
shifting out of the {\em Chandra} bandpass (see lower right panel of Figure
\ref{fig:hr_rmag_z_1}~and \S~\ref{sec:selection}). To quantify this effect, 
we consider the 114 BLAGN at $z<1.5$; only 12\% of this subset 
have \nh $<10^{22}$ cm$^{-2}$, compared to 29\% for BLAGN 
at all redshifts. For the 62 BLAGN with
$z<1$, the fraction is even lower (9\%). This systematic trend also likely
explains the tendency for many of the most luminous BLAGN in Figure
\ref{fig:nhhist}~to be the most obscured. SEXSI, as with any flux-limited
survey, has a redshift-luminosity relation (Figure \ref{fig:lum_z}): our most
distant sources are the most luminous sources, and this systematic effect
makes them also appear most obscured.

Furthermore, the column density measurements can be affected by changes in the
Galactic \nh~(which is fixed during the spectral fitting): small underestimates
in the true Galactic \nh~can give rise to overestimates of the intrinsic column density. We note that, for consistency with all previous work, we have used for
Galactic absorption column density the $N_{HI}$ derived from 21 cm
observations \citep{Dickey:90}. This common practice is
incorrect, however; the X-ray absorption column density is two to three times
the $N_{HI}$ value, since it is the heavy elements, rather than hydrogen, that
absorb X-rays, and they are present in the molecular and ionized phases
of the ISM as well as in the atomic phase \citep[e.g.,][]{Iyengar:75}. 
Doubling or tripling the Galactic \nh~value can reduce
the inferred intrinsic column density by a large factor, particularly
for high-redshift sources, and particularly for sources from fields with
relatively large Galactic column densities. Indeed, we find the fraction
of high-\nh~BLAGN monotonically increases as the Galactic column density
increases, from 25\% for the 10 fields with $N_H<2 \times 10^{20}$ cm$^{-2}$, 
to 28\%
for the eleven fields with $2<N_H<7 \times 10^{20}$ cm$^{-2}$, 
to 32\% for the four 
fields with $N_H\sim 9 \times 10^{20}$ cm$^{-2}$, 
to 70\% for the two fields with 
$N_H\sim 20 \times 10^{20}$ cm$^{-2}$. 
Any use of the \nh~distribution in this, and 
all other, survey paper(s) for quantitative purposes must take the uncertainty
in Galactic column density and the insensitivity of measurements for 
high-redshift sources into account.

The Doppler-broadened emission lines exhibit a distribution of widths; for the purposes 
of consistency, SEXSI puts a strict cutoff between narrow- and broad-lined 
sources at 2000 km s$^{-1}$, following  
\citet{Veilleux:87}. Some sources have
a line or lines that are just above this width cut and thus are classified as 
BLAGN, although their other properties may be more similar to a typical NLAGN. 
Our BLAGN
classification includes sources with broad-absorption lines blue-shifted with 
respect to the object redshift (BALQSO) or broad-absorption at the source redshift. Of the nine
sources 
with \nh$>10^{23}$ cm$^{-2}$, four of the sources are noted as possible BALQSO's, 
a type of quasar associated with high
\nh~values \citep[e.g.,][]{Gallagher:02}.

\subsection{Narrow-lined AGN}
\label{sec:nlagn}

The NLAGN, comprising 8\% of our sample, tend to be both the most obscured objects and
highly luminous. Of our 32 NLAGN, five had no optical 
counterpart in our imaging --- four 
of these were blind pointings at the X-ray positions in fields with $R_{\rm 
limit}=23-24$, while the fifth was from our shallowest field with $R_{\rm 
limit}=21.1$.  The 27 NLAGN with optical counterpart photometry have 
$\langle R \rangle = 22.0 \pm 1.6 $, while the BLAGN have $\langle R \rangle=20.9$.

As mentioned in \S~\ref{sec:redshiftandclass}, the redshift distribution of 
the NLAGN is broad, extending over the redshift range $z=0-4$.
The NLAGN are the only narrow-lined 
sources in the sample with $z>1.5$. This fact is largely a consequence
of a two-part selection effect:  1) bright, high-ionization UV emission lines
are availably only for objects at higher $z$, while we reach the redshift desert 
for typical galaxy emission lines found in ELG at $z>1.5$, and 2) for redshifts
sufficiently high that \lya~moves into the optical window, normal galaxies
would be too faint both optically and at X-ray energies in our moderate-depth,
wide-area survey. In addition, the higher mean luminosity of the NLAGN makes
them visible to higher redshift.

For the NLAGN, we find an average (obscured) luminosity near 
$10^{44} \lumin$:  $\langle \log{L_{\rm 2-10~keV}} \rangle = 43.8 \pm 0.6$. 
Fifteen sources have obscured quasar luminosities, with $L_{\rm 2-10~keV} > 10^{44}
\lumin$ and, of these, 73\% are found at $z>1.5$. 
Twelve out of these fifteen galaxies have $R>23$, while two have limiting $R$ magnitudes
of 21.1 and 22.9, and one has $R=21.7$. 
Of the 
sources with lower luminosities, two have luminosities near $10^{42} \lumin$. 
The other fifteen have $43.0 < \log{L_{\rm 2-10~keV}} < 44.0$. The non-quasar luminosity 
($L_{\rm x} < 10^{44} \lumin$) sources are all found at redshifts below $z=1.5$, with $0.3 < z < 1.3$. The 
luminosities presented in the catalog and referred to in this paragraph 
are all {\em obscured} \hardrange\ rest-frame luminosities, uncorrected for the intrinsic obscuring 
column. Since the majority of the NLAGN are obscured, their intrinsic
luminosities are larger. 
At $z=0$ and \nh $<10^{23}$ cm$^{-2}$ unobscured 
luminosity will be increased by a factor of $\simlt 2$, while 
at \nh $=10^{24}$ cm$^{-2}$ the unobscured X-ray luminosity increases 
by a factor of $\sim 10$. At higher redshifts this
increase is not as large since the lower energy X-rays 
that are most subject to absorption shift out of the observed 
frame; at $z=1$ ($z=2$) the \nh $<10^{23}$ cm$^{-2}$ unobscured 
luminosity is increased by a factor of $\sim 1.2$ ($\sim 1.04$)
and the \nh $=10^{24}$ cm$^{-2}$ unobscured luminosity
increases by $\sim 2.6$ ($\sim 1.36$). 
Using unobscured luminosites, 17 (53\%) of NLAGN have quasar
luminosities.

\subsection{Emission-line galaxies}
\label{sec:elg}

One of the surprising discoveries made by the various {\em Chandra} and
{\em XMM-Newton} surveys conducted to date is the large population 
of X-ray sources that lack AGN signatures in their optical spectra, and yet have X-ray 
luminosities too high to be powered by stellar emission alone. 
In \S~\ref{sec:nature_of_elg}~we discuss the nature of these sources; here 
we explore the properties of the ELG found in our sample.
Table \ref{tbl:lines}~shows the typical emission lines detected in our
ELG spectra (eg., \oii, \oiii, etc.). In addition, it shows that 
45\% of the 148 ELG with the requisite wavelength coverage show \cahk~absorption
and 42\% of the 146 ELG with requisite wavelength coverage show the D4000 break.

Of the 168 ELG, we find an average (obscured) luminosity nearly 
an order of magnitude lower than for the NLAGN:
$\langle \log{L_{\rm 2-10~keV}} \rangle = 43.14 \pm 0.05$ 
(this rises to $\log{L_{\rm x}}=43.25 \pm 0.04$ if we exclude
the eleven sources with $L_{\rm x} < 10^{42} \lumin$ which, in principle,
could be starburst galaxies -- but see \S 10). 
Thirteen sources (8\%) have (obscured) quasar luminosities, $>10^{44}  \lumin$. 
The majority of the ELG tend to be obscured and, thus, their intrinsic 
luminosities are larger; considering unobscured luminosities, 16 (9\%) ELG 
have quasar luminosities.

Twelve identified extragalactic SEXSI sources have $L_{2-10~{\rm keV}} < 10^{42} \lumin$;
eleven of the twelve are classifed as ELG with $0.09 < z < 0.34$. In \S~\ref{sec:starformation}
we discuss these sources and the possibility that their X-ray emission is 
starburst-dominated instead of AGN-dominated.

\subsection{Absorption-line galaxies}
\label{sec:alg}

Absorption-line galaxies are not found in great numbers in our survey. 
Although nearly half of our ELG spectra show absorption features, such as 
CaHK absorption and/or the D4000 break, those sources also have 
emission lines, most frequently \oii. With 98\% of our spectroscopy from
the Keck 10 m telescopes, our ability to detect faint lines is greater
than in surveys that use smaller telescopes and thus our identification 
statistics may be skewed towards ELG. In addition, some surveys
\citep[e.g., ChaMP --][]{Silverman:05}~classify sources as ALG even if
emission lines are detected.

Of the 438 sources with redshift and classification information, only 8 ($< 2$\%) 
are identified as ALGs. Of these eight, two are flagged as having only
a tentative line identification, meaning their
redshift and class identification is likely but not secure. In addition, two other
ALGs are flagged as falling just outside the X-ray-to-optical
counterpart search area (see Table \ref{tbl:newrmags}). The 
distances of these two objects from the edges of their 
respective search areas are $\sim 1$\arcsec, among the larger offsets found
in Table~\ref{tbl:newrmags}.
One more source identified in the catalog is flagged as an ALG that is identified 
using the spectrum of 
one of two optical sources within the X-ray-to-optical search area --- we would
need additional data (e.g., a spectrum of the other source or an on-axis 
{\em Chandra} observation) to determine the true counterpart identification. 

Setting aside these special cases, we find that there are are only three sources
with a secure identification as an ALG --- less than 1\% of the identified
sources. This fraction is lower than in some other surveys, as deiscussed 
further in \S~\ref{sec:z_comp}.

\subsection{Line-free spectra}
\label{sec:line-free}

Of the 477 spectra we collected, 39 exhibit roughly power-law continuum 
emission with no detectable line emission. Here we explore the 
possibility that the sources are BL Lac objects -- AGN that have 
X-ray and radio emission but show no emission lines in their optical 
spectra -- and the alternative notion that they are the higher-redshift end of the 
ELG distribution. 

\subsubsection{BL Lac Contribution}

Prominent among the serendipitous sources found in the first wide-area
X-ray imaging survey -- the {\em Einstein Medium Sensitivity Survey} 
\citep[{\em EMSS}:][]{Gioia:90,Stocke:91} -- were BL Lac objects
which comprised $\sim 6\%$ of the point sources detected.
Subsequent radio and X-ray surveys for BL Lacs have left
the population statistics of this relatively rare AGN class
somewhat uncertain owing to the non-Euclidean \lognlogs\
relation for X-ray-selected objects, and the debate over
the relative proportions of the X-ray-bright and radio-bright segments
of the population. The two recent X-ray surveys using, respectively,
the whole {\em ROSAT All-Sky Survey (RASS)} and the Greenbank radio catalog \citep{Laurent-Muehleisen:99}, 
and the {\em ROSAT}/VLA North Ecliptic Pole survey \citep{Henry:01}~define
the \lognlogs~relation for X-ray-selected objects down to a flux
of $1-3 \times 10^{-13} \fluxu$ in the \softrange~band.
Thus, reaching the SEXSI flux limit of $10^{-15} \fluxu$ requires an extrapolation of nearly
two orders of magnitude and a shift from soft X-rays to hard X-rays. Our detection of BL Lacs, or a lack thereof,
could be constraining.

Adopting the mean power law slope of $\Gamma=2.2$ derived from a large 
collection of RASS-detected BL Lacs by \citet{Brinkmann:97}~and assuming
only Galactic absorption, the \hardrange~flux should be 95\% of the \softrange~band 
flux. Using the SEXSI coverage of 1 deg$^2$ at $10^{-14} \fluxu$
and 0.1 deg$^2$ at $2.5 \times 10^{-15} \fluxu$ (Figure 
1 in Paper I) with this flux correction factor, and extrapolating the 
\lognlogs~for X-ray-detected BL Lacs from Figure 5 of
\citet{Henry:01}~($\alpha \sim 0.7$), we expect between 0.5 and 1 BL Lacs
to appear in SEXSI. Increasing the \lognlogs~slope by 0.1 (well within the
uncertainties) more
than doubles this number. Furthermore, if we use the radio number counts
for BL Lacs at 1.0~mJy from Figure 5 of \citet{Giommi:99}, we 
would also expect roughly one source in our survey area (although, again, the
extrapolation required is nearly two orders of magnitude in radio flux density).
Note that while the extreme, high-energy-peaked BL Lacs discussed in
\citet{Giommi:99}~make up only $\sim 2$\% of this expected radio population, the
fact that our survey goes nearly three orders of magnitude deeper in X-ray flux 
means that essentially all of the radio-selected objects should be detected. 

Using large-area, public radio surveys
we checked for radio emission from our 39 sources that exhibit line-free spectra.
FIRST (15) and NVSS (24) radio images were examined for each of these sources; 
no sources were detected in the FIRST images to a $3\sigma$ limit of 0.75~mJy. 
Of the NVSS images, 19 had upper limits of $\sim 1.4$~mJy, while three were
in a single noisy field with upper limits closer to 5~mJy. One source 
(J125306.0-091316) is within the contours of an extended NVSS source, but the
low resolution of that survey ($\sim 50$\arcsec) makes it difficult to 
establish an association. The final source, J022215.0+422341, has a 2.8~mJy 
NVSS source within 10\arcsec\ (1.5$\sigma$), and may represent the expected 
$\sim 1$ radio-loud BL Lac source in our survey area. This general lack of radio counterparts 
leads us to conclude that most of these 39 line-free SEXSI sources are not 
BL Lacs.

\subsubsection{ELGs in the redshift desert?}
\label{sec:elg_in_z_desert}

Since it is highly likely that the majority of these line-free objects are not
BL Lacs, what are they? In Figure \ref{fig:cont_plots}, we examine the notion that these objects
are predominantly ELGs in which the \oii~line has slipped beyond the
wavelength coverage of our spectra. 
We estimate a redshift for each continuum-only source by assuming that
\oii~falls just longward of the optical spectral range for each source.
Six of the objects have spectra with
limited wavelength coverage, and could fall within the redshift distribution
of the other ELGs at $z<1.5$; the remaining 34 objects would have to be
at higher redshifts (see lower left panel of Figure \ref{fig:cont_plots}). 

As would be expected if their mean redshift is higher, 
their magnitude distribution is shifted toward fainter values: 85\% are
fainter than $R=22$ while only 57\% of the ELGs are this faint (comparing 
only sources with $R_{\rm limit} > 22$). 
The median $R$-magnitude for the ELG is 22.2, while for the continuum-only 
sources, $R_{\rm median}>23$ -- the measurement is limited by our 
imaging depths. The $f_{\rm x}$ distributions of the two
samples are statistically indistinguishable, while the $HR$ distribution of
the lineless objects is slightly softer, consistent with the fact that 
we can more easily detect unobscured objects at higher redshift. The values of
$L_{\rm x}$ derived using the estimated redshifts 
are consistent with the rest of the ELG distribution, although, again, since the
sources are by definition at higher redshift but have similar \hardrange~fluxes,
the median value of $L_{\rm x}$ is higher (see lower panel in Figure \ref{fig:cont_plots}).
We determine that these continuum-only sources are {\em consistent} with being 
the high-redshift end of the ELG population. 

\citet{Treister:04}~combine X-ray luminosity functions with spectral
energy distributions of AGN to model the X-ray and optical
distributions of X-ray sources from the GOODS survey and find that
the predicted distribution for $R<24$ sources is consistent with the
GOODS spectroscopically-identified redshift distribution. The sources
that remain spectroscopically unidentified are predicted to be 
either optically faint, obscured sources nearby or in the redshift desert, 
consistent with our notion that the SEXSI line-free sources are part of the 
redshift-desert ELG population.

The population statistics of the SEXSI sample change if we include these  
39 line-free spectra with the ELG sample: 44.8\% $\pm$ 3.1\% BLAGN, 42.7\%
$\pm$ 3.0\% ELG, 6.9\% $\pm$ 1.2\% NLAGN, and 1.7\% $\pm$ 0.6\% ALG. Adopting
this assumption, the BLAGN and ELG fractions are the same within 1
$\sigma$, at $\sim 43-44$\% each.

\subsection{Stars}
\label{sec:stars}

Of the 969 X-ray sources covered by our optical images, seven are associated
with bright ($8.5<R<14$) stars, all of which were previously detected by
ROSAT (although several are apparently identified here for the first time);
these seven sources are labeled as `star' in Table~\ref{tbl:catalog}. All have colors of
spectral types G to M and hardness ratios at the softest end of the distribution
($-0.73<HR<-0.95$); six out of seven have hard-band X-ray-to-optical flux ratios
in the range $10^{-2.5}$ to $10^{-4}$, typical of the upper end of the stellar
$\lxlo$ distribution \citep[cf.][]{Pizzolato:03,Feigelson:04}. The
seventh source has $\fxfo \sim -1.8$ which is extraordinarily high;
however, since it is one of the five brightest sources in our entire survey,
the identification with a $12^{th}$ magnitude star is likely correct.
The very small fraction of stars ($<1$\%) we detect is simply a consequence of
our hard-band selection criterion.

Of the 468 spectra we obtained for the fainter optical counterparts, an 
additional eleven objects have stellar spectra at zero redshift. 
One of these is a $20^{th}$ magnitude object with strong H and He emission lines
superposed on TiO bands, characteristic of a cataclysmic variable. 
The spectrum for this source is shown in Figure \ref{fig:cv}. The
observed velocity is $-274$ km s$^{-1}$ marking it as a candidate for a rare halo 
CV, although the high velocity may simply mean the system was at an extremum
in orbital phase at the time of our observation.
The remaining stellar objects are likely to all be chance 
coincidences. Five are fainter than $20^{th}$ magnitude, have hardness ratios 
$HR>-0.5$, and have $\fxfo > -0.3$, e.g., values quite atypical of stellar
X-ray sources. In order to eliminate the
possibility that a giant hard X-ray flare was responsible for the detected 
source, we examined the light curves for all five sources and found no evidence 
of dramatic variability. We calculated the probability of chance
coincidence for each source using stellar and X-ray number counts and
error radii for each field, and found that all had probabilities $\gsim 10$\%;
in a survey with 27 fields, all are comfortably consistent with being chance
alignments of a foreground star with a faint background source that is the
true origin of the X-ray emission.

The remaining five stellar spectra are for counterparts with magnitudes
$16.7<R<18$. CXOSEXSI J152151.6+074651, the softest of the five stars, 
showed evidence of variability,
with a rapidly declining count rate in the first 5\% of the observation. 
Two of the others lie outside the formal
error circles (see Table \ref{tbl:newrmags}), making it likely they are chance coincidences. 
The remaining two sources have extreme $\fxfo$ values 
($>-1$, using a hard-band $f_{\rm x}$) and $HR>-0.3$; it is unlikely that 
these stars are the true counterparts.

\section{Selection Effects and Sample Completion}
\label{sec:selection}

Like all surveys with {\em Chandra}, the steep rolloff in effective
area above $\sim5$~keV limits the range of column densities that can
be probed by SEXSI. Our sample includes only sources detected in the 2 --
7~keV band, limiting our ability to identify sources with $\log{N_H}
\gsim 23.4$ at $z \lsim 1$.  In addition, we become less able to
constrain $N_H$ for high-redshift sources, where the absorption cutoff
for typical columns shifts out of the soft band (to $E \lsim
0.3$~keV).  For a $z=2$ source this happens for columns $N_H
\lsim 10^{22}$ cm$^{-2}$.  This effect can be seen in Figure~\ref{fig:hr_rmag_z_1}, 
which shows the hardness ratio distribution in different redshift
bins.  In the highest bin ($z > 2$), the hardness ratios tend
towards $-0.5$. 
Figure \ref{fig:nh_z}~shows \nh~versus $z$ and illustrates
the inability to constrain well the \nh~measurement of high-$z$, 
low-column density sources. 

Compared to some ``hard-selected" surveys, we are somewhat biased
against steep-spectrum X-ray sources.  We have focused our followup
effort on sources with independent hard-band detections.  By
comparison, some surveys compile source lists by
searching full-band images and then count a source as a hard-detection if
it has positive counts in the hard-band image
\citep[e.g.,][]{Stern:02b, Yang:04}; this will happen $\sim 50$\%
of the time due to statistical fluctuations in the background when 
no hard-band counts are recorded.  These
catalogs will therefore include a higher fraction of
steep-spectrum sources.

Completeness at our followup magnitude limits varies with source
class and redshift range. The procedures we use to classify sources
depend on specific lines, so that the redshift and $R$ magnitude
ranges over which we can properly identify sources depend on the
source type.  BLAGN are relatively easy to identify even at our
typical followup limit of $R = 23 - 24$, since their broad lines
constitute a significant fraction of the total $R$ band luminosity.
NLAGN, ELG, and ALG are all more challenging to identify at the faint
end, since their typical line-to-continuum ratios are smaller.  The ELG, which
are identified only from nebular lines such as \oii, \oiii, etc., have a
large ``redshift desert'' from $z\sim 1.4 - 2.2$, where the \oii\ has
shifted into the IR and \lya\ has yet to shift into the optical from
the UV.  Our 39 sources with continuum-only emission
have X-ray and optical properties consistent with the notion that they
are ELG in the redshift desert
(\S~\ref{sec:elg_in_z_desert}). 
For most ELG, the optical counterparts are roughly
consistent with L* host galaxies and thus their $R$-band magnitude
increases predictably with redshift (see \S~\ref{sec:non_stellar_emission}), 
unlike BLAGN, for example, where the optical luminosity is dominated by AGN 
emission and thus $R$ is related to $L_{\rm x}$.
Thus, there are likely many more ELGs at higher redshifts that we
have not spectroscopically followed up due to faint $R$-band 
counterparts; the ELG we do identify at $z \simgt 1$ 
typically have little continuum emission.

Since we use two different instruments for spectroscopy, we have
investigated the extent to which their different quantum efficiencies
as a function of wavelength may have affected source identification.
The blue arm of LRIS (LRIS-B) has good sensitivity
further into the near-UV than does the blue side of DEIMOS. There was
one case where we had both a DEIMOS spectrum and an LRIS
spectrum of a source which suggested different classifications: the
DEIMOS spectrum showed only ELG lines, while the LRIS spectrum had a
broad \mgii~line on the blue side which led to the final
classification as a BLAGN. This seems to be an isolated case in our
sample, although there are few sources for which we have both LRIS and
DEIMOS coverage.

We checked the statistics of the sources we identified with these two
instruments, ignoring the Doublespec data from the Palomar 200 inch (5
m) since it only comprises 2\% of the sample and includes only bright
sources. We find no significant difference in classification
statistics between the 137 sources classified by DEIMOS and the 280
sources classifed by LRIS.  With DEIMOS, we find $48.9 \pm 6.0$\% (67) BLAGN 
while in the LRIS spectra, $48.2 \pm 4.1$\% (135) are BLAGN; DEIMOS and LRIS 
identify $40.1 \pm 5.4$\% (55
sources) and $37.5 \pm 3.7$\% (105) ELGs, respectively.  Both
instruments classify a small number of targets as ALG (both $ < 2$\%),
and find a small number of stars ($\sim 3 - 5$\%).  

The only marginal difference we find is in the rate of identifying
NLAGN.  Using LRIS, we identify 26/280 such sources ($9.3 \pm 1.8$\% ),
while with DEIMOS we find only only 6/137 sources ($4.4 \pm 1.8$\%).
The NLAGN DEIMOS may be missing would be classified as ELG. This
difference probably results from the superior LRIS-B sensitivity,
which allows faint, narrow, high-ionization UV-lines such as \ciii\ or
\civ\ to be detected at observed wavelengths below $\simeq 5000$ \AA\,
enabling LRIS to properly identify
NLAGN in the redshift range $1 \simlt z \simlt 2$.  Also,
the \nevsingle\ line does not shift into the band covered by DEIMOS until
$z\sim 0.3 - 0.4$ and, as a consequence, we should also expect more
low-$z$ NLAGN with LRIS. These trends are seen in the data -- of the
six NLAGN identified with DEIMOS, only one has $1 < z< 2$, a source
identified by its \nevsingle\ line at $z=1.28$.  There are no
DEIMOS-identified NLAGN below $z=0.8$ and half of the sources are found
at $z>2$. LRIS, on the other hand, identifies eight (31\% of its)
NLAGN at $1 < z < 2$ and another eight at $z\simlt 0.5$.

Figure~\ref{fig:completeness}~gives an indication of our spectroscopic
completeness (defined as the fraction of spectroscopically-identified
sources as compared to the number of \hardrange~SEXSI sources) as a
function of several quantities: \hardrange~flux, $R$-magnitude,
$\fxfo$, and $HR$.  {\em Chandra} targets that were eliminated from
the X-ray catalog (e.g., nearby galaxies, quasars, extended target
cluster emission - see Paper I) are not considered SEXSI sources,
while X-ray point sources near target clusters are included in the
catalog.  To best illustrate the selection effects, separate from
incompleteness due to lack of follow-up, we plot data only from the
seventeen fields where we have a substantial fraction of spectroscopically-identified
sources (see Table \ref{tbl:fields}). All fields were included
in this subset if they had 50\% completeness; we added six
fields that have $<$50\% completeness when a high fraction of $R<24$
sources are identified. When considering only photometrically identified
$R<24$ sources the fraction of spectroscopically-identified sources
ranges from 67\% to 100\%.  These $R<24$ completeness numbers
depend on the particular $R_{\rm limit}$ of each field since some
imaging does not reach $R=24$ (see Paper II).  These seventeen fields
contain 725 \hardrange~SEXSI sources and 375 (52\%) spectroscopically identified
sources (86\% of all spectroscopic ID's presented in this article). 
The fields contain 445 photometrically identified $R<24$
counterparts (84\% of the photometrically identified $R<24$ sources
in these fields have spectroscopic ID's).
Figure \ref{fig:completeness}~shows histograms of all 725 X-ray
sources from these fields (open histogram), sources with optical
spectroscopic $z$ and class (shaded black), and sources with optical
follow-up but continuum only -- no $z$ or class (shaded gray).
The hatched histogram shows sources with $R>R_{\rm limit}$. 

The first plot of Figure \ref{fig:completeness}~shows that the
spectroscopic completeness in \hardrange~flux is relatively even,
45\% -- 70\% complete for $10^{-13.5} < $ \hardfluxrange~$ < 10^{-15}
\fluxu$, where the majority of the sources lie. The last plot shows
that the distribution of spectroscopic identifications in $HR$ is
$\sim$40\% -- 65\% for the majority of sources, with 80\% --
100\% identification rates for the softest bins.  The middle two plots
show that, as expected, the fraction classified decreases towards fainter
$R$-magnitudes, from 80\% near $19 \simlt R \simlt 22$, to $\sim$30\%
at $23<R<24$ (note that this bin has the highest number of
sources). The third plot illustrates the distribution in $\fxfo$. The
identification rate again is near 80\% for low $\fxfo$ and then falls
to $<$40\% for $\fxfo > 0.5$, which is expected due to the
difficulty identifying sources at faint optical fluxes.

Figure \ref{fig:hr_rmag_z_1}~shows scatter plots of $HR$ versus
$R$-magnitude for five redshift bins.
As discussed above, the
average $HR$ becomes smaller as redshift increases since at higher redshift the
$HR$ is less sensitive to changes in \nh. The first three boxes
($z<1.5$) contain the majority of the open circles (non-BL sources)
due to the large ELG population. The correlation between ELG
$R$-magnitude and $z$ is apparent -- for $z<0.5$ the majority of
non-BL sources are spread from $16\simlt R \simlt 22$, while for $0.5
<z<1$ the spread shifts to $20 \simlt R \simlt 24$, and for
$1<z<1.5$ the sources are almost all found at $R>22$.  The non-BL
sources at $z>1.5$ (all NLAGN) on average have $R>22$. The
BLAGN (filled circles) lie predominantly at low $HR$s with a spread in
$R$-magnitude; some are optically bright even at high-$z$ (typical of
unobscured quasars/AGN).

\section{Global Characteristics of the Sample and Comparison with
Other Surveys}
\label{sec:global_comp}

\subsection{Redshift Distribution}
\label{sec:z_comp}

The SEXSI sample confirms several of the basic conclusions of other
survey work.  First, as can be seen in Figure \ref{fig:lum_z}, 
very few low-redshift
AGN have high rest-frame X-ray luminosities.  Only 22/203 (11\%) of
sources with $z < 1$, and 1/65 (1.5\%) with $z < 0.5$ have unobscured $L_{\rm x} \geq
10^{44}$~erg~s$^{-1}$.  The difference in survey volume cannot alone explain 
this trend. Of the high-luminosity $z < 1$ sources, the
majority (65\%) are BLAGN.  Even accounting for the smaller volume surveyed
at low-redshift, if the X-ray to bolometric luminosity of
this sample is typical of AGN (if there is not an uncharacteristically
high fraction of the accretion luminosity emitted at longer
wavelengths), then this reflects a dearth of high-mass, high-accretion-rate
sources at low redshift.  This was hinted at in the deep surveys
\citep{Barger:01}, albeit from small survey volumes.  \citet{Steffen:04}~found this effect
in the 0.4~deg$^2$ CLASXS survey, and we confirm it here with SEXSI,
which samples the high-luminosity, low-$z$ population with a high
degree of completeness over $\sim2$~deg$^2$.

Figure~\ref{fig:z_comp}~shows the source redshift distributions found
by SEXSI, ChaMP~\citep{Silverman:05}, CLASXS~\citep{Steffen:04}, 
CYDER~\citep{Treister:05}, and
HELLAS2XMM~\citep{Fiore:03}. Broad-lined AGN are shown by filled
histograms, while non-broad-lined sources are left unshaded.  To
provide the best comparison to our sample of hard-band-selected
sources, we have eliminated sources from CLASXS and CYDER 
which have significant
soft or broad-band detections, but where the hard flux is determined
from a small number of counts (S/N $< 2$), and should therefore be
considered an upper limit.  To obtain similar hard-band detection
significance to SEXSI, we impose a cut at \hardfluxrange $< 2 \times
10^{-15} \fluxu$, removing approximately 50 sources from the CLASXS
sample.  CLASXS reaches a similar magnitude limit for spectroscopy as
SEXSI.  We adopt the ChaMP hard sample from \citet{Silverman:05} that
includes {\em Chandra} sources selected to have S/N $>2$ in the 2.5 --
8 keV band.  ChaMP obtained optical spectroscopic classifications for
44\% (220 sources) of their sample of 497 hard-band {\em Chandra}
sources, primarily for sources with counterparts having $R < 22$.

All the surveys demonstrate that the BLAGN population peaks at higher
redshift than the sample as a whole.  NLAGN and ELG peak at $z <
1$ in all samples, and show rapid evolution up to $z \sim 0.7$.  It
should be noted that for $z \gsim 1$, incompleteness due to the
faintness of optical counterparts is a significant factor, so the
decline in source density above this redshift is likely an artifact of the
spectroscopic survey depth.  As can be seen from the second panel of
Figure~\ref{fig:hr_rmag_z_1}, almost all SEXSI non-BLAGN sources above
$z = 1$ have $20<R<24$, with half above $R=22$, so that our survey
limit of $R = 23 - 24$ implies we are increasingly incomplete above
this redshift. ChaMP's shallower spectroscopic magnitude cutoff ($R\simlt 22$) results in
the sharper redshift cutoff of the ChaMP non-BLAGN population
compared to SEXSI (ChaMP finds no non-BLAGN above $z=0.8$, whereas the
SEXSI NLAGN spread to above $z=3$ and ELG/ALG to $z \approx 1.5$). SEXSI, 
CYDER 
and CLASXS, with similar survey depths for spectroscopy, find similar
redshift distributions.

The ChaMP BLAGN $z$-distribution is broader and flatter than that of SEXSI,
with a larger fraction of broad-lined sources at $z > 2$. Again, this
is largely a result of selection effects. The ChaMP X-ray source
population is on average softer than ours, due to the fact that they
select sources from full-band images and include any source with a
hard-band S/N $\geq 2$.  Combined with the predominance of BLAGN in the
$z > 1$, $R < 21$ sample (see Figure~\ref{fig:r_z}), this results in a
larger relative fraction of BLAGN. 

In addition, we note that ChaMP finds a significantly larger fraction
of ALG (7\%). This is simply a matter of nomenclature. They classify as ALG 
sources with absorption lines and a D4000 break whether or not they exhibit weak
emission lines. Our ALG are strictly sources in which only absorption features
are detected. Since 50\% of our 168 ELG do show absorption features, our
sample would contain {\it more} ALG than ChaMP if we adopted their definition;
our greater ability to detect weak lines explains any discrepancy.

\subsection{Obscured Sources}
\label{sec:obscured_srcs}

To explore the fraction of obscured sources, we split our sample at
\lognh $=22$. Figure~\ref{fig:frac_obsc_flux}~shows the fraction of
obscured sources (\lognh $>22$) as a function of {\em unobscured} \hardrange~flux 
(fluxes corrected for intrinsic \nh~obscuration) for spectroscopically 
identified sources from
SEXSI (filled circles), GOODS CDFN (triangles; E. Treister, private communication),
and ChaMP \citep[diamonds;][]{Silverman:05}. 
The datapoints are calculated using the survey catalogs, 
binned into flux ranges shown
by the vertical dashed lines at the bottom of the plot. Each SEXSI,
ChaMP, and GOODS CDFN datapoint is offset slightly along the x-axis
for clarity and the number of sources in each bin is indicated
at the bottom.

Both SEXSI and ChaMP $N_H$ values are calculated using X-ray spectral
fitting, which also provides individual (asymmetric) errors for each 
column density.
We would like to calculate a fraction of obscured sources and its
associated error
taking into account both the
individual \nh\ errors from the spectral fits and 
Poisson counting statistics.
To calculate the fraction of obscured sources for each flux bin  
we first generate, for each individual source in that flux bin, 
a Monte Carlo distribution of 1000 
\nh\ values. 
To account for the asymmetric error bars provided by the spectral fitting, 
we generate two Gaussian distributions, one with a standard deviation
equal to $\sigma^{+}$ and one with a standard deviation 
equal to $\sigma^{-}$. The two distributions are then patched
together to make a single, asymmetric Gaussian distribution.
ChaMP reports 90\% confidence error bars instead of 1 $\sigma$, as SEXSI
does, so we estimate that the ChaMP $\sigma$ is equal to 1/1.65 times
the 90\% confidence limit.   Because the purpose of this exercise is to 
calculate a fraction of sources above and below \lognh$=22$, we do 
not need to take into account \nh\ values in these distributions that, 
for example, fall below 0 or are otherwise unphysical. By construction, 
each distribution has 500 $N_H$ values above and below the best-fit 
value.

We then use the Monte Carlo $N_H$ values to calculate the mean and
standard deviation of obscured fractions (\lognh$>22$) in each 
flux bin. These are plotted in Figures \ref{fig:frac_obsc_flux}~and 
\ref{fig:hr_v_xspec_comp}.
The plotted error bars are the 1 $\sigma$ errors that result
from adding the standard deviation described above in quadrature with 
Poisson counting error ($\frac{\sqrt{N}}{N}$, where $N$ is the number
of sources in each bin). 
For many sources, the column density and associated errors constrain 
the source to lie either solidly above or below \lognh$=22$; thus 
the individual \nh\ errors contribute negligibly
to the uncertainty in the fraction of obscured sources; the
errors from the Poisson counting statistics strongly dominate the 
overall error bars.  

The GOODS CDFN sources are calculated from hardness ratios and thus do not
have similar errors from the fits. For these sources we simply calculate 
a fraction of obscured sources by counting the number with \lognh$>22$ in 
each bin and
dividing by the number of sources in that bin. The errors presented are
solely from Poisson counting statistics, which provide an adequate
comparison since the counting errors dominate in the SEXSI and ChaMP calculation.

In Figure~\ref{fig:frac_obsc_flux}~the SEXSI data shows an obscured 
fraction consistent with $\sim 0.5$ for all
flux ranges, 
with a marginally
significant decrease with increasing flux. In the second through 
fourth bins, which
include sources from $3 \times 10^{-15} - 3 \times 10^{-14} \fluxu$,
the range over which the \hardrange~\lognlogs~ changes slope, there
are 314 SEXSI sources, providing a tight constraint on the obscured
fraction. All three of these fractions fall between 
$0.5$ and $0.6$. 

The GOODS CDFN data shows a higher fraction of obscured sources, 
though most of the flux bins contain few sources (since the majority
of GOODS sources fall below our flux limit) and thus have large
errors. Since GOODS uses hardness ratios to calculate their \nh\
values we explored the difference between the SEXSI $N_H$ values when 
calculated as described using the XSPEC fits, and when using the 
$HR$. Both methods are described in more detail in \S~\ref{sec:catalog}.
Figure~\ref{fig:hr_v_xspec_comp} shows the fraction of obscured 
SEXSI sources calculated using absorbed (observed) \hardrange\ fluxes.
The black filled circles are calculated using the method described
above, with the \nh\ values from the spectral fits. The lighter gray 
points show the fraction of obscured sources when we use \nh\
values calculated from $HR$'s. There is an obvious difference:
the $HR$-calculated fractions of obscured sources are consistently
higher, across the entire flux range. In the bin from 
$\sim 3 \times 10^{-15}- 1 \times 10^{-14} \fluxu$, where we have
149 sources, the two values are not even consistent to within 1 $\sigma$. 
This discrepancy arises mainly from sources with 
the lowest column densities. For sources 
with a significant ($\simgt 5 \times 10^{21}$ cm$^{-2}$) column density from 
the X-ray spectral fit, the \nh\ values from the two methods are 
typically consistent within errors. The large discrepancy arises from sources
where the XSPEC fit finds no significant obscuration (`$<$' in the catalog), though 
the 1 $\sigma^{+}$ high value does include significant obscuration 
(even \lognh$>22$) in 
some cases --- see Figure~\ref{fig:nh_z}. For such sources, the $HR$-calculated value 
is often $>10^{22}$ cm$^{-2}$. This difference seems to result from details of the 
calculation. For example, a source with many counts near 0.3 keV will 
affect the XSPEC fit differently than the $HR$-based \nh\ calculation. 
To calculate $HR$, as described in more detail in Paper I, we extract
counts from 0.3--2.1 keV and from 2.1--7 keV and transform each value using 
the appropriate exposure map into the standard band (\softrange\ and \hardrange) 
fluxes. 
Our exposure maps correction assumes
$\Gamma=1.5$. We then use these fluxes to calculate the \nh\ values.
Thus, a source with many counts near 0.3 keV 
but fewer at 2 keV will result in an increased 
flux for the entire \softrange\ band, while in the X-ray 
spectral fit we know that those counts are actually at 0.3 keV. 
The $HR$ based \nh\ values will have large individual errors, and
these are not taken into account on the plots.
Thus, the discrepancy of the fraction of obscured sources in 
Figure~\ref{fig:frac_obsc_flux}~is not wholly unexpected.

The ChaMP data show consistently smaller fractions of
obscured sources in all flux ranges, although the discrepancy is
highly significant only for the flux bin from 
$\sim 3 \times 10^{-15} - 1 \times 10^{-14} \fluxu$.  
The lower obscured fractions in
ChaMP results from the predominance of BLAGN, which, as described
above, arises from the brighter magnitude limit of their spectroscopic
followup.  As a population, the BLAGN have lower obscuration than the
NLAGN and ELG.  This selection effect is illustrated in
Figures~\ref{fig:hflux_hist}~and \ref{fig:hr_hist}.  From our sample,
79\% of BLAGN have $R<22$ and 52\% have $R<21$.  Combining these with
the \nh~histogram split by source type, one can see that a brighter
optical spectroscopic followup limit will cause an \nh~distribution
with fewer obscured sources.  Not only will there be more BLAGN
compared to other source types, but of the BLAGN found, a higher
fraction are unobscured sources brighter in $R$.

Figure \ref{fig:frac_obsc_lum}~shows the fraction of obscured sources
as a function of unobscured luminosity for sources with spectroscopic redshifts. 
The luminosities are
calculated using unobscured fluxes and the calculation of each fraction and 
associated error are determined as described for Figure
\ref{fig:frac_obsc_flux}. The SEXSI sample (filled circles)
shows a constant fraction of obscured sources 
of $\sim 0.5$.
The difference with ChaMP
can once again be explained by the difference in spectroscopic 
followup depth. SEXSI's most obscured sources tend to be NLAGN and 
ELG (see Figure \ref{fig:nhhist}). 

Of the total sample of NLAGN, 26/32 (81\%) are 
obscured (\nh $>10^{22}$ cm$^{-2}$) and of those with the highest 
luminosities ($L_{\rm x} > 10^{44}\lumin$) all fifteen (100\%) are obscured. 
Of the more numerous ELG, 119/162 (73\% $\pm$ 7\%) 
are obscured. For the quasar-luminosity ELG, 11/13 (85\%) are obscured. 
These numbers are both consistent with the obscured fraction of NLAGN. These sources
make up about half of the total SEXSI AGN sample, and thus they 
contribute heavily to the obscured fractions seen in Figure \ref{fig:frac_obsc_lum}. 
They have median magnitudes 
of $R \simgt 22$; following up brighter sources only misses 
most of these low-luminosity, obscured AGN.

\subsection{Obscured Sources with Quasar Luminosities}
\label{sec:qso2}

According to unified AGN models \citep{Antonucci:93}, hard X-ray surveys
should find significant numbers of type-2 quasars (e.g., 
quasars viewed edge-on, through significant amounts of absorbing
material). These will be identified as X-ray sources with large \nh,
$L_x \gsim 10^{44} \lumin $, and with narrow-lined optical
counterparts.  Confirming candidate type-2 quasars is, however,
difficult. The two {\em Chandra} deep fields have found fourteen
narrow-lined sources with quasar X-ray luminosities
\citep{Norman:02,Barger:03,Dawson:03, Szokoly:04}. Of these, only
\citet{Dawson:03}~has published an infrared spectrum confirming that
lines, such as \halpha, that are redshifted out of the optical, are
narrow.

In the SEXSI survey we find 17 sources with {\em unobscured} rest-frame
luminosities above $10^{44} \lumin $ which we classify as NLAGN, and
16 which we classify as ELG. Of these 33 sources, 32 (97\%) have
\nh$>10^{22}$ cm$^{-2}$. Nine of these, all NLAGN, have $z > 2$.  This
$z > 2$ type-2 quasar density is roughly consistent with the 1 -- 2
type-2 quasars per deep {\em Chandra} field predicted by \citet{Stern:02b},
although it is unlikely that all of our candidates are true
type-2 quasars \citep[e.g., see][]{Halpern:99,Stern:02a}.
We note that by comparison to our 33 narrow-lined quasars, 
ChaMP find no luminous,
narrow-lined quasars. This difference is mainly
attributable to their shallower spectroscopic coverage. Of our
narrow-lined quasars, only five have $21<R<22$ while the rest have
$R>22$ -- and, in fact, most NLAGN quasars have $R>23$.

\vspace{1cm}

\section{The Nature of the Emission-Line Galaxies}
\label{sec:nature_of_elg}

SEXSI has identified a substantial population of X-ray luminous
($10^{41} - 10^{44}$~erg s$^{-1}$) sources with optical spectra
lacking both high-ionization lines and evidence for a non-stellar
continuum.  Such sources, with typical redshifts $z < 1$, are found in
most {\em Chandra} and {\em XMM} surveys, in particular in the deep
fields \citep[see][and references therein]{Brandt:05}.  
The nature of this population is somewhat uncertain.
\citet{Moran:02}~suggest that most are akin to Seyfert galaxies
where dilution by the host-galaxy light hinders detection of the
high-ionization lines. In some cases, these high-ionization lines
may also be weak due to partial
obscuration.  Some ELG, however, have optical spectra of high
signal-to-noise, implying the AGN signatures are extremely weak or
absent \citep{Comastri:02}.  A number of suggestions as to the
nature of these sources have been made. At the low-luminosity end of
the distribution, some may be powered by starburst activity; 
\citet{Yuan:04} suggest that some may be AGN with radiatively
inefficient accretion flows. Alternatively, they may be AGN that are
entirely obscured (over 4$\pi$~sr) so that ionizing photons cannot
escape the nuclear region \citep{Matt:02}. A few may also be BL Lac objects
\citep{Brusa:03}.

In this section we discuss constraints that we can place on the
SEXSI sample of ELG.

\subsection{Low-luminosity ELG; Powered by Star Formation?} 
\label{sec:starformation}

Starburst galaxies exhibit low-ionization, narrow emission lines and
produce significant hard X-ray fluxes, and so must be considered as
potential contributors to the low-luminosity end of our ELG population. 
Their X-ray emission arises from a 
combination of hot gas heated in supernova remnant shocks and the population
of high-mass X-ray binaries whose compact components are produced in these 
supernovae. Starburst galaxy radio emission is predominantly synchroton emission
from cosmic rays accelerated in these same remnants; since the galaxy residence 
time for the cosmic rays is comparable to the lifetime of the X-ray binary 
population, the X-ray and radio luminosities are correlated 
\citep{Ranalli:03}.
The most
luminous local starburst is NGC 3256; detailed spectral analysis by
\citet{Moran:99}~of this galaxy showed that it has a \hardrange~X-ray
luminosity of $2.5\times 10^{41} \lumin$, produced by a star
formation rate of $\sim 40M_{\odot}$ yr$^{-1}$. \citet{Helfand:01} use
the radio source \lognlogs~relation in conjunction with the specific
X-ray luminosity per O-star and the ratio of radio to X-ray
luminosities in starbursts to predict a surface density for \hardrange~
X-ray sources attributable to starburst galaxies of 2.2 deg$^{-2}$ at
the SEXSI flux threshold of $2\times 10^{-15} \fluxu$.

More recent work by \citet{Ranalli:03}, \citet{Grimm:03}, and
\citet{Gilfanov:04}~explicitly use \hardrange~luminosity as a
star-formation rate indicator and reach similar conclusions. With a
total sample of 37 local star-forming galaxies, NGC 3256 remains the
most luminous, although two of six candidate star-forming
galaxies selected by \citet{Ranalli:03}~from the {\em Chandra/Hubble} Deep
Field have inferred luminosities of $\sim 3\times 10^{42} \lumin$, 
implying star formation rates of several hundred solar
masses per year. These authors predict higher surface densities of
$10-20$ deg$^{-2}$ at $2\times 10^{-15} \fluxu$ and
$1-2.5$ deg$^{-2}$ at $1\times 10^{-14} \fluxu$. Folding
these higher predicted densities with our areal coverage curve from
Paper I predicts that we should have $1-3$ such objects in our 2 deg$^2$
sample of hard X-ray sources.

A total of 11 ELGs in our sample have hard X-ray luminosities
$L_{\rm x}<10^{42} \lumin$, and another 11 have
$10^{42}<L_{\rm x}<10^{42.5} \lumin$ (this upper limit would
imply a star formation rate of $\sim 600 M_{\odot}$ yr$^{-1}$ according to
the \citet{Ranalli:03}~calibration).  We examined radio images for all 22
galaxies from the FIRST \citep{Becker:95}~and NVSS \citep{Condon:98}~surveys.
Of the 17 sources that fell within the FIRST survey
limits, 15 have upper limits of $f_{20~{\rm cm}}<0.75$~mJy and one (in a
slightly noisy field) has a limit of $\sim 1$~mJy. The final source,
J170423.0+514321, is coincident with a bright (17~mJy) barely resolved
double (or core-jet) source, clearly marking it as an AGN. For the
five sources appearing only in the NVSS, four have upper limits of
$1.5-2$~mJy, while the fifth, J030532.6+034301, is coincident with a
subthreshold source with a flux density of a little over 2~mJy
(although a slight positional offset suggests the possibility of a
chance coincidence).

When plotted on the $L_{\rm x}-L_r$ correlation plot of \citet{Ranalli:03}, both
the two possible detections described above and the upper limits
place galaxies factors of $3-10$ above the correlation (i.e., they are
too X-ray bright for their radio luminosities -- or upper limits thereon --
to be starburst galaxies). A few of the ELG with $L_{\rm x}>10^{42.5} \lumin$ 
are also coincident with weak radio sources. However, their X-ray luminosities
fall more than an order of magnitude above the \citet{Ranalli:03}~$L_{\rm x}/L_r$ 
correlation for starbursts.  Thus, consistent with the
expectations of the starburst surface density described above, we
conclude that few if any of the ELGs in our sample have X-ray
luminosities dominated by star formation -- essentially all must
be powered by accretion.

\subsection{Non-Stellar Emission}
\label{sec:non_stellar_emission}

The hypothesis that ELG are dominated in the optical by galaxy light
rather than non-stellar emission can be tested to some degree by
plotting $\fxfo$~as a function of $\log{(L_{2-10~{\rm keV}})}$
\citep{Fiore:03}.  Figure~\ref{fig:fxfo_lum2} shows this for BLAGN
(top panel), and sources which lack broad emission lines (bottom
panel) for the SEXSI sample, together with sources from HELLAS2XMM and
CDFN.  The BLAGN, clearly dominated by accretion luminosity, are
clustered at high luminosity, whereas the non-BLAGN sources show a
correlation between $\fxfo$ and $\log{(L_{2-10~{\rm keV}})}$.
\citet{Fiore:03} argue that the correlation between the two quantities
indicates that the optical light is largely dominated by $\sim$L* host
galaxy light in the non-broad-lined sources, approximately independent
of AGN luminosity. This is seen by the relatively small (one decade)
scatter in optical flux seen over a large range (four decades) in
X-ray flux. This correlation is due to a relationship in $R-z$ that
is independent of $L_{\rm x}$; 
Figure \ref{fig:r_z}, the $R-z$ plot, shows that the $R$-magnitude of 
the ELG (triangles) varies predictably with $z$ (with some scatter). 
\citet{Bauer:04}~show that this $\fxfo - L_{\rm x}$ relationship 
does not hold for $R\simgt 24$, the optical brightness at which the 
$R-z$ track of \citet{Fiore:03}~and simple galactic
evolution tracks begin to differ noticeably.

Figure \ref{fig:fxfo_lum2_nlagn} plots the \citet{Fiore:03}~correlation for NLAGN
and ELG separately.  The ELG generally fall on the best-fit line from
\citet{Fiore:03}, but the NLAGN tend to fall below the line. This
suggests that NLAGN have brighter X-ray-luminosity-normalized-optical
magnitudes than do ELG.  This would be consistent with NLAGN having a
smaller fraction of their AGN optical emission obscured from view.
Again, this result is not unexpected, as Figure \ref{fig:r_z}~shows 
that NLAGN do not have the same $R-z$ relationship as ELG;
instead many of the NLAGN have redshifts much higher than would
be expected if their $R$-magnitude were simply dominated by L* galaxy
light.
Figure \ref{fig:fxfo_lum2_nlagn} also suggests that searching only the highest $\fxfo$
sources for type 2 quasars, as suggested by \citet{Fiore:03}, will
miss some of the highest-luminosity NLAGN accessible via optical
spectroscopy with current telescopes.

\subsection{Search for Faint High-Ionization Lines}
\label{sec:faint_lines}

In this section we discuss difficulties in detecting weak high-ionization lines
over the SEXSI redshift range, and investigate to
what extent narrow-line AGN signatures may be present, but not
detected in individual spectra.  In addition, we ask if significant
numbers of BLAGN could be identified as ELG due to host-galaxy dilution.  

Dilution of AGN light by emission from the host galaxy, and the
resulting difficulty in detecting weak high-ionization lines, is
certainly an important factor in optically classifying the population
of ELG. \citet{Moran:02}~obtained wide-slit integrated spectra (including both
the nuclear region and the galaxy) of well-studied,
nearby Seyfert 2 galaxies to mimic observations of distant sources
and found that eleven of eighteen galaxies would not be considered
Seyfert 2's based on their integrated spectra: the nuclear emission
lines had diminished flux compared to the stellar lines, so that the
line-flux ratios ({\em e.g.}, \nii/\halpha, \oiii/\hbeta) were
consistent with the values observed in \htwo\ regions or starburst
galaxies, not with those of Seyfert 2's. These sources were all at low
redshift, so that optical spectroscopy covers a different observed
wavelength range than is the case for the large majority of our
objects; these line-flux ratio diagnostics are not available for 
most of our sources.

SEXSI classifies ELG based 
on failure to detect broad or narrow
high-ionization lines, rather than on quantitative line-flux ratio measurements.
Here we explore the line-detection statistics as a qualitative indicator
of line-flux ratios and source class diagnostic.
The relatively
broad redshift range and significant stellar continuum makes \hbeta~
difficult to detect; we identify it in only 21\% of 113 ELG with
appropriate wavelength coverage (Table \ref{tbl:lines}).  In two-thirds of the 103 lower-$z$
ELG we do detect \oiii. High \oiii/\hbeta~ratios are found in Sy2
spectra but also in \htwo~region-like galaxy spectra;
without another line ratio such as \nii/\halpha~or
\siiplus/\halpha~we are unable to securely classifiy the optical spectrum 
as a NLAGN.
151 ELG spectra include
\neiii~coverage, but only 6\% show the emission line. Strong \neiii~is
also found in Sy2 spectra, but again multiple line-ratios are needed
to secure the Sy2 classification. \neiii~is a
weaker line in Seyfert 2 spectra, and was shown to be easily erased by
dilution by \citet{Moran:02}; thus, the low detection rate of \neiii~is
not constraining.

The line that primarily allows us to identify a source as a NLAGN
rather than an ELG over much of our redshift range is \nevsingle. 
This line is redshifted into the optical window over the redshift
range $0.1 \simlt z \simlt 1.8$, the range within which most of the
ELG in our sample lie.  The other typical high-ionization UV lines do
not shift into the optical until $z \sim 1.1$ (\ciii) and $z \sim 1.6$
(\civ). Thus, it is detection of \nevsingle, or the lack thereof,
that most often places a source in the NLAGN or ELG subsample.

As shown in Table \ref{tbl:lines}, 80\% of the NLAGN with
\nevsingle~access have \nevsingle, while only 20\% of our BLAGN
exhibit this line. The fraction of the NLAGN in which \nevsingle~is
present should be treated as an upper bound, since in the range
$z<1-2$ we cannot identify the source {\it unless} \nevsingle~is
present. Thus there are many instances where a NLAGN has no
\nevsingle~detection in its spectrum, consistent with the idea that
\nevsingle~may be hidden, and, due to wavelength coverage, a significant
fraction of NLAGN will be classified as ELG since only 
\nevsingle~is useful for making the distinction.

In contrast, we conclude that few BLAGN are classified as ELG due to
finite wavelength coverage.  89\% of our BLAGN with wavelength
coverage of the \mgii\ line do show broad \mgii~emission. \mgii~shifts
into the optical near $z\sim 0.4$, thus the \mgii~line is accessible
in the majority of the ELG spectra but broad emission is not detected.
Since the \mgii~ detection rate is so high, we conclude that BLAGN are
not easily misclassified as ELG, even at $z \simlt 1$, when the broad,
high-ionization lines farther in the UV (eg., \ciii, \civ) do not lie
in our spectral range. A note of caution is in order, however; 
\citet{Glikman:04}~have shown 
that dust-reddened quasars which exhibit only narrow lines in
their optical spectra, can have broad Paschen lines in the near-IR.

Our basic conclusion is that many of the ELG may be classified as
NLAGN in higher S/N spectra and/or with wider wavelength coverage,
but that few are broad-lined sources we have misclassified.  This conclusion is 
consistent with the \nh~distribution of the ELG (85\% of which 
have \nh$>10^{22}$ cm$^{-2}$). To further test this
hypothesis we have stacked a group of ELG optical spectra to increase
the signal-to-noise, searching for \nevsingle~emission that is the
hallmark of NLAGN over much of our redshift range. 

Figure \ref{fig:stacked_elg}~presents the spectrum created by 
stacking 21 ELG spectra obtained with LRIS on the Keck I Telescope. 
Sources from our March and June 2002 observations were used such that 
the spectra were obtained with the same spectrometer configuration. 
Of these spectra, we chose 21 sources with $0.7<z<1.2$. This redshift 
range ensures that the \nevsingle~emission line will fall on LRIS-R 
from $5800 - 7500$ \AA\ (observed frame).  The stacking procedure used 
standard stacking commands in IRAF. First each spectrum is shifted to 
its rest frame using {\tt dopcor} and then these spectra are combined with 
{\tt scombine} using median weighting.  The weights of each spectrum were 
close to 0.05 as would be expected for equal weighting: no single 
individual bright spectrum dominated the stacked spectrum. 

The stacked spectrum shows the features typical in individual ELG 
at these wavelengths: the strong \oii~emission line, the \cahk~absorption 
lines and the D4000 continuum break. The stacked spectrum also shows 
additional absorption features such as \hten, \hnine, and \hzeta.  
\neiii~emission, which is seen in only 6\% of the ELG spectra, 
is well-detected in the stacked spectrum. Most importantly, however, 
is the detection of \nevsingle. This emission is produced by 
highly-ionized Ne, confirming the presence 
of a strong AGN X-ray ionizing continuum and consistent with the
idea that our ELG population contains a significant fraction of 
NLAGN.

\vspace{3cm}

\section{Sources Associated with Target and Non-Target Galaxy Clusters}
\label{sec:clusters}

Few cluster galaxies have been determined to have obvious AGN signatures; in this section 
we explore the population of spectroscopically-identified SEXSI AGN associated 
with known galaxy clusters. The pre-{\em Chandra}
best estimate was that 1\% of cluster galaxies host AGN~\citep{Dressler:83}. This estimate came
from laborious optical spectroscopic studies of galaxies.  {\em Chandra's} unprecedented angular
resolution in the \hardrange~band allows identification of X-ray emitting galaxies that appear
in images near the cluster center. While some of these sources will be unassociated AGN,
searching these sources for cluster-member AGN is much more efficient than searching in the 
optical for AGN signatures, since the optical source density is high compared to the X-ray 
source density; furthermore, as we have discussed 
throughout this paper, many of the X-ray sources assumed to be AGN from their high X-ray 
luminosities will have no obvious optical indication of 
nuclear activity. 

\citet{Martini:02}~obtained spectra of the optically-bright ($R <20$) counterparts to
\hardrange~{\em Chandra} sources near A2104, a well studied $z=0.154$ cluster, and found that
at least 5\% of the cluster galaxies (6 sources) had X-ray fluxes consistent with AGN activity, while only
$\sim$1\% showed AGN activity in the optical in agreement with the earlier 
estimates derived via optical surveys. We explore this finidng further here
using our larger sample of cluster X-ray sources.

\subsection{Target Clusters}
\label{sec:target-clusters}

Of the SEXSI fields, fifteen (56\%) have galaxy clusters as targets. Three of these are nearby
clusters ($z=0.014 - 0.045$), two are at $0.18<z<0.22$, and the rest are at $z>0.43$.  In Paper I
we flagged sources when they fell within 1 co-moving Mpc of the cluster center as 
projected on the sky. The area
associated with this 1 Mpc radius region and the sources within the area (the flagged sources)
were excluded from our \lognlogs~calculation to avoid biasing our background sample.  The SEXSI spectroscopic followup, however, establishes source 
redshifts, enabling the determination of cluster membership. Table~\ref{tbl:clusters} lists the SEXSI cluster fields and details the
spectroscopic completeness among cluster-flagged sources and the spectroscopically identified X-ray emitting cluster members. 

Of the three nearby clusters for which the 1 Mpc radius covered more than the 
entire {\em Chandra} field-of-view, we have spectra for 43 of the 103 sources, and none are at the cluster redshift. This result is likely a consequence 
our spectroscopic followup strategy which skipped the brightest sources
(likely cluster members) to focus on background sources. 

Of the twelve higher-$z$ cluster fields, we have taken 41 spectra 
of sources that fall near the cluster in our image 
and have discovered that ten of the flagged sources are at the target cluster 
redshift (see Table~\ref{tbl:clusters}). 
Of these ten, only three show high-ionization
optical emission lines characteristic of active galaxies, although all ten have 
\hardrange~luminosities of $L_{\rm x} \simgt 10^{43}$ erg s$^{-1}$, suggesting an active nucleus. 
Of the ten confirmed members, the BLAGN have luminosities of 
$\log{L_{\rm x}} = 43.3,~43.8,~{\rm and}~44.5$ with obscuring column densities 
of $N_H<10^{21.2}, 0, ~{\rm and}~10^{21.5}$ cm$^{-2}$, respectively. The ELG have 
$42.9 < \log{L_{\rm x}} < 44.0$ and $22.6 < \log{N_H} < 23.4$. The ALG has
$\log{L_{\rm x}} = 43.4$ and $\log{N_H} = 21.9$.  All of the non-broad-lined 
sources are considered obscured ($\log{N_H} >22$), save the ALG just below
the $\log{N_H} = 22$ cutoff.
In addition to the ten sources within 1 Mpc (projected), we identify three 
additional sources at the target cluster redshift lying from $1-2$ Mpc (projected) from the cluster center. These three sources are all ELG with 
$L_{\rm x}>10^{43}$ erg s$^{-1}$.

Of the thirteen sources listed in Table \ref{tbl:clusters}~within $\sim2$ Mpc of the target clusters we find that only 23\% $\pm$ 13\% are BLAGN (3 sources). 
One of these BLAGN has broad lines barely over our 2000 km s$^{-1}$ cut; this
radio source, MG1 J04426+0202, was the target of the CL 0442+0202 {\em Chandra} exposure
\citep[see][]{Stern:03}; thus, we exclude this field from the analysis. We then
find that only 17\% $\pm$ 12\% (two of twelve sources)
of the confirmed cluster AGN have broad lines. 
The target clusters
for which we have identified member AGN have redshifts in the range
$0.46 < z < 1.27$, where ELG's are still readily detectable. 
If we naively look at the SEXSI \hardrange~source class statistics 
including only sources in that
$z$ range, we find 37\% $\pm$ 4\% (82 sources) BLAGN and 52\% $\pm$ 4\% (114 sources) ELG. Though our cluster member AGN sample is relatively small, 
they hint that the fraction of BLAGN is lower in cluster AGN than in the SEXSI
background sample.

\subsection{Non-Target Clusters Identified in \citet{Holden:02}}
\label{sec:nontarget-clusters}

In addition to the thirteen target cluster member AGN we also report on two additional clusters from 
\citet{Holden:02}; we have identified two cluster member AGN (ELG) in each cluster.
The ``Notes" column and associated footnotes of Table~\ref{tbl:clusters} references these serendipitous clusters, 
with redshift determinations from spectroscopic followup of our
\hardrange~sources.  Appendix A of \citet{Holden:02}~reports two new X-ray emitting groups 
or low-mass clusters of galaxies, discovered as extended sources, in SEXSI fields RX J0910 and RX J1317; 
their study included no
optical spectroscopic followup. We have spectra of~\hardrange~SEXSI sources nearby these non-target clusters.  

For SEXSI field RX J0910, we have identified two ELG at $z\simeq 1.1$, $\sim$2\arcmin~from the
position of CXOU\_J091008.6+541856. The X-ray analysis of \citet{Holden:02}~suggests two
redshift possibilities due to a $T, z$ degeneracy: $z = 0.68 \pm 0.06$ or $z =
1.18^{+0.08}_{-0.07}$. The two SEXSI ELG, CXOSEXSI\_J090954.0+541752 and
CXOSEXSI\_J090955.5+541813, are at $z=1.101$ and $z=1.102$ and 2.4\arcmin~and 2.0\arcmin~
away (projected 1.2 Mpc and 1.1 Mpc) from the position of CXOU\_J091008.6+541856.  These
sources are 3.2 Mpc projected (6.17\arcmin) from the target-cluster of the field, RX J0910+5244 ($z=1.11$).
Our spectroscopic redshifts at $z\sim 1.1$ agree within 1 $\sigma$ of the higher-$z$ 
prediction of \citet{Holden:02}.

The other serendipitous cluster is in SEXSI field RX J1317, which had a target of cluster RX
J1317+2911 ($z=0.805$). The {\em Chandra} data analysis of CXOU\_J131654.2+291415 by
\citet{Holden:02}~suggests a cluster with $T=2.9^{+3.1}_{-2.1}$ keV and $z=0.42^{+0.14}_{-0.10}$. 
We have identified two
\hardrange~sources that fall from $1.7$\arcmin$-4.2$\arcmin~of the reported position and have redshifts around $z\sim 0.58$. The sources,
CXOSEXSI\_J131700.2+291307 and \_J131706.2+291058 at $z= 0.580~{\rm and}~0.579$, 
fall $\sim$0.8 Mpc and 2.0 Mpc (projected) from the
CXOU\_J131654.2+291415, an indication of a cluster at $z\sim 0.58$.  This redshift agrees within $\sim 1~\sigma$ of the prediction in \citet{Holden:02}.

Including these four additional cluster AGN in our comparison of cluster member
class statistics to those of the SEXSI background sample, we find a fraction of cluster BLAGN to be even lower: 12\% $\pm$ 8\%. This fraction again is calculated excluding SEXSI field CL 0442+0202. 
Comparing to the SEXSI background sample from $0.46 < z < 1.27$ which contains 37\% $\pm$ 4\% BLAGN, we find a stronger indication (albeit still short 
of $3~\sigma$)
that the AGN cluster-member sample has fewer BLAGN than does the background sample.

\section{Summary}
\label{sec:summary}

We have presented a sample of 477 spectra of \hardrange~{\em Chandra}
sources.
Of our 438 spectroscopically-identified sources with counterpart magnitudes 
$R \simlt 24$, we confirm with high
significance a number of results found in other surveys.  We find that
few AGN at $z < 1$ have high rest-frame X-ray luminosities,
reflecting a dearth of high-mass, high-accretion-rate sources at low
redshift. In addition, our sample of broad-lined AGN peaks at
significantly higher redshift ($z > 1$) than do sources we
identify as emission-line galaxies.  We find that 50\% of our sources
show significant obscuration, with $N_H > 10^{22}$~cm$^{-2}$,
independent of intrinsic luminosity.  We have identified nine narrow-lined AGN
with $z > 2$ having quasar luminosities ($L_{\rm x} > 10^{44} \lumin$).
This is consistent with predictions based on unified AGN models.

We have investigated in some detail the nature of the large sample of
168 sources which we classify as emission-line galaxies.  These X-ray
luminous objects (most with $L_{\rm x}>10^{42} \lumin$) have optical
spectra lacking both high-ionization lines and evidence for
non-stellar continuum.  We conclude that few of these sources, even at
the low-luminosity end, can be powered by starburst activity.  By
stacking 21 spectra for sources in a similar redshift range in order
to increase the signal-to-noise, we are able to identify \nevsingle~emission, 
a clear signature of AGN activity.  This demonstrates that the
majority of these sources are Seyfert~2 galaxies, where the
high-ionization lines are diluted by stellar emission and/or extincted by dust.

\acknowledgements

We are grateful to Michael Cooper and Jeffrey Newman
of UC Berkeley for their
help with the DEIMOS data reduction pipeline. 
The analysis pipeline was 
developed at UC Berkeley with support from NSF grant AST-0071048. 
We thank Patrick Broos of Penn State for his generous help with 
{\tt acis\_extract} 
and Chandra Fellow Franz Bauer for a careful reading of the 
manuscript and many helpful suggestions.
We also thank Ed Moran, Mark Davis, and Alison Coil for
collaboration on the CL 0848+4454 spectroscopy.
We wish to recognize and acknowledge the significant cultural
role that the summit of Mauna Kea plays within
the indigenous Hawaiian community.  We are fortunate to have the
opportunity to conduct observations from this mountain. 
In addition, this research has made use of  NASA's Astrophysics Data System
Abstract Service
and the NASA/IPAC Extragalactic Database
(NED) which is operated by the Jet Propulsion Laboratory, California
Institute of Technology, under contract with NASA.
This work has been supported  by NASA NAG5-6035 (DJH), 
as well as by a small {\em Chandra} archival grant.  
MEE gratefully acknowledges support from the NASA Graduate
Student Research Program.
The work of DS was carried out at the Jet Propulsion Laboratory,
California Institute of Technology, under a contract with NASA.

\clearpage

\begin{deluxetable}{lllrrrrrrrr}
\setlength{\tabcolsep}{0.02in}
\tablecaption{SEXSI Fields.}
\tablecolumns{11}
\tablewidth{0pc}
\tablehead{
\colhead{} & \colhead{} & \colhead{} & \colhead{} & \colhead{\# 2--10 keV} & \colhead{\# 2--10 keV} & \colhead{\# with} & \colhead{\# with} & \colhead{\# with} & \colhead{} & \colhead{} \\
\colhead{ } & \colhead{R.A.} & \colhead{Decl.} &  \colhead{Exp.} & \colhead{srcs in} & \colhead{srcs in} & \colhead{phot. ID} & \colhead{phot. ID} & \colhead{phot. ID} & \colhead{\#} & \colhead{\#} \\
Name & \colhead{(J2000)} & \colhead{(J2000)} & \colhead{[ks]} & \colhead{X-ray area\tablenotemark{a}} & \colhead{opt area\tablenotemark{b}} & \colhead{$R<22$} & \colhead{$R<23$} & \colhead{$R<24$\tablenotemark{c}} & \colhead{with $z$} & \colhead{with cont.\tablenotemark{d}}}
\startdata
NGC 891       & 02 22 33 &  +42 20 57  &  51 & 21 &    19 &   10  &  13  &  14  &   6  &  3  \\
AWM 7         & 02 54 28 &  +41 34 47  &  48 & 28 &    24 &   11  &  14  &  16  &  14  &  2  \\ 
XRF 011130    & 03 05 28 &  +03 49 59  &  30 & 33 &    33 &   17  &  21  &  24  &  20  &  2  \\ 
NGC 1569      & 04 30 49 &  +64 50 54  &  97 & 33 &    31 &   18  &  19  &  19  &  16  &  3  \\ 
3C 123        & 04 37 55 &  +29 40 14  &  47 & 26 &    25 &   13  &  15  &  15  &   7  &  3  \\
CL 0442+0202  & 04 42 26 &  +02 00 07  &  44 & 45 &    37 &    2\tablenotemark{e}  &   2\tablenotemark{e}  &   2\tablenotemark{e}  &   7  &  0  \\
CL 0848+4454  & 08 48 32 &  +44 53 56  & 186 & 81 &    72 &   25  &  37  &  49  &  47  &  4  \\ 
RX J0910      & 09 10 39 &  +54 19 57  & 171 & 92 &    89 &   25  &  43  &  55  &  40  &  4  \\
\enddata
\tablenotetext{a}{This column gives numbers from the entire X-ray catalog from Paper I.}
\tablenotetext{b}{This column gives numbers from the portion of the X-ray catalog that has adequate imaging coverage. This column excludes sources from the X-ray catalog that lack imaging coverage (optflag=0; Paper II) or fall near a bright source such that the optical counterpart is not identified and the $R_{\rm limit}$ cannot be determined (optflag=6; Paper II).}
\tablenotetext{c}{This column gives the number of {\em confirmed} sources with $R<24$. Additional sources may have $R<24$ in cases where $R_{\rm limit}<24$.}
\tablenotetext{d}{Sources with spectra that show continuum-only -- no emission lines or absorption features -- and thus no redshift or classification was determined.}
\tablenotetext{e}{Field CL 0442+0202 has shallow $R$-band imaging ($R_{\rm limit}=21.1$). Since the numbers listed in this table are {\em confirmed} optical counterparts, this row indicates few $R$-band identifications.}
\label{tbl:fields}
\end{deluxetable}

\clearpage
\LongTables
\begin{landscape}

\begin{deluxetable}{lrrrrrcccrclcccccc}
\tablecaption{SEXSI optical spectroscopy catalog.}
\pagestyle{empty}
\tablecolumns{18}
\tablewidth{0pc}
\tablehead{
\colhead{ }  & 
\multicolumn{5}{c}{\underline{~~~~~~~~~~~~~~~~~~~~~~~~~~~X-ray Data~~~~~~~~~~~~~~~~~~~~~~~~~~~}} &
\multicolumn{8}{c}{\underline{~~~~~~~~~~~~~~~~~~~~~~~~~~~~~~~~~Optical Counterpart Data~~~~~~~~~~~~~~~~~~~~~~~~~~~~~~~~}}  &
\colhead{ } & \colhead{ } & \colhead{ } & \colhead{ } \\
\colhead{CXOSEXSI\_}  & 
\colhead{$\alpha_{\rm x}$ (J2000)} & 
\colhead{$\delta_{\rm x}$ (J2000)} & 
\colhead{OAA} & 
\colhead{$S_{\rm 2 - 10}^{(-15)}$} & 
\colhead{HR} & 
Flag &  
\colhead{$R$} & 
\colhead{$R_{\rm limit}$} & 
$\log{\frac{f_{\rm x}}{f_{\rm o}}}$ &
\colhead{$z$} & 
Class & 
Notes &
\colhead{$\Delta\lambda_{\rm rest}$} & 
\colhead{$\log{L_{\rm x}^{(2 - 10)}}$} & 
\colhead{$\log{{\rm N}_{\rm H}}$} &  
\colhead{$\log{{\rm N}_{\rm H}^-}$} & 
\colhead{$\log{{\rm N}_{\rm H}^+}$} \\
\colhead{(1)} & \colhead{(2)\tablenotemark{a}} & \colhead{(3)\tablenotemark{a}} & 
\colhead{(4)} & \colhead{(5)} & \colhead{(6)\tablenotemark{b}} &
\colhead{(7)\tablenotemark{c}} & \colhead{(8)} & \colhead{(9)} & 
\colhead{(10)} & \colhead{(11)} & \colhead{(12)\tablenotemark{d}} & 
\colhead{(13)\tablenotemark{e}} & \colhead{(14)\tablenotemark{f}} & 
\colhead{(15)}  & \colhead{(16)\tablenotemark{g}} & \colhead{(17)\tablenotemark{g}} & \colhead{(18)\tablenotemark{g}} }
\startdata
J022142.6+422654 &  02 21 42.75 & +42 26 53.5 &   9.49 &   23.10 & $-$0.33 & 1 & 18.73 & 24.3  & $-$0.64 & 0.310 & BLAGN &  \nodata & 3553$-$7500 & 42.9 &                       $21.6$\tablenotemark{j} &  \nodata &  \nodata \\
J022205.0+422338 &  02 22 05.08 & +42 23 37.7 &   4.24 &    8.73 &  0.84\tablenotemark{h} & 1 & 20.37 & 24.3  & $-$0.41 & 0.622 &   ELG &  \nodata & 2953$-$6165 & 43.1 &                                        $23.7$ &     $23.5$ &     $23.9$ \\
J022210.8+422016 &  02 22 10.93 & +42 20 16.1 &   2.17 &    3.63 & $-$0.46 & 1 & 21.57 & 24.3  & $-$0.31 & 3.563 & BLAGN &  \nodata & 1174$-$2321 & 44.3 &                                        $22.1$ &      $ < $ &     $22.7$ \\
J022215.0+422341 &  02 22 15.12 & +42 23 41.0 &   3.15 &   15.40 & $-$0.41 & 1 & 22.52 & 24.3  &  0.70 &  \nodata &  \nodata &  \nodata & 4515$-$9740\tablenotemark{i} &  \nodata &  \nodata &  \nodata &  \nodata \\
J022215.1+422045 &  02 22 15.19 & +42 20 44.5 &   1.32 &   42.90 & $-$0.42 & 1 & 17.33 & 24.3  & $-$0.94 & 0.155 &   ELG &  \nodata & 4454$-$8982 & 42.5 &                                         $ < $ &      $ < $ &     $20.0$ \\
J022215.5+421842 &  02 22 15.63 & +42 18 41.7 &   2.46 &    6.79 &                   0.28 & 1 & 20.33 & 24.3  & $-$0.54 & 1.063 &   ELG &   K & 2695$-$5225 & 43.5 &                                        $21.8$ &     $21.3$ &     $22.1$ \\
J022219.3+422052 &  02 22 19.40 & +42 20 51.6 &   0.54 &    6.38 &                   0.37 & 1 & 23.82 & 24.3  &  0.83 &  \nodata &  \nodata &  \nodata & 5000$-$10225\tablenotemark{i} &  \nodata &  \nodata &  \nodata &  \nodata \\
J022224.3+422139 &  02 22 24.45 & +42 21 38.4 &   0.91 &   64.60 & $-$0.44 & 1 & 17.41 & 24.3  & $-$0.73 & 1.181 & BLAGN &  \nodata & 2152$-$4539 & 44.6 &                                        $20.8$ &     $20.3$ &     $21.1$ \\
\enddata
\end{deluxetable}

\clearpage
\end{landscape}

\begin{deluxetable}{lrrrrrrrcr}
\tablecaption{Photometry for sources with spectral ID outside the formal match area.}
\tablecolumns{10}
\tablewidth{0pc}
\tablehead{
\colhead{CXOSEXSI\_}  & 
\colhead{$\alpha_{\rm x}$ (J2000)} & 
\colhead{$\delta_{\rm x}$ (J2000)} & 
\colhead{$R$} & 
\colhead{$\sigma_{\rm R}$} & 
\colhead{$\Delta \alpha$} & 
\colhead{$\Delta \delta$} & 
\colhead{$\Delta$r} & 
\colhead{Rad.} & 
$\log{\frac{f_{\rm x}}{f_{\rm o}}}$  \\ 
\colhead{(1)} & 
\colhead{(2)\tablenotemark{a}} & 
\colhead{(3)\tablenotemark{a}} & 
\colhead{(4)} & 
\colhead{(5)} & 
\colhead{(6)} & 
\colhead{(7)} & 
\colhead{(8)} & 
\colhead{(9)\tablenotemark{b}} & 
\colhead{(10)} }
\startdata
J030527.0+035528 &  03 05 26.97 & +03 55 29.3 &  17.61 & 0.17 &   2.5 &   0.1 &  2.5 &  1.5 & $-$1.50  \\
J084822.5+445854 &  08 48 22.62 & +44 58 55.9 &  21.08 & 0.13 & $-$2.3 & $-$0.4 &  2.3 &  1.9 & $-$0.50  \\
J084823.9+445852 &  08 48 23.94 & +44 58 53.8 &  22.01 & 0.13 & $-$1.8 & $-$0.6 &  1.9 &  1.8 &   0.12  \\
J084824.8+445740 &  08 48 24.83 & +44 57 41.3 &  20.52 & 0.13 &   2.2 & $-$0.5 &  2.3 &  1.5 & $-$0.79  \\
J084846.4+444830 &  08 48 46.46 & +44 48 32.0 &  20.55 & 0.13 &   2.2 &   0.8 &  2.4 &  1.5 & $-$0.43  \\
J084931.3+445549 &  08 49 31.38 & +44 55 50.5 &  21.20 & 0.13 &   2.3 &   1.3 &  2.7 &  1.7 & $-$0.22  \\
J090955.5+541813 &  09 09 55.55 & +54 18 15.4 &  22.21 & 0.13 & $-$2.7 &   0.4 &  2.8 &  1.7 &   0.09  \\
J091012.7+541205 &  09 10 12.74 & +54 12 07.8 &  22.49 & 0.13 & $-$1.1 &   3.5 &  3.6 &  2.8 &   0.68  \\
\enddata
\tablenotetext{a} {X-ray positions are corrected for average X-ray to optical offset (to correct {\em Chandra} astrometry).}
\tablenotetext{b} {The search radius, in arcsec, used for X-ray-to-optical source matching (Paper 2). This radius may be compared with the value in Column (8), the offset from the X-ray source to the optical source in the spectroscopy catalog (Table~\ref{tbl:catalog}).}
\label{tbl:newrmags}
\end{deluxetable}

\clearpage

\begin{deluxetable}{lrrrrrrrrr}
\tablecaption{Optical spectroscopic line detections.}
\tablecolumns{10}
\tablewidth{0pc}
\tablehead{  & \multicolumn{3}{c}{\underline{~~~~~~~~BLAGN~~~~~~~~}} & \multicolumn{3}{c}{\underline{~~~~~~~~NLAGN~~~~~~~~}} & \multicolumn{3}{c}{\underline{~~~~~~~~~~ELG~~~~~~~~~~}} \\ 
   & \colhead{\# w/} &  \colhead{\# w/} & \colhead{\%} & \colhead{\# w/} & \colhead{\# w/} & \colhead{\%} & \colhead{\# w/} & \colhead{\# w/} & \colhead{\%} \\
\colhead{Feature} & \colhead{$\Delta\lambda$\tablenotemark{a}} & \colhead{Det.} & \colhead{Det.}  & \colhead{$\Delta\lambda$\tablenotemark{a}} & \colhead{Det.} & \colhead{Det.}  & \colhead{$\Delta\lambda$\tablenotemark{a}} & \colhead{Det.} & \colhead{Det.}
 }
\startdata
\civ         & 105 & 66  & 62\% &     13 & 10 & 76\% &       0 & \nodata  & \nodata \\
\ciii        & 129 & 83 & 64\%  &     16 & 10 & 62\% &       0 & \nodata & \nodata \\
\mgii        & 169 & 151 & 89\% &     18 & 4 & 22\% &     112 & 13 & 11\% \\
\nevsingle   & 126 & 26 & 20\% &      21 & 17 & 80\% &     147 & 0 & 0\% \\
\oii         & 119 & 54 & 45\% &     21 & 21 & 100\% &     151 & 115 & 76\% \\
\neiii       & 110 & 19 & 17\% &         19 & 12 & 63\% &       151 & 10 & 6\% \\
\cahk        & 103 & 20 & 19\% &        19  & 7 & 36\% &   148 & 67 & 45\% \\
D4000 break  & 102 & 10 & 9\% &   19 & 8 & 42\% &    146 & 59 & 40\% \\
\hbeta       & 58 & 26 & 44\% &    14 & 7 & 50\% &        113 & 24 & 21\% \\
\oiii        & 50 & 41 & 82\% &     14 & 13 & 92\% &       103 & 68 & 66\% \\
\enddata
\tablenotetext{a}{These columns (\# w/ $\Delta\lambda$) indicate the number of sources
with spectral coverage of the given spectral feature.}
\label{tbl:lines}
\end{deluxetable}

\clearpage
\begin{deluxetable}{llcllll}
\tablecaption{Cluster Fields: SEXSI Sources Associated with Known Clusters}
\tablecolumns{6}
\tablewidth{0pc}
\tablehead{
Field & \colhead{cluster} & \colhead{\# flagged total\tablenotemark{a}}                    & confirmed       & additional  &           \\
Name & \colhead{$z$}      & \colhead{(\# with spectra)}& cluster members & cluster members\tablenotemark{b}     & Notes}
\startdata
AWM 7         &  0.017 &    28 (14) &                        &  & \\ 
3C 123        &  0.218 &    10 (3)  &                        &  & \\
CL 0442+0202  &  1.11  &     8 (4)  &  BLAGN\tablenotemark{c}&  & \\
CL 0848+4454  &  1.27  &     7 (5)  &  ELG, ELG              &  & \\
RX J0910      &  1.11  &    13 (6)  &  ELG                   & ELG ($\sim$1.1 Mpc) & cluster at $z \sim 1.10$, 2 AGN ID'd\tablenotemark{d}\\
HCG 62        &  0.014 &    39 (20) &                        &  & \\
RX J1317      &  0.805 &     6 (4)  &  ELG, BLAGN            &  & cluster at $z\sim0.58$, 2 AGN ID'd\tablenotemark{e}\\
BD 1338       &  0.640 &     5 (1)  &                        &  & \\
\enddata
\tablenotetext{a} {A source was flagged in Paper I when it fell within an area potentially less than 1~Mpc from the target cluster center, and excluded from the~\lognlogs\ calculation to avoid including target cluster members. With the addition of redshift data (\# of flagged sources with spectroscopic redshifts are indicated in parenthesis), we can definitively identify or exclude cluster membership. Confirmed cluster members are flagged in the spectroscopic catalog, Table~\ref{tbl:catalog}.}
\tablenotetext{b} {This column inicates sources within $\sim2$ Mpc of the cluster center -- the particular distances are noted in parenthesis. These sources are not cluster flagged in the Catalog. Their ID's are CXOSEXSI\_J091040.8+542006, \_J162315.4+263506, and \_J171714.5+671136.}
\tablenotetext{c} {Cluster member is target radio galaxy, MG1 J04426+0202 (MG 0442+0202), a borderline NLAGN from our spectrum. See \citet{Stern:03}.}
\tablenotetext{d} {Apendix A of~\citet{Holden:02}~(paper on RX J1317+2911, RX J1350+6007)
notes that there is an X-ray group, CXOU\_J091008.6+541856, at $z = 0.68 \pm 0.06$ or $z = 1.18^{+0.08}_{-0.07}$ (T, $z$ 
degeneracy).
We find two ELG (CXOSEXSI\_J090954.0+541752 and \_J091008.6+541806) at $z=1.101$ and $z=1.102$, 
1.2 Mpc and 1.0 Mpc away from their reported position, respectively. See \S~\ref{sec:nontarget-clusters}.}
\tablenotetext{e} {Appendix A of~\citet{Holden:02}~notes that there is a cluster/group, CXOU\_J131654.2+291415, 
at $z=0.42^{+0.14}_{-0.10}$. We find two ELG (CXOSEXSI\_J131700.2+291307 and \_J131706.2+291058) at 
$z\sim 0.58$, 0.8 MPc and 2.0 Mpc from the reported position. See \S~\ref{sec:nontarget-clusters}.}
\label{tbl:clusters}
\end{deluxetable}
\clearpage

\begin{figure}
\epsscale{1.0} 
\plotone{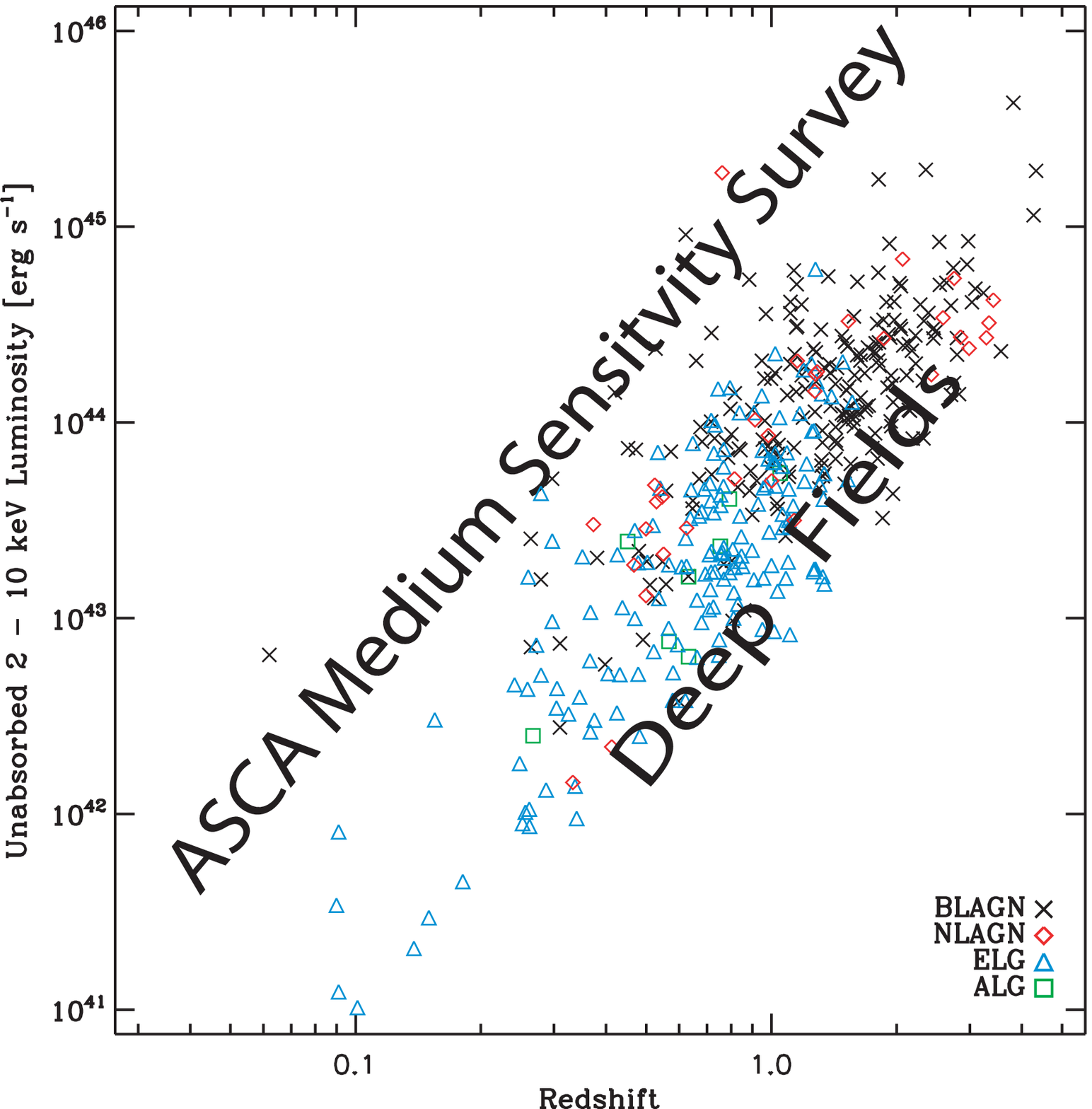}
\caption{Redshift vs. luminosity for SEXSI sources, with optical
spectral classification indicated. The approximate phase space
covered by the ASCA Medium Sensitivity Survey \citep{Akiyama:03}~and 
the {\em Chandra} Deep Fields 
({\em e.g.,} CDF-N -- \citealt{Alexander:03,
Barger:03}; CDF-S -- \citealt{Rosati:02, Szokoly:04}) is
illustrated with text. The
luminosity plotted is the instrinsic, unobscured luminosity in the
rest-frame \hardrange~band. See \S~\ref{sec:catalog} for a description
of the unobscured luminosity calculation.}

\label{fig:lum_z}
\end{figure}
 
\clearpage

\begin{figure}
\plotone{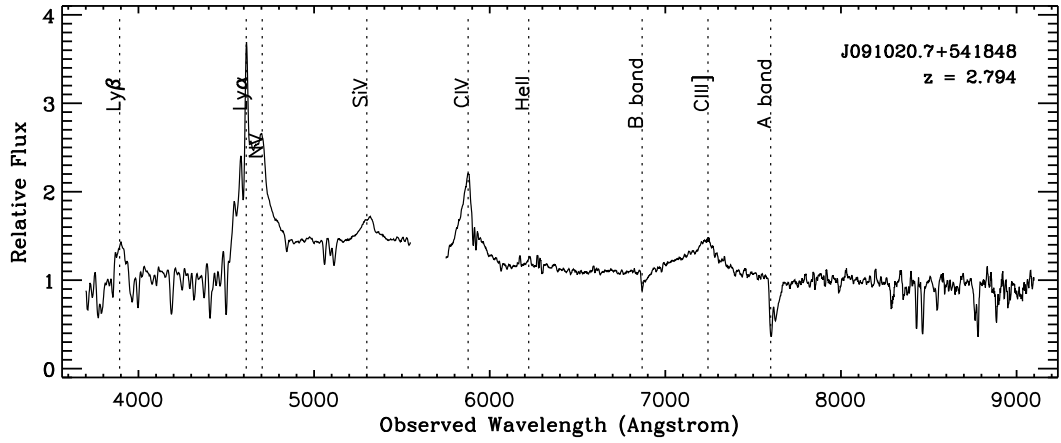}
\caption{Example of one of the 212 SEXSI BLAGN. This source has $z=2.794$. 
Note the broad, high-ionization emission lines typical of Type 1 Seyferts
and quasars. This spectrum was obtained with LRIS, using a 5600 \AA\ dichroic.
The absorptions at 7600 \AA\ (A-band) and 6850 \AA\ (B-band) are telluric in nature.}
\label{fig:blagn}
\end{figure}

\clearpage

\begin{figure}
\plotone{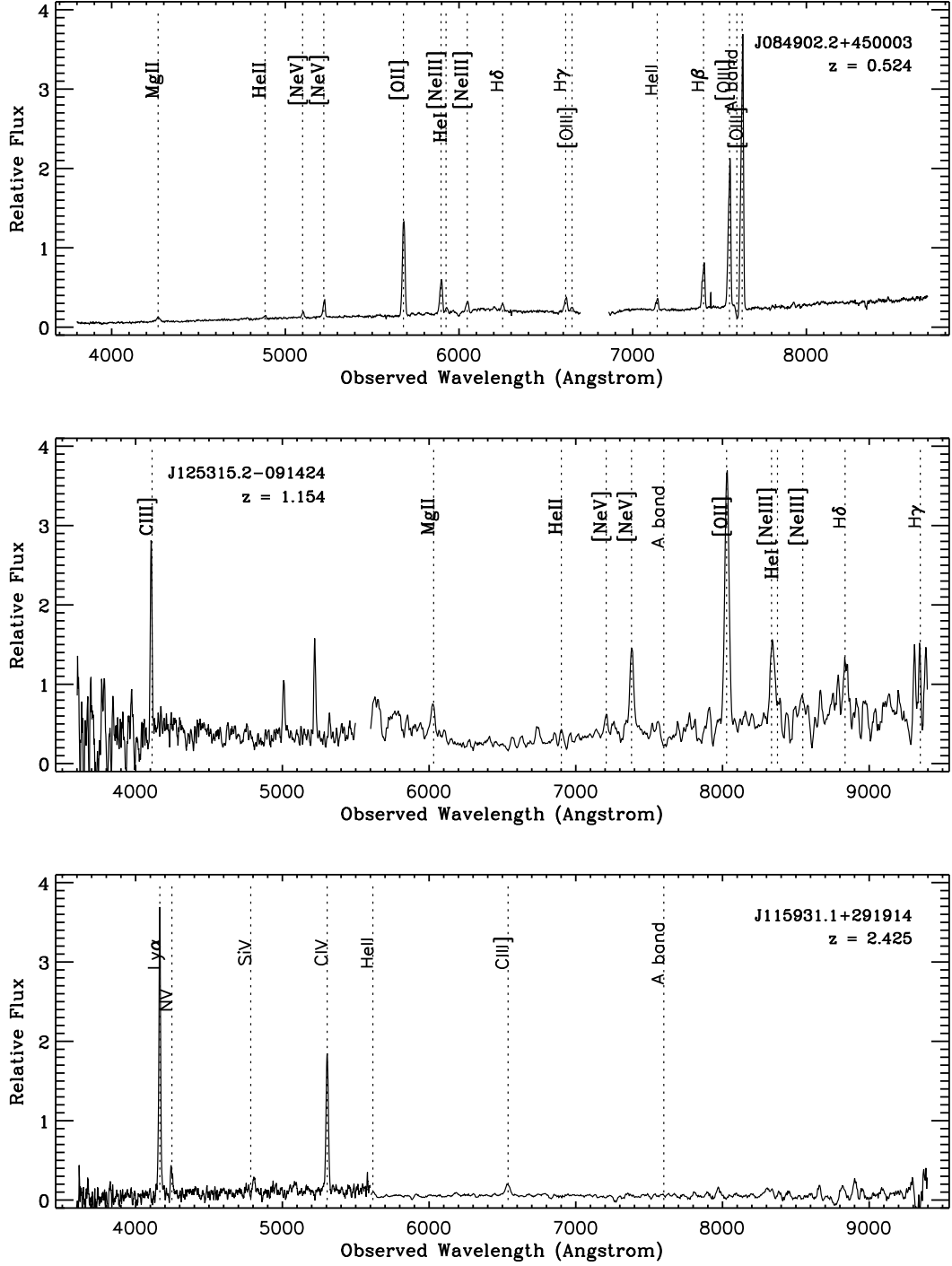}
\caption{Example spectra of NLAGN at $z=0.5$, $z=1.2$, and $z=2.4$. Note 
the narrow, high-ionization lines. For the lowest-$z$ source (top panel)
the classification depends on the \nevsingle~detection, while the 
higher-$z$ sources have narrow, high-ionization, UV
emission lines.}
\label{fig:nlagn}
\end{figure}

\clearpage

\begin{figure}
\plotone{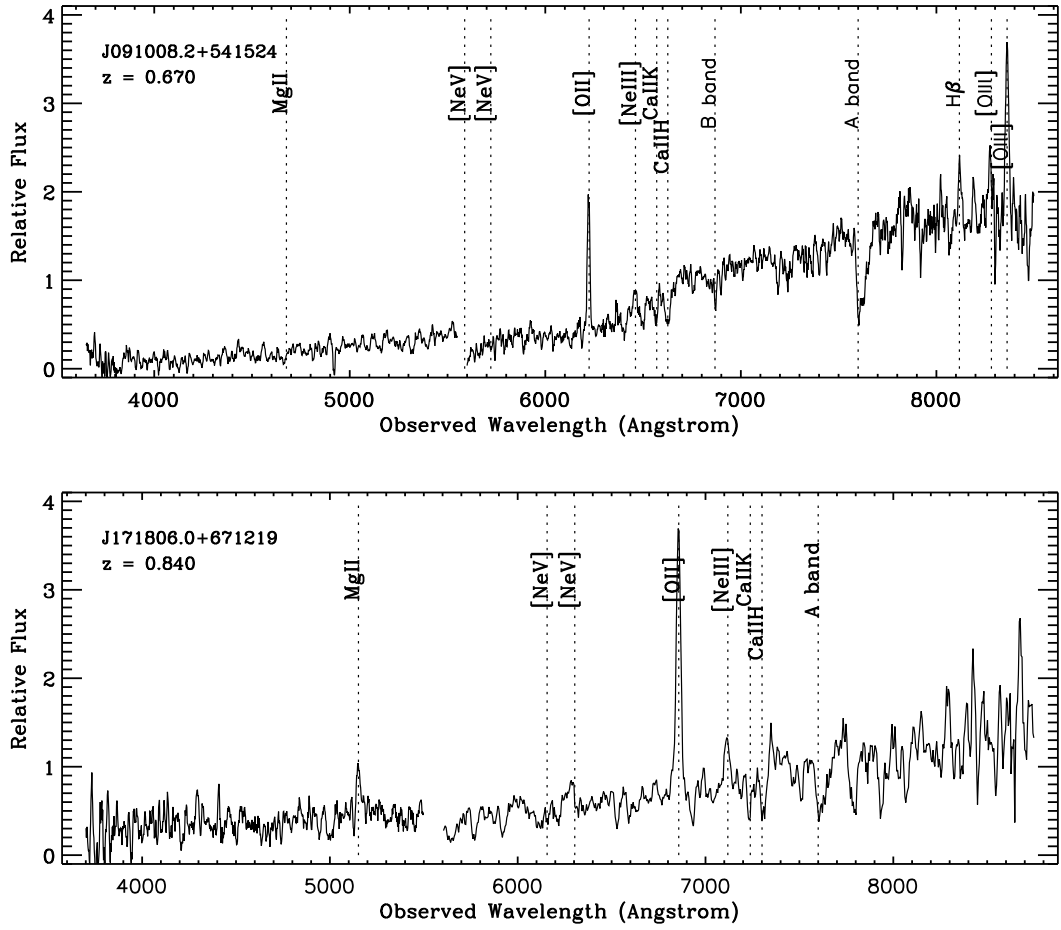}
\caption{Example spectra for two of the 168 SEXSI ELG. The sources
show narrow emission and absorption lines typical of normal galaxies and
lack \nevsingle.}
\label{fig:elg}
\end{figure}

\clearpage

\begin{figure}
\plotone{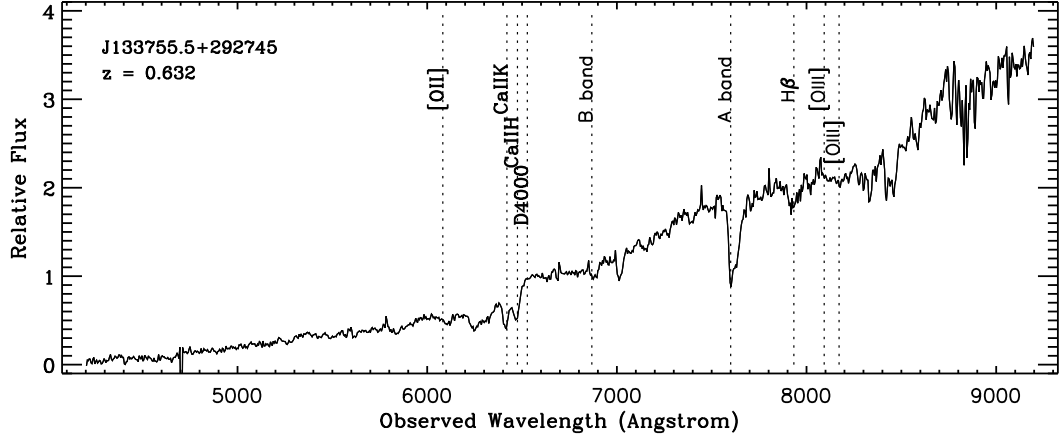}
\caption{Example of one of the eight SEXSI ALG. The ALG are identified
by \cahk~absorption and the break at 4000 \AA, typical of early-type 
galaxies. No emission features are detected.}
\label{fig:alg}
\end{figure}

\clearpage
\begin{figure}
\epsscale{1.0} 
\plotone{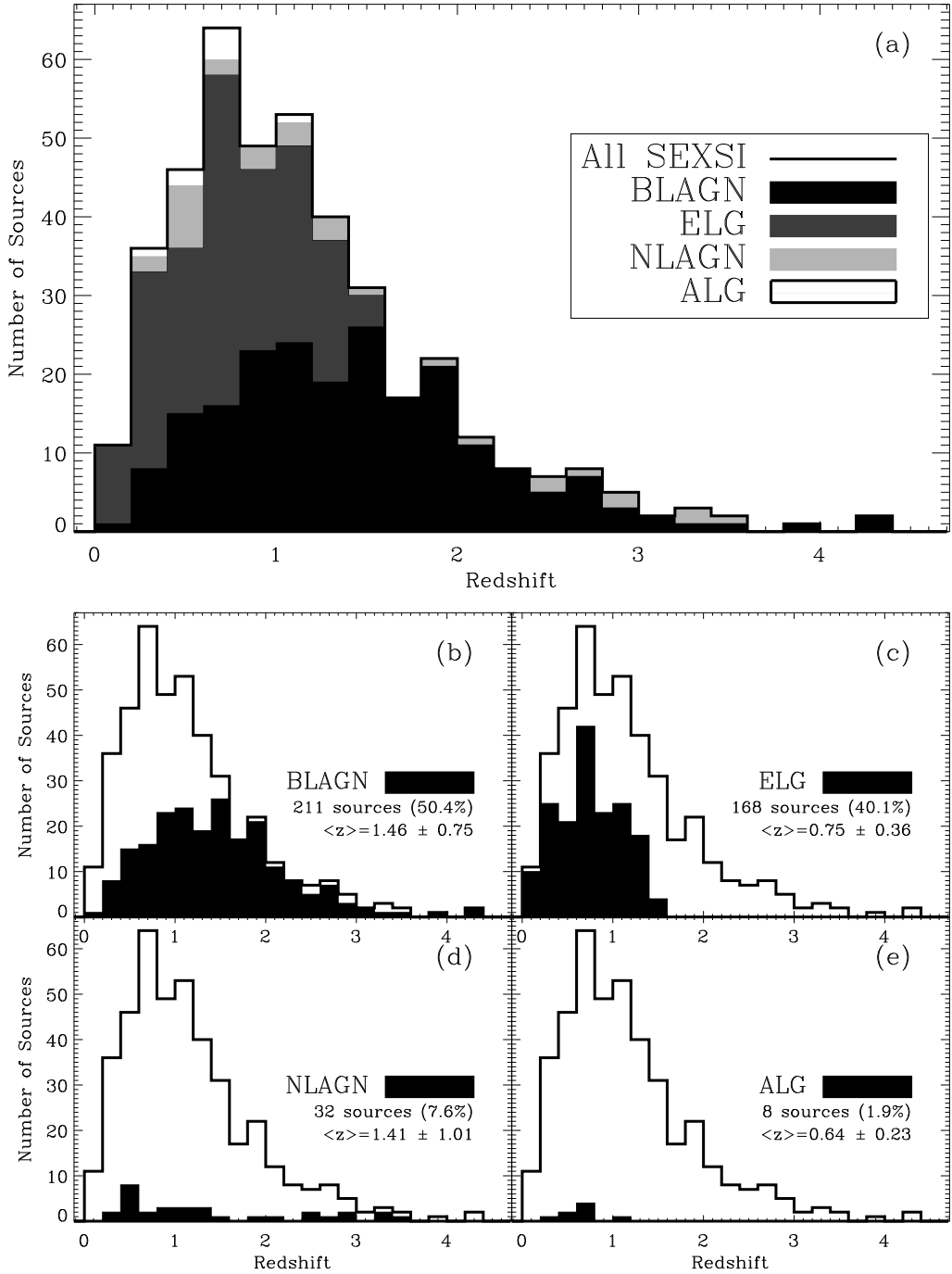} 
\caption{Redshift histogram for the 419 sources with spectroscopic
redshifts presented in Table~\ref{tbl:catalog}, excluding the stars
(at $z=0$). Panel (a) shows the entire histogram with shading according
to class. Panels (b) -- (e) show the same histogram with each individual
source class highlighted in black. These plots emphasize that the sample
is dominated by the BLAGN, with a broad redshift distribution, and the
ELG, dominant at lower-$z$. The NLAGN have a $z$-distribution most similar
to the BLAGN; the NLAGN are the only narrow-lined sources with $z>1.5$.}
\label{fig:z_hist}
\end{figure}

\clearpage
\begin{landscape}
\begin{figure}
\epsscale{1.0} 
\plotone{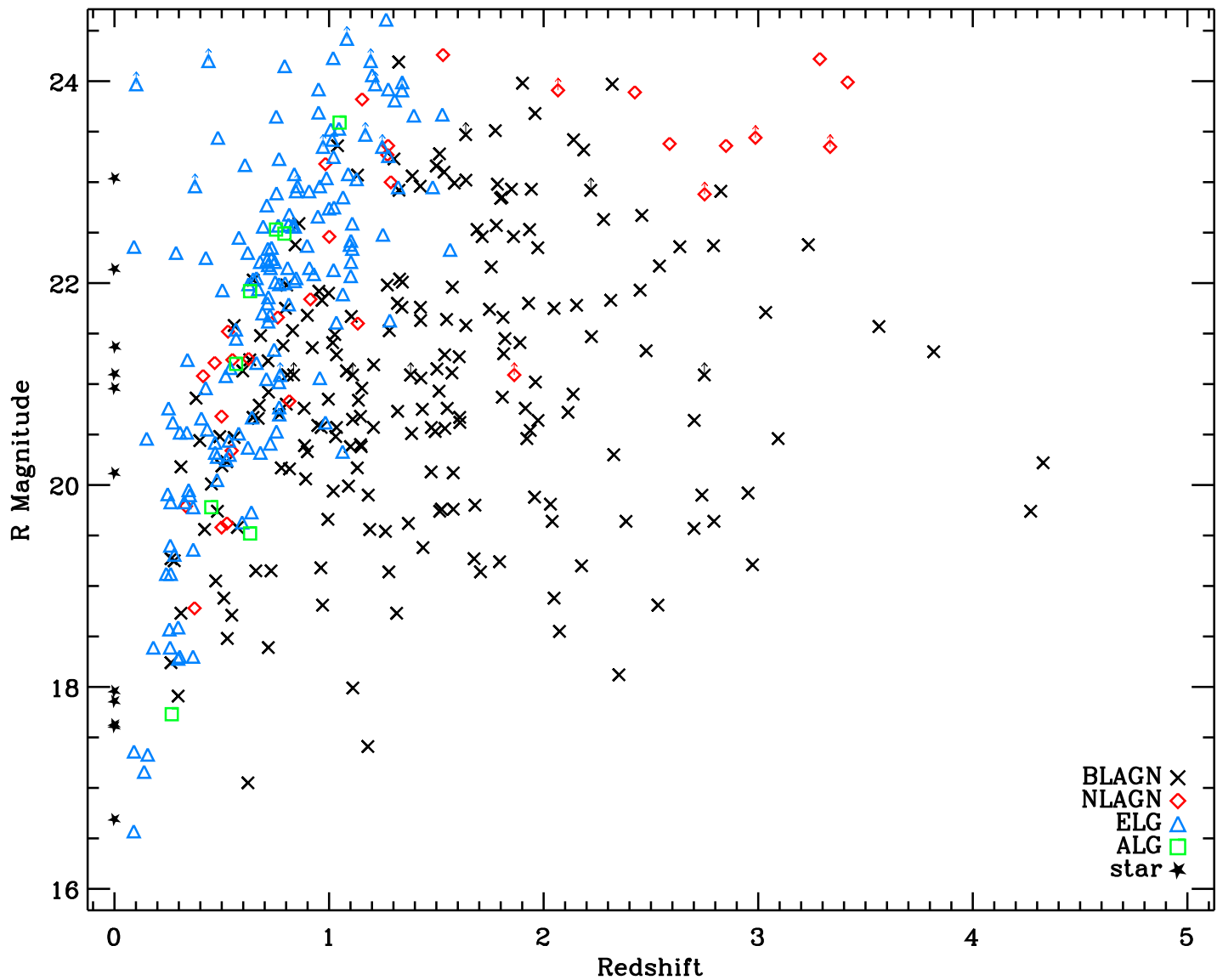} 
\caption{$R$-magnitude versus redshift for spectroscopically-identified sources. The distinct regions in $R-z$ parameter space covered by each class type is illustrated. The BLAGN have the largest spread in $z$ and also occupy a large spread $R$, though they are not found in as great number at the faintest $R$ fluxes. The NLAGN are also spread widely in $z$, but tend to be the nearer the faint end of the $R$ distribution. The ELG appear with $z\simlt 1.5$ and are mainly found to have $R>20$. The few ALG have a distribution similar to the ELG.
Stars with $R>18$ are most likely chance coincidences, and not real X-ray counterparts.}
\label{fig:r_z}
\end{figure}
\clearpage
\end{landscape}

\begin{figure}
\epsscale{1.0} 
\plotone{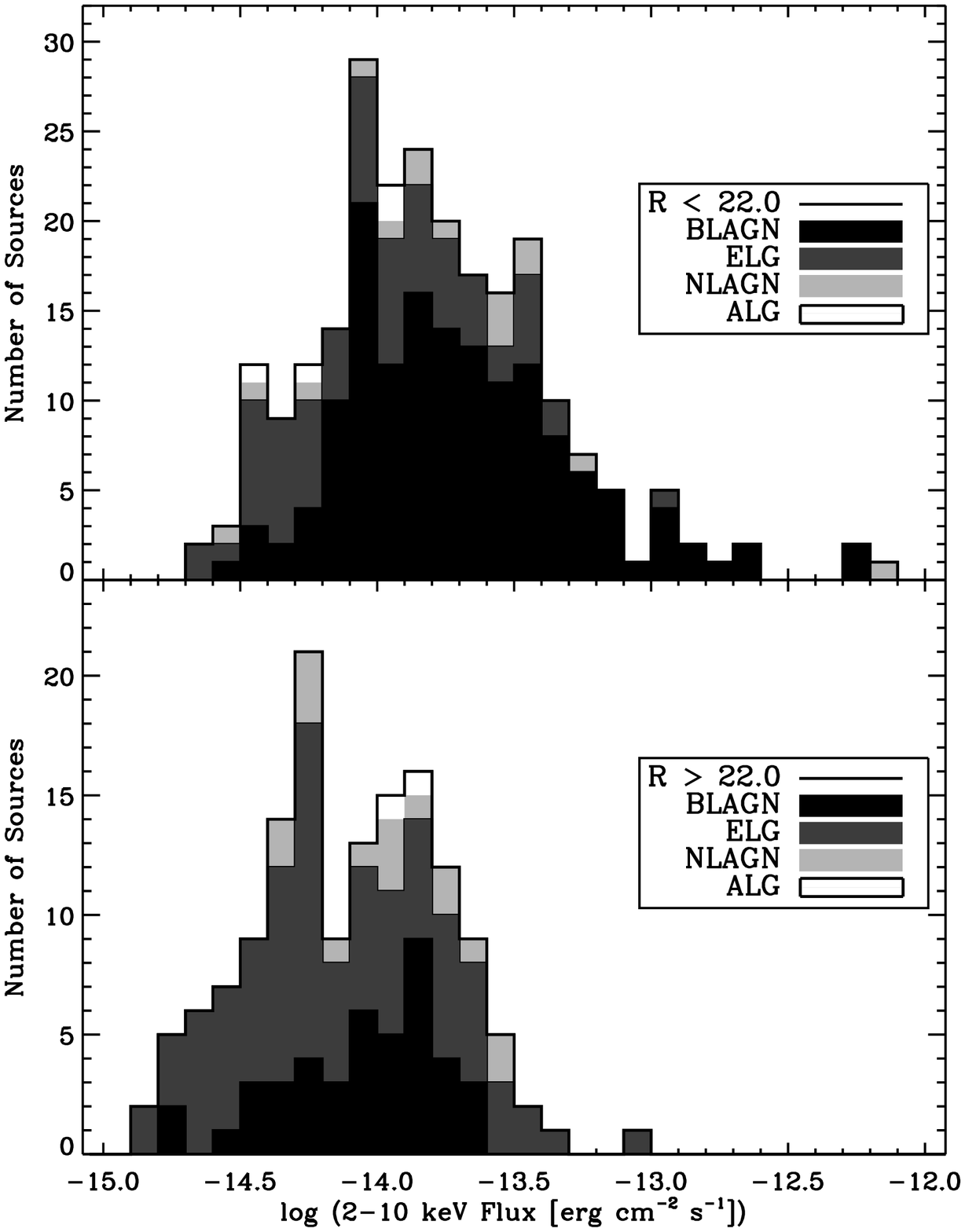} 
\caption{The \hardrange\ flux histogram of spectroscopically-identified sources,
excluding stars, split at $R=22$. The optically-brighter sources 
($R<22$), presented in the top panel, show a broad peak between 
\hardfluxrange $\sim 10^{-14} \fluxu$ and 
\hardfluxrange $\sim 10^{-13.5} \fluxu$. These $R<22$ sources
are dominated by BLAGN. The bottom panel shows the optically-fainter 
sources ($R>22$). The hard-flux histogram is shifted
to lower fluxes and includes many ELG as well as BLAGN, ALG, and 
NLAGN. (Only sources with $R_{\rm limit}>22$ are included in this
plot. This cut eliminates few sources since most SEXSI 
imaging has $23<R_{\rm limit}<24$.)}
\label{fig:hflux_hist} 
\end{figure}

\clearpage
\begin{figure}
\epsscale{1} 
\plotone{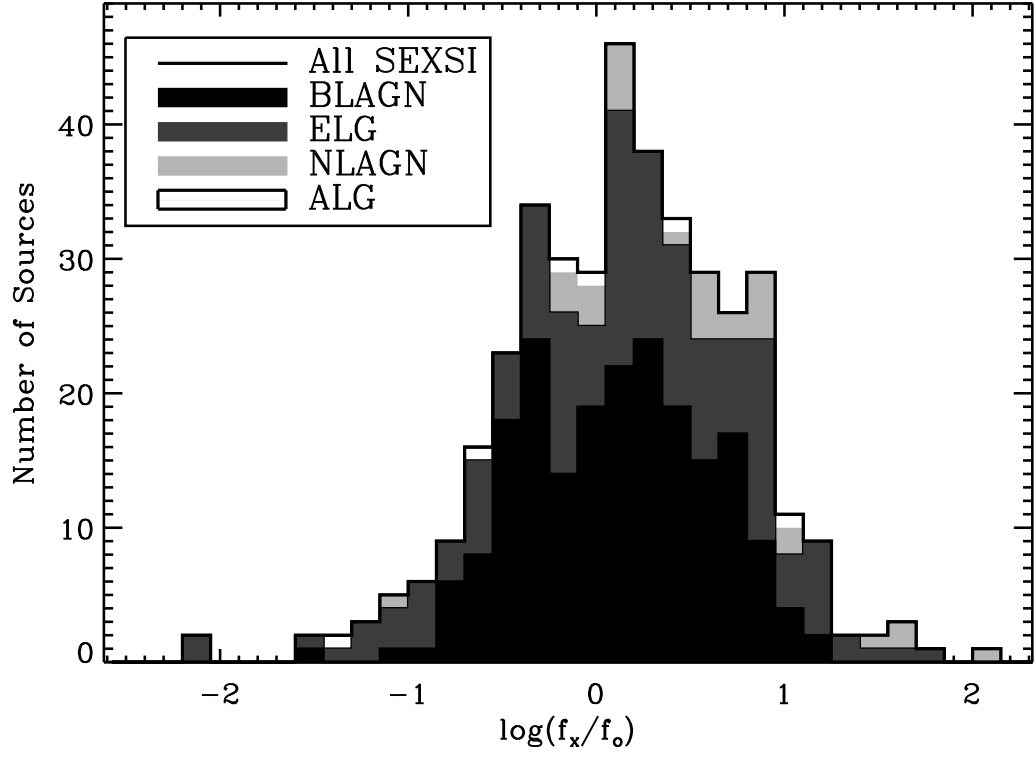} 
\caption{Histogram of $\fxfo$~distribution of spectroscopically-identified sources ($f_{\rm x} = $\hardfluxrange). Most sources are found between $-1 < \fxfo < 1$, while the NLAGN tend towards higher values of $\fxfo$. At $\fxfo>1$, all but five of the twenty-six sources lack broad lines.}
\label{fig:fxfo_hist} 
\end{figure}

\clearpage
\begin{figure}
\epsscale{1.0} 
\plotone{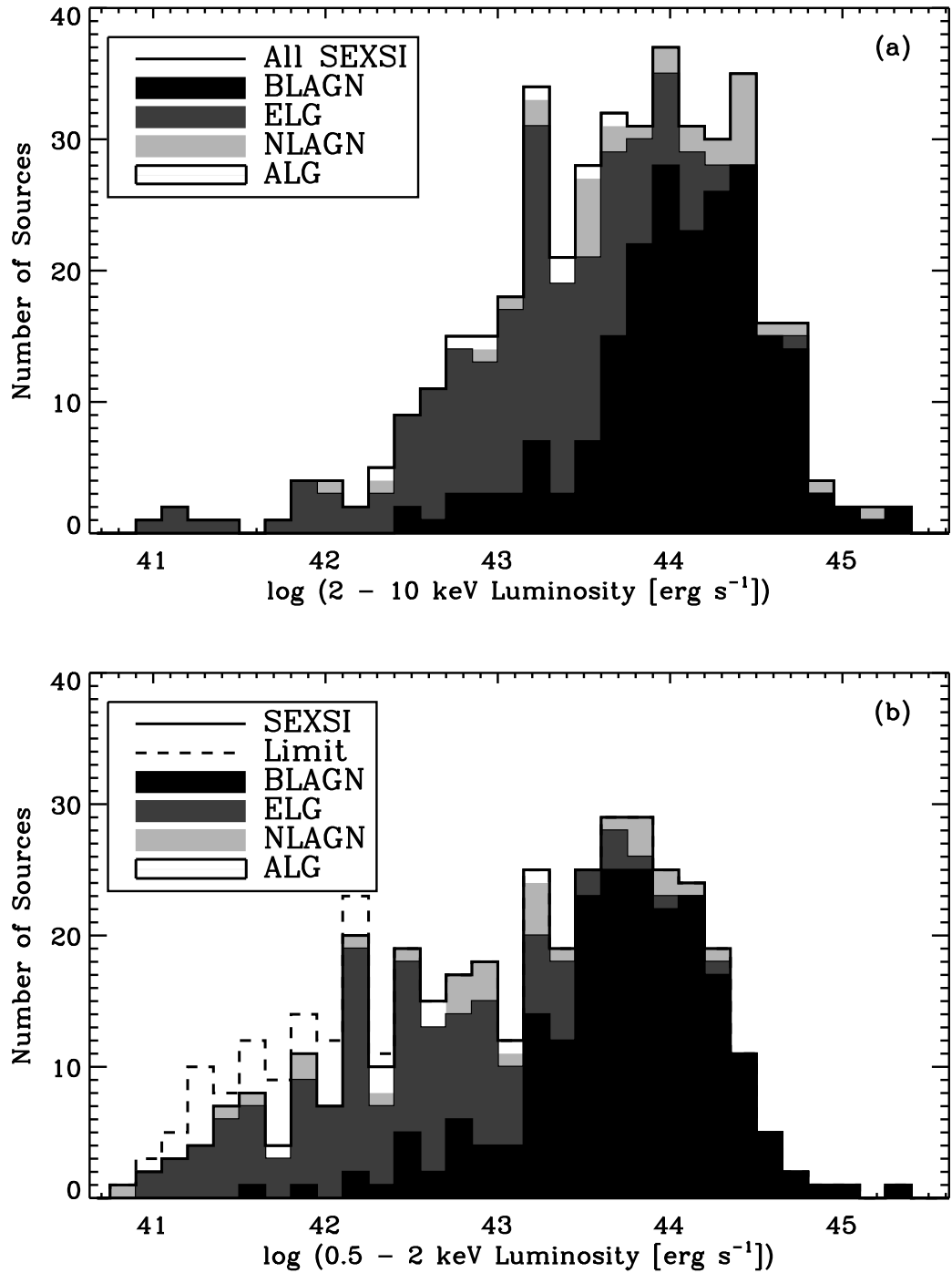} 
\caption{X-ray luminosity histograms, uncorrected for intrinsic
absorption at the source. The top panel shows the \hardrange\
luminosity distribution, while the bottom panel shows the \softrange\ 
distribution.  The dashed histogram in the bottom panel indicates
the 32 sources with upper limits to $L_{0.5-2.0~{\rm keV}}$; the majority of these
sources (26/32) are ELG while 4 are NLAGN.}
\label{fig:lum_hist} 
\end{figure}

\clearpage
\begin{landscape}
\begin{figure}
\epsscale{1.0} 
\plotone{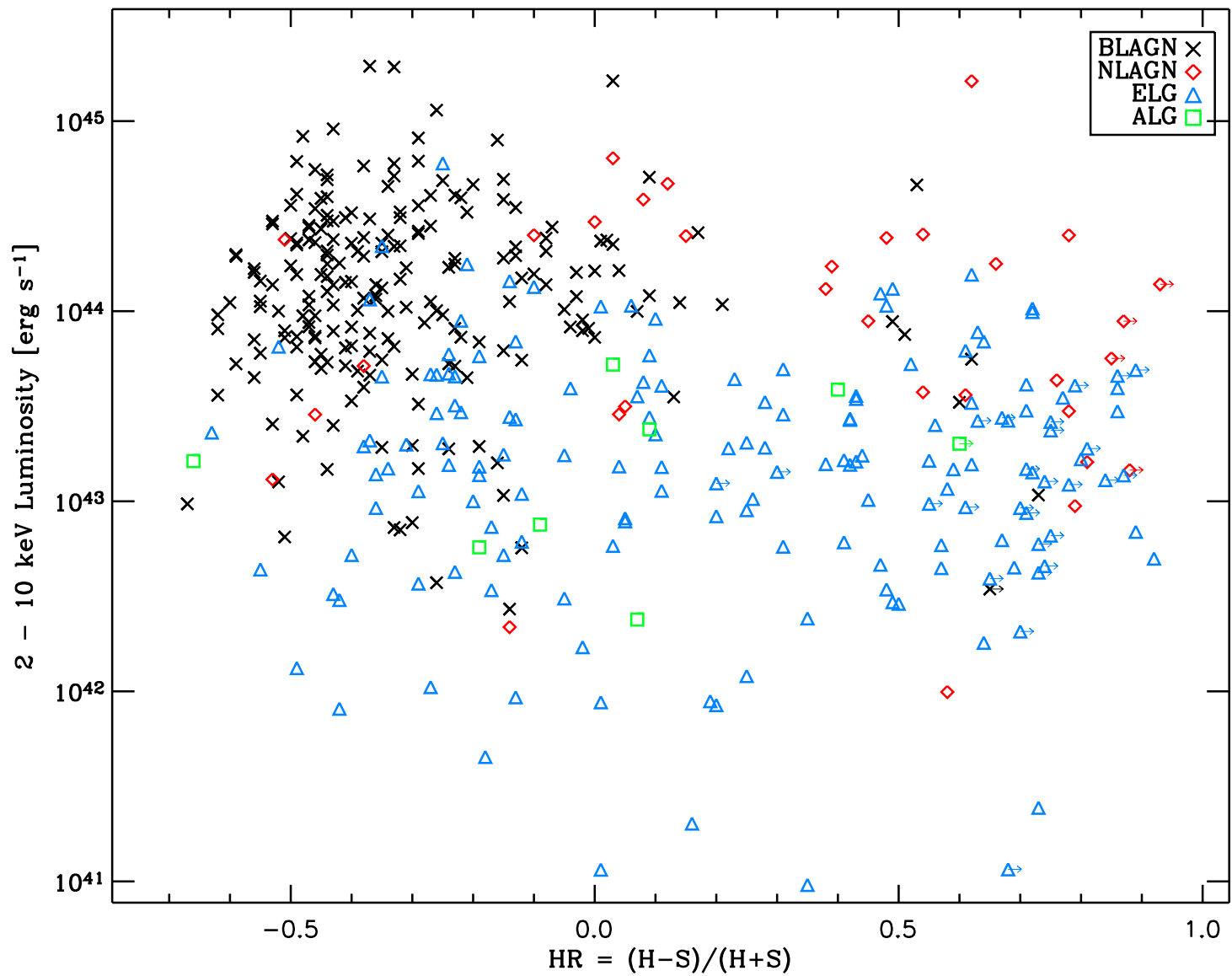} 
\caption{\hardrange\ luminosity vs. $HR$. Crosses indicate
BLAGN -- note that they mainly fall at high luminosity and 
$HR<0$. NLAGN are shown in diamonds; most have $HR>0$, 
consistent with the notion that obscuration at the source is involved.
Emission-line galaxies (triangles) have a wide spread in $HR$, 
as do the absorption-line galaxies (squares).}
\label{fig:lum_hr}
\end{figure}
\clearpage
\end{landscape}

\begin{figure}
\epsscale{1.0} 
\plotone{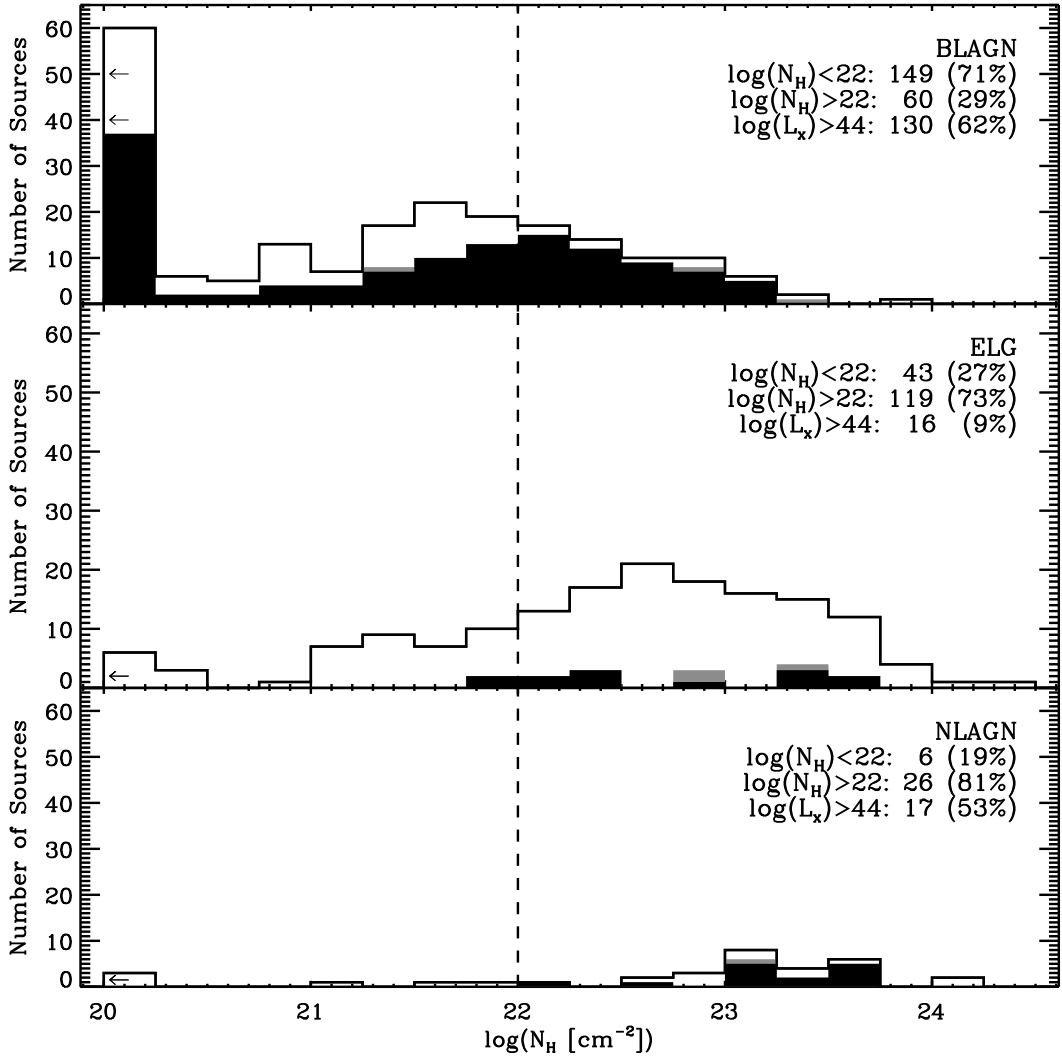} 
\caption{\nh~histogram for BLAGN, ELG, and NLAGN. The small number of
ALG are omitted. Sources with a best-fit \nh\ value below \lognh$=20$
are placed in the bin at 20.
The black fill indicates sources with obscured \hardrange\ rest-frame luminosities
above $10^{44} \lumin$, wile gray indicates sources with unobscured luminosities
above $10^{44} \lumin$.
The dashed line shows our adopted
break, at \lognh$=22$, between
obscured sources and unobscured sources. Note that while about half of the SEXSI sources
have $N_H>10^{22}$ cm$^{-2}$, it is the BLAGN that dominate the unobscured 
distribution. The majority of the ELG are obscured as are the NLAGN to an even greater
extent.}
\label{fig:nhhist}
\end{figure}

\clearpage
\begin{figure}
\epsscale{1.0} 
\plotone{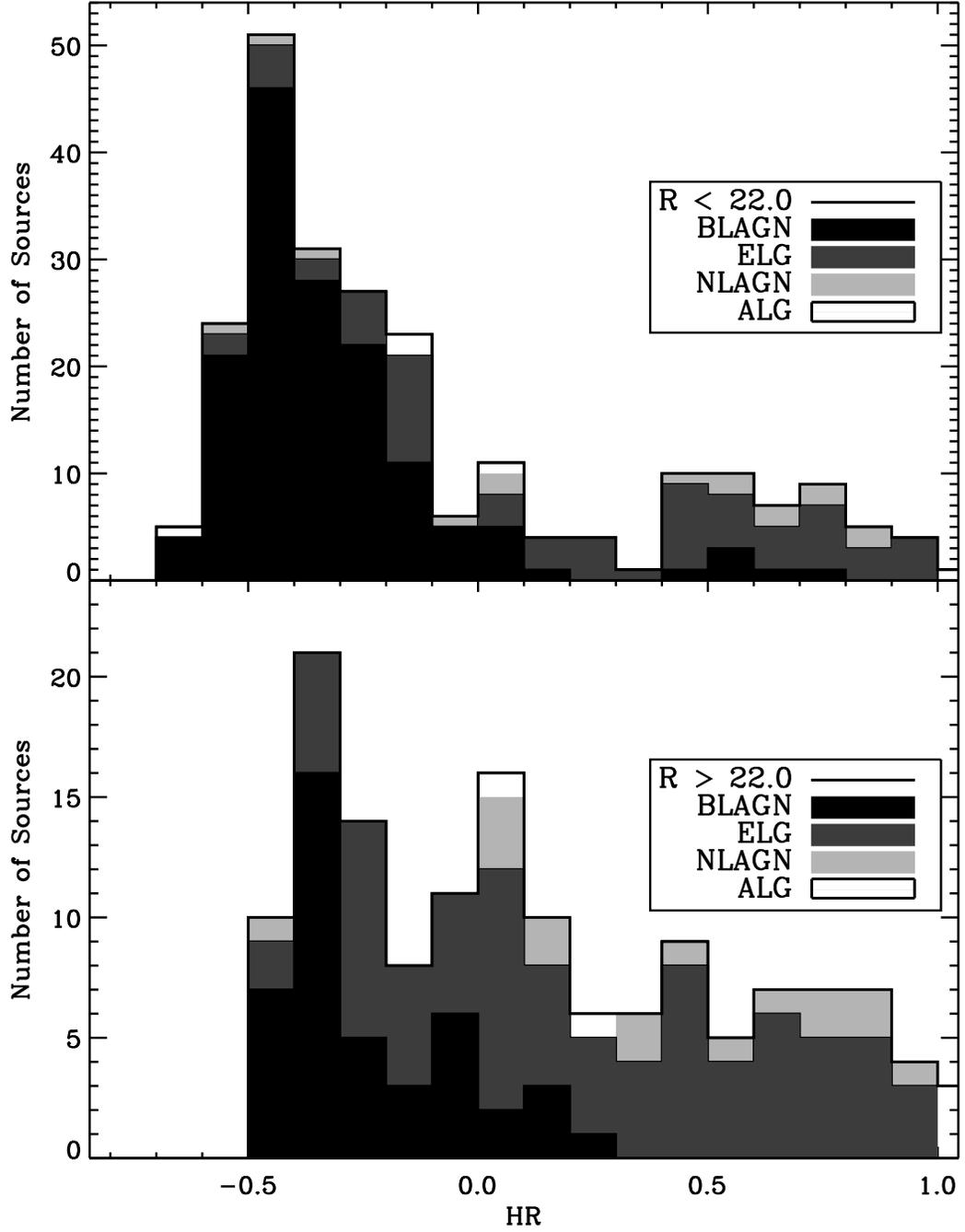}
\caption{Hardness ratio histogram of spectroscopically-identified sources, 
excluding stars,
split at $R=22$. The top panel shows the optically-brighter sources
($R<22$). The black-filled peak near $HR = -0.4$ represents the 
broad-lined sources, which dominate the optically-brighter population 
of \hardrange\ SEXSI sources. The lower panel shows the optically-fainter 
($R>22$) sources which are, on average, much harder and have 
a broader $HR$ distribution. This group of sources is a mix
of spectral classes: there are BLAGN, but in addition there are many 
ELG, ALG, and NLAGN.}
\label{fig:hr_hist}
\end{figure}

\clearpage
\begin{figure}
\epsscale{1.0} 
\plotone{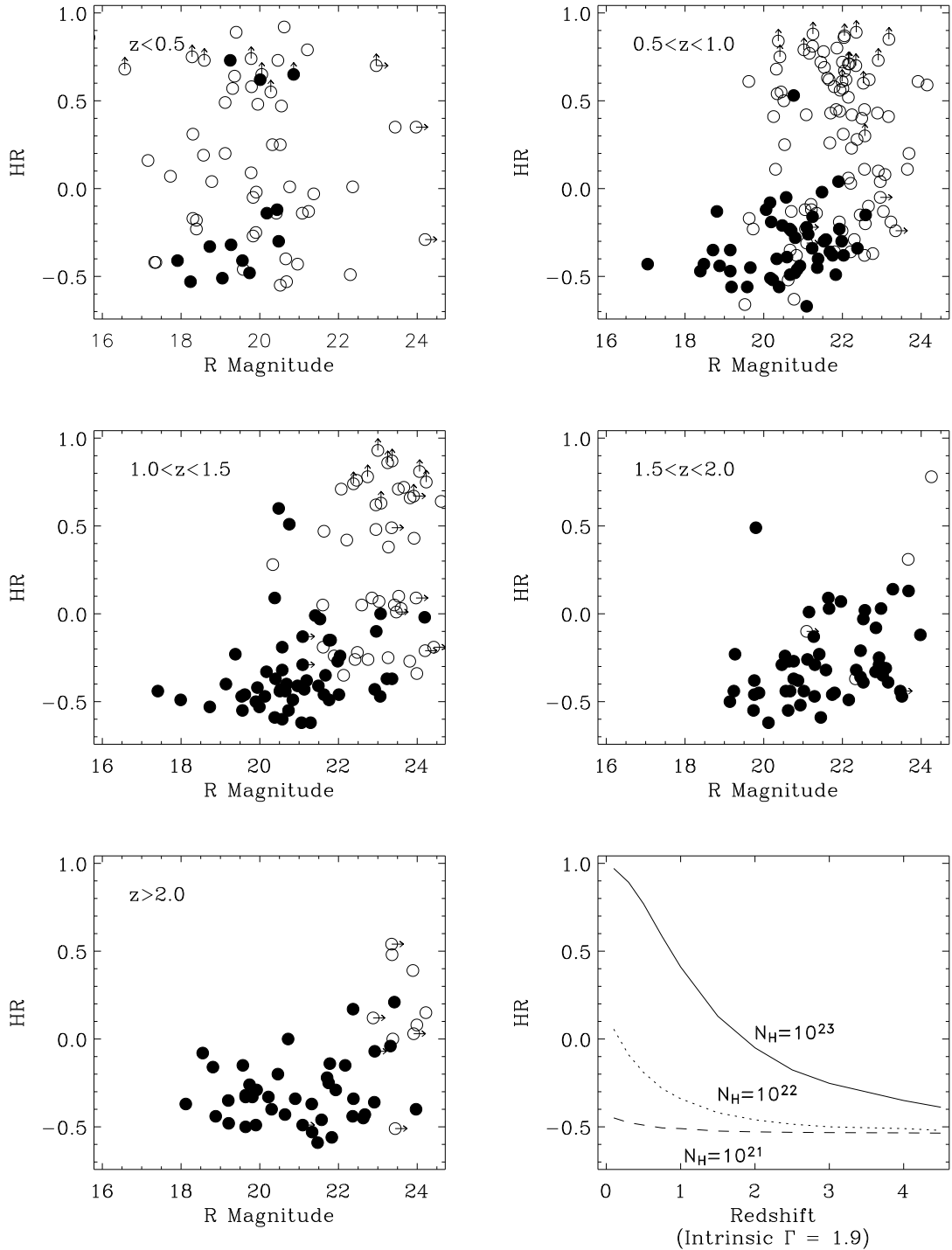} 
\caption{$HR$ versus $R$-magnitude for five redshift ranges. Broad-lined sources (BLAGN)
are shown with filled circles, while non-BL sources (NLAGN, ELG, ALG) are open. 
Arrows indicate limits to the $HR$ (upward pointing) and $R$-magnitude
(right pointing). The abrupt drop in the number of non-BL sources in the highest-$z$ panels ($z>1.5$) is apparent, and caused both by \oii~shifting out 
of the optical band,
and the inability to spectroscopically identify faint sources. The bottom 
right panel shows $HR$ versus $z$ for three typical values of intrinsic
obscuring column density given a source power-law index $\Gamma=1.9$, for 
reference. The column densities indicated on the plot are in units of 
cm$^{-2}$.}
\label{fig:hr_rmag_z_1}
\end{figure}

\clearpage
\begin{figure}
\epsscale{1.}
\plotone{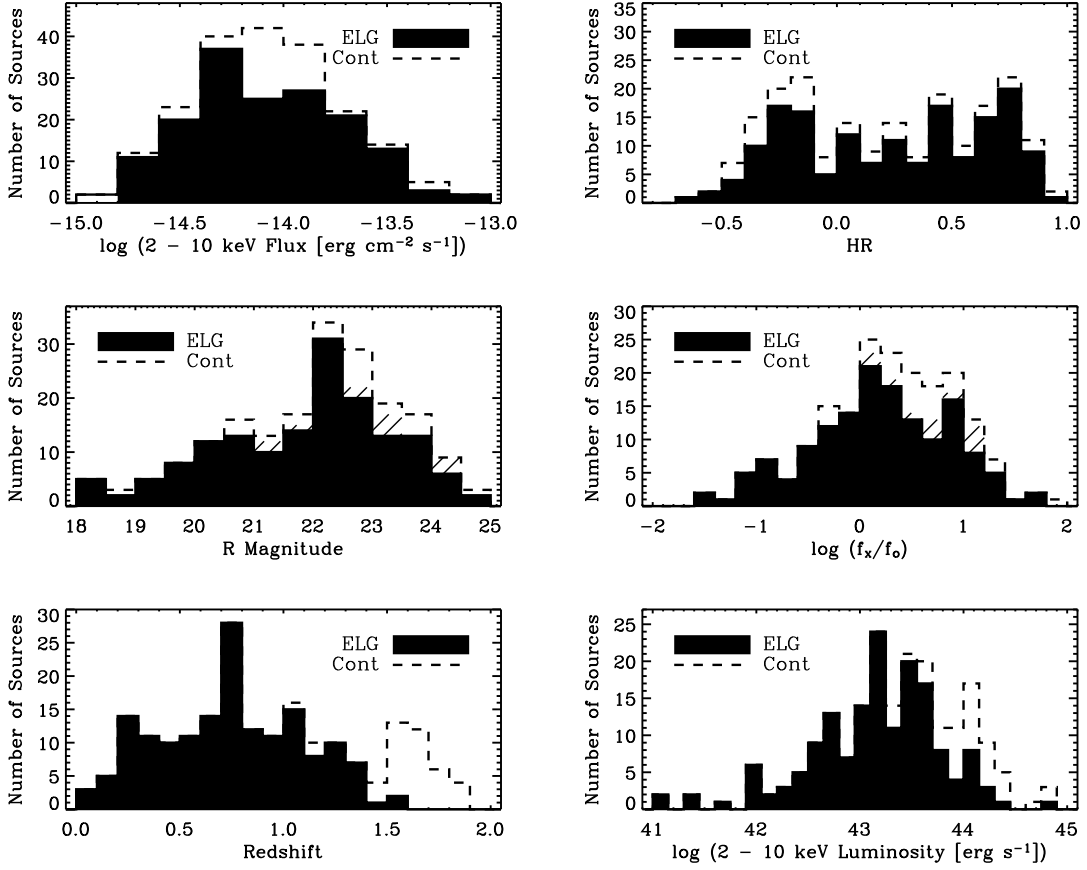}
\caption{The ELG population compared to sources
that have spectra with a continuum lacking detected emission or absoprtion features.
The top panels show the \hardrange~flux, $HR$, $R$-magnitude, and $\fxfo$ distributions of 
the ELG (filled black) and the continuum-only
sources (dashed line). The hatched regions show the continuum-only sources
that have $R=R_{\rm limit}$. The bottom two panels show
the spectroscopic redshift distribution and corresponding $L_{\rm x}$ distribution for the 
ELG (filled black). These distributions are compared to distributions of 
the continuum-only
sources whose assigned ``redshift limits" were calculated assuming that
\oii~falls just longward of the optical spectral range for each source.}
\label{fig:cont_plots}
\end{figure}

\clearpage
\begin{figure}
\epsscale{1.0}
\plotone{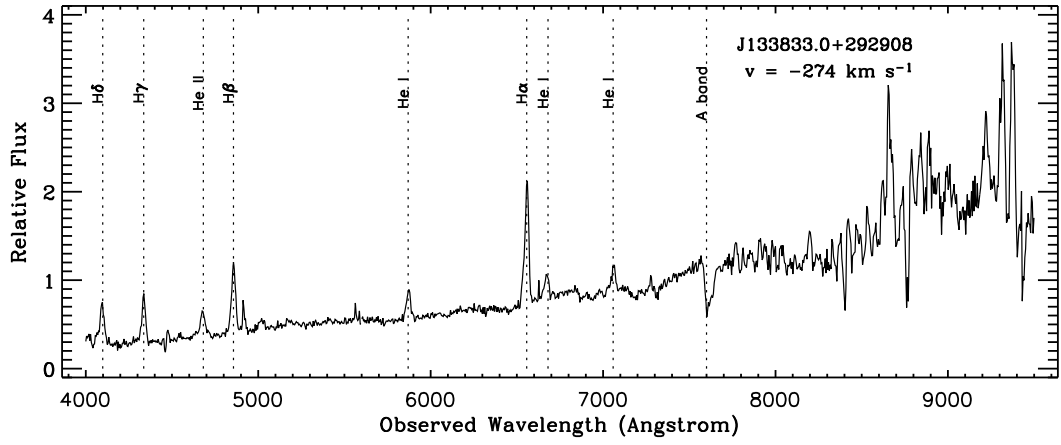} 
\caption{A cataclysmic variable (CV) that shows a blueshift of 274 km s$^{-1}$. This source 
has \hardfluxrange $= 3.25 \times 10^{-14} \fluxu$, $HR= -0.18$, and $R=20.1$.
The measured velocity indicates either that the star is at an extremum in its orbit or 
that it is a rare halo-CV.}
\label{fig:cv}
\end{figure}

\clearpage
\begin{landscape}
\begin{figure}
\epsscale{1.0}
\plotone{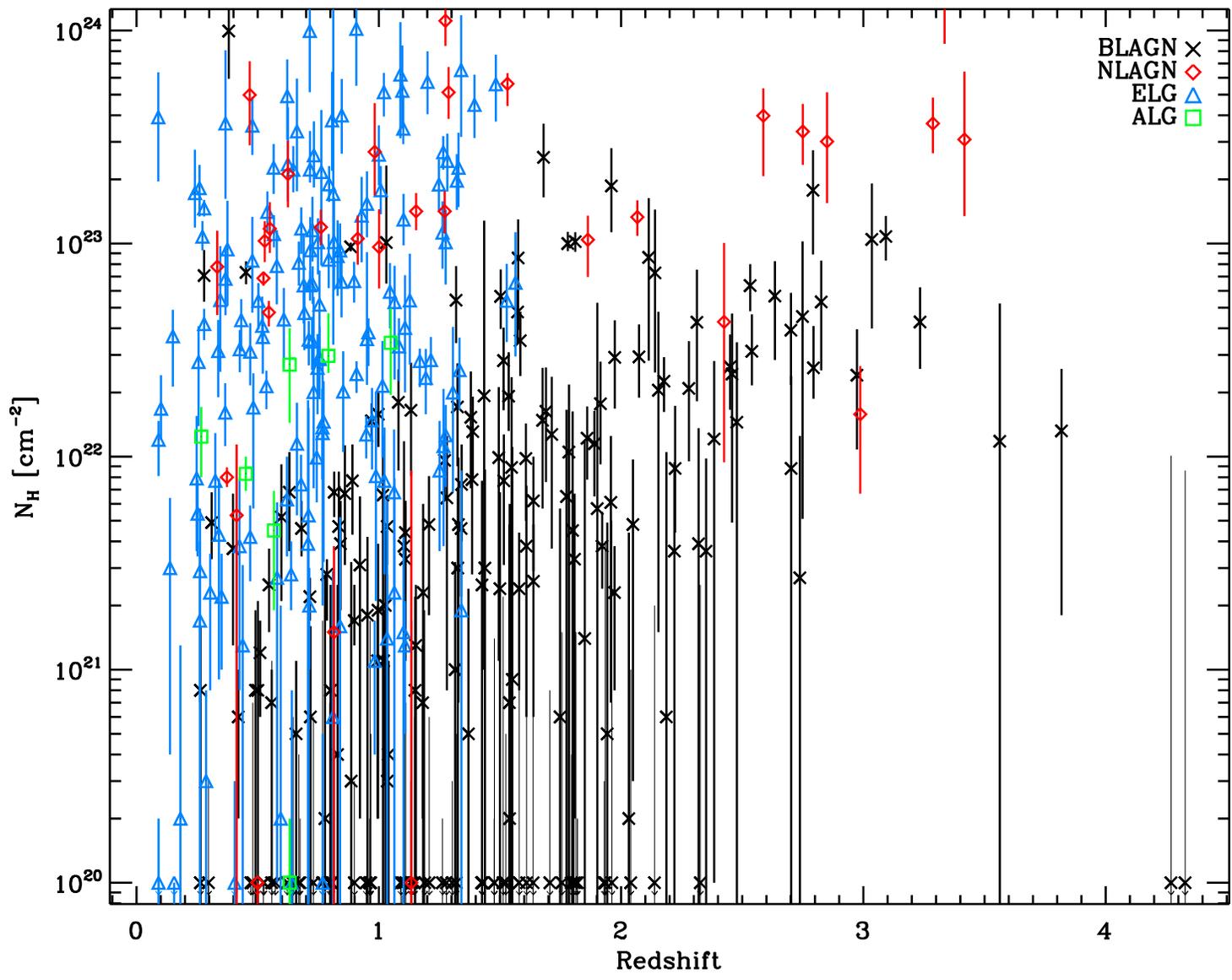} 
\caption{\nh~versus redshift. 1 $\sigma$ error bars from the XSPEC 
fits are presented; the 14 sources with \nh\ values determined from the
$HR$ (see \S~\ref{sec:catalog}) are omitted. Sources with a best-fit
\nh\ value below  $10^{20}$ cm$^{-2}$  are placed at $10^{20}$  cm$^{-2}$ with 
a downwards-pointing arrow. If the 1 $\sigma$ upper-bound to the \nh\
is above $10^{20}$  cm$^{-2}$ then the error bar shows on the plot; conversely
if the upper-bound is also below $10^{20}$  cm$^{-2}$ no error bar is present.}
\label{fig:nh_z}
\end{figure}
\clearpage
\end{landscape}

\begin{figure}
\epsscale{1.} 
\plotone{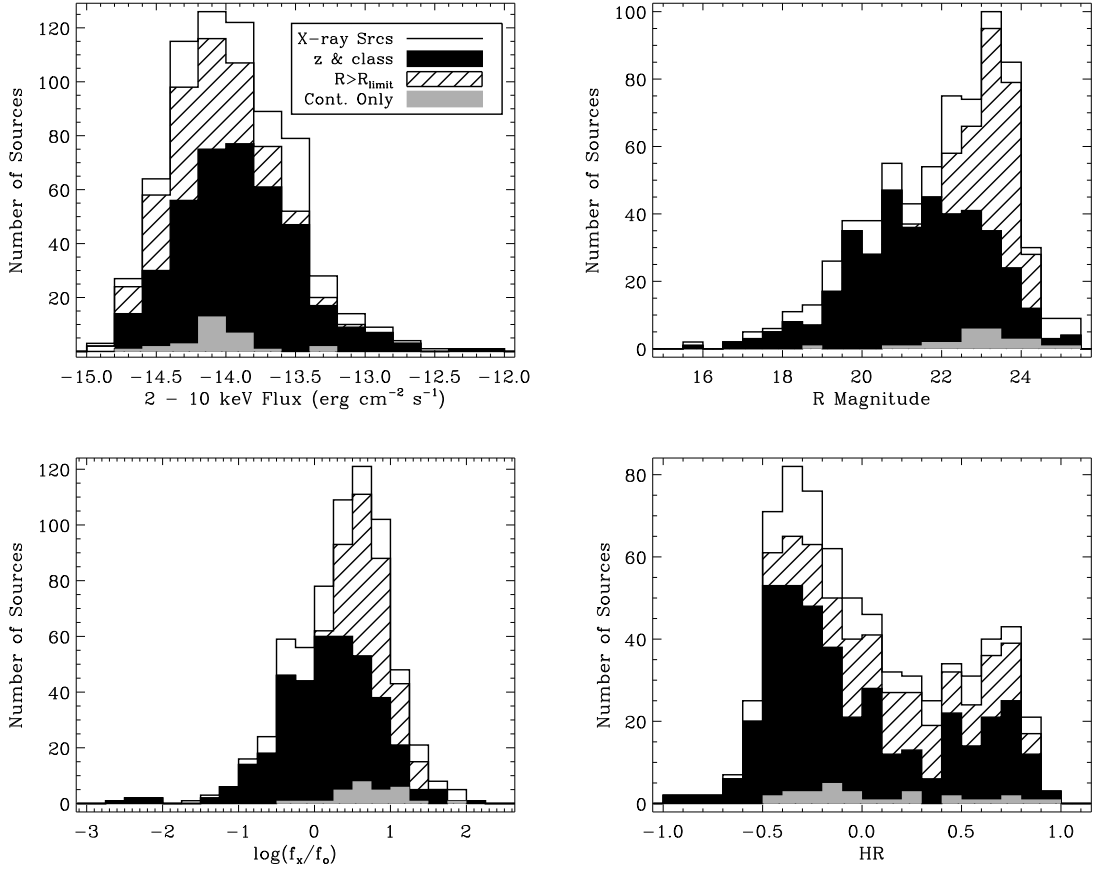} 
\caption{Histograms of \hardfluxrange, $R$, $\fxfo$, and $HR$ from seventeen fields where we have extensive spectral coverage.  Each panel shows a histogram of all SEXSI \hardrange~sources from these fields. Sources with a spectroscopic redshift and classification are filled with black and sources with a spectra that show continuum only are shaded in gray. The hatched part of each histogram indicates sources with no photometrically identified optical counterpart ($R>R_{\rm limit}$). For the $R$-magnitude and $f_{\rm x}/f_{\rm o}$ plot the $R$-magnitude plotted is $R_{\rm limit}$. Sources with no shading or hatch marks have optical photometric ID's but no spectroscopic followup.}
\label{fig:completeness}
\end{figure}

\clearpage

\begin{landscape}
\begin{figure}
\epsscale{1.0}
\plotone{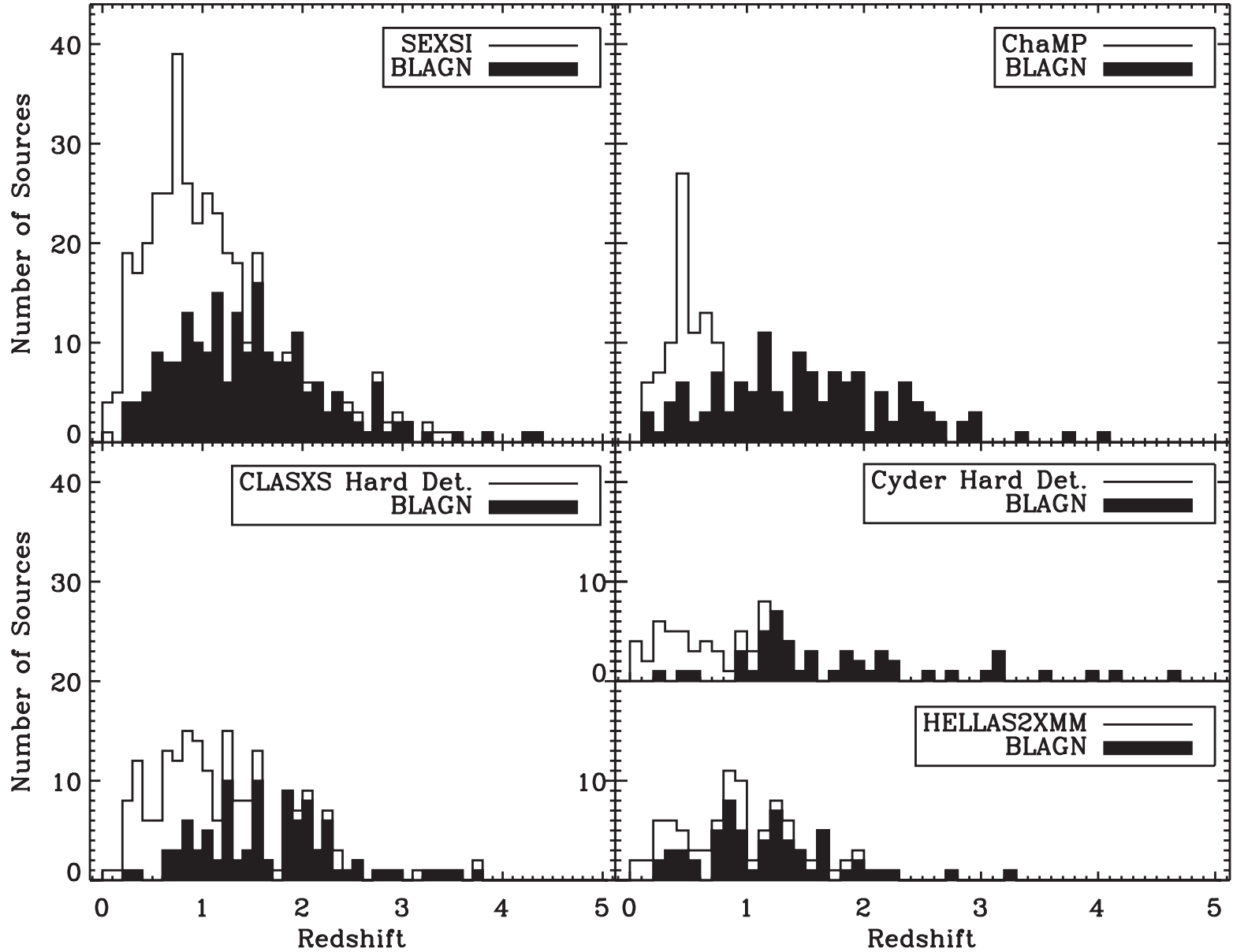}
\caption{Redshift distribution of 
SEXSI, ChaMP~\citep{Silverman:05}, 
CLASXS~\citep{Steffen:04}, 
 CYDER~\citep{Treister:05}, 
and HELLAS2XMM~\citep{Fiore:03}. Broad-lined AGN are represented by
filled histograms, while non-broad-lined sources are left unshaded. 
We plot only {\em hard-band-selected} sources; to this
end we have eliminated sources with \hardfluxrange $< 2 \times 10^{-15} \fluxu$.
See \S \ref{sec:z_comp}~for details.}
\label{fig:z_comp}
\end{figure}
\clearpage
\end{landscape}

\begin{figure}
\epsscale{1.0}
\plotone{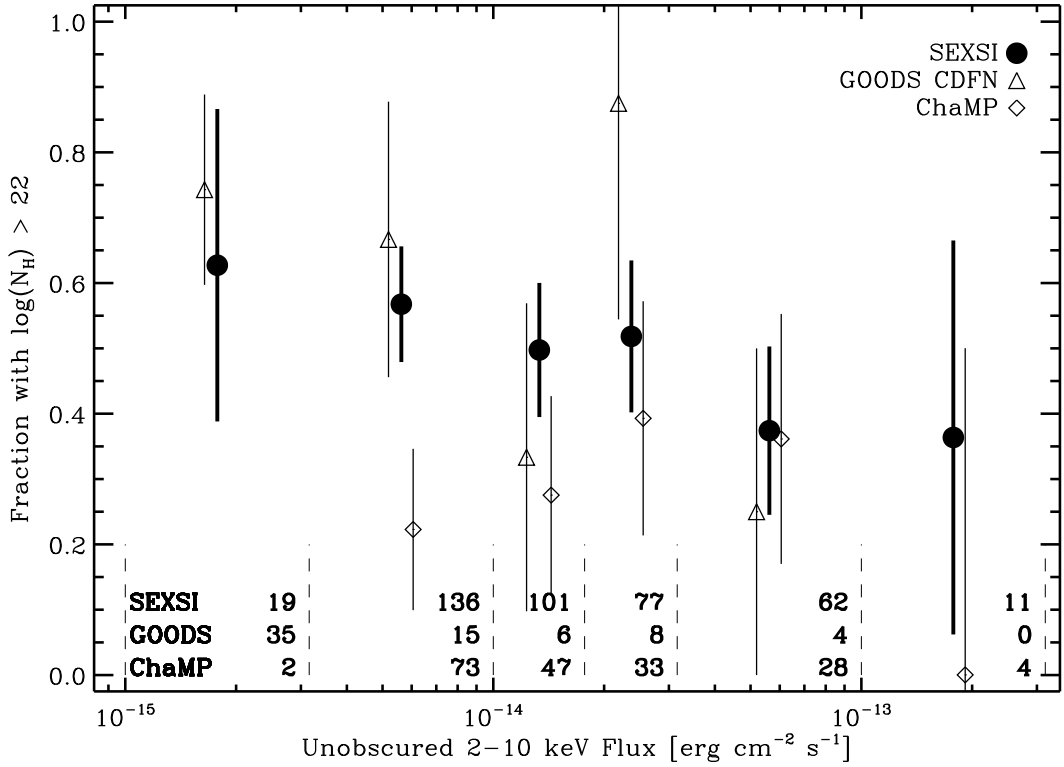}
\caption{Fraction of obscured (\lognh $>22$) sources 
as a function of {\em unobscured} \hardrange~flux for the spectroscopically 
identified samples from SEXSI (filled circles), 
GOODS CDFN (triangles; E. Treister, private communication), 
and ChaMP \citep[diamonds;][]{Silverman:05}. 
The datapoints are calculated using the available catalogs, 
binned into flux ranges shown by the vertical dashed lines 
at the bottom of the plot. The errors on the GOODS CDFN fractions
are 1 $\sigma$, calculated from Poisson counting statistics, 
while the SEXSI and ChaMP error bars incorporate the individual 
\nh\ error bars from the spectral fits in addition (see 
\S~\ref{sec:obscured_srcs}~for details).  The numbers printed near
the bottom of each bin show the number of sources in each bin for 
each survey. Datapoints are
offset slightly along the x-axis for clarity. 
The GOODS CDFN data lacks sources in the highest flux bin so we omit the datapoint at 
$\sim 2 \times 10^{-13} \fluxu$. Conversely, ChaMP has only two sources below
$\sim 3 \times 10^{-15} \fluxu$ and we omit the lowest flux datapoint.}
\label{fig:frac_obsc_flux}
\end{figure}

\begin{figure}
\epsscale{1.0}
\plotone{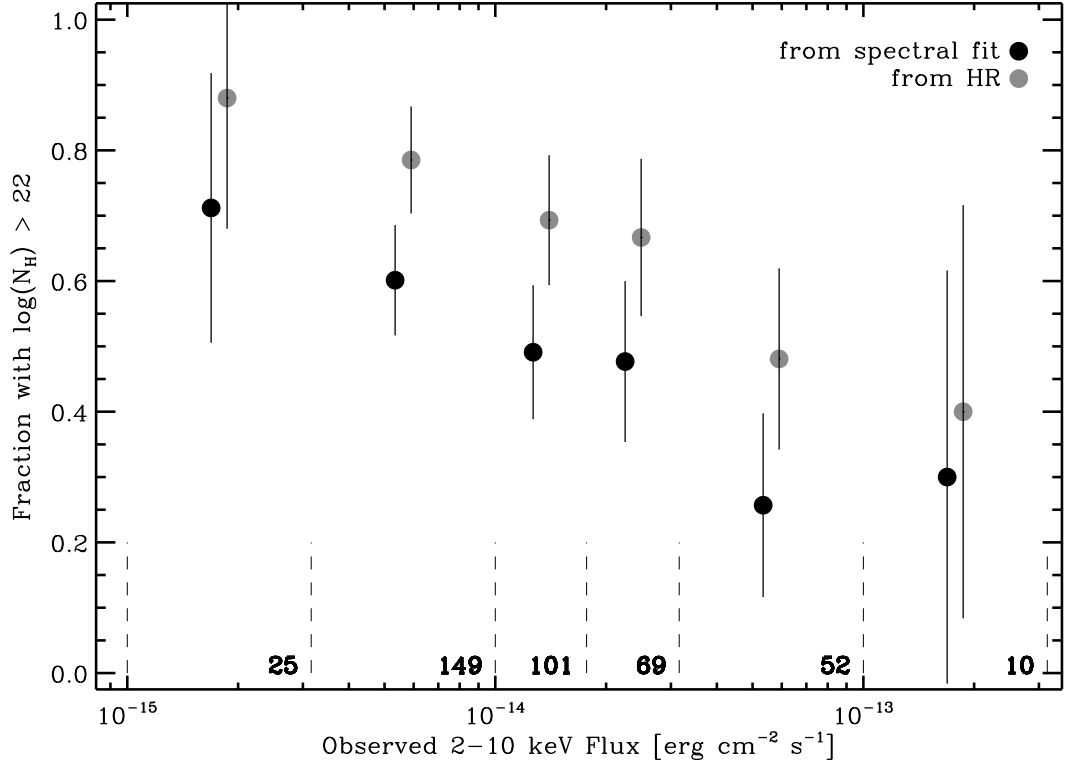}
\caption{Comparison of \nh~calculation methods. The fraction of 
obscured sources as a function of observed (absorbed) 
\hardrange\ flux for SEXSI sources with \nh\ calculated by
X-ray spectral fitting (black), as presented in the source catalog
and throughout this article, and from hardness ratios (gray).
The $HR$ derived obscured fractions tend to be significantly 
higher. See \S~\ref{sec:obscured_srcs} for discussion.
The number of sources in each bin are shown in text along the bottom.}
\label{fig:hr_v_xspec_comp}
\end{figure}

\begin{figure}
\epsscale{1.0}
\plotone{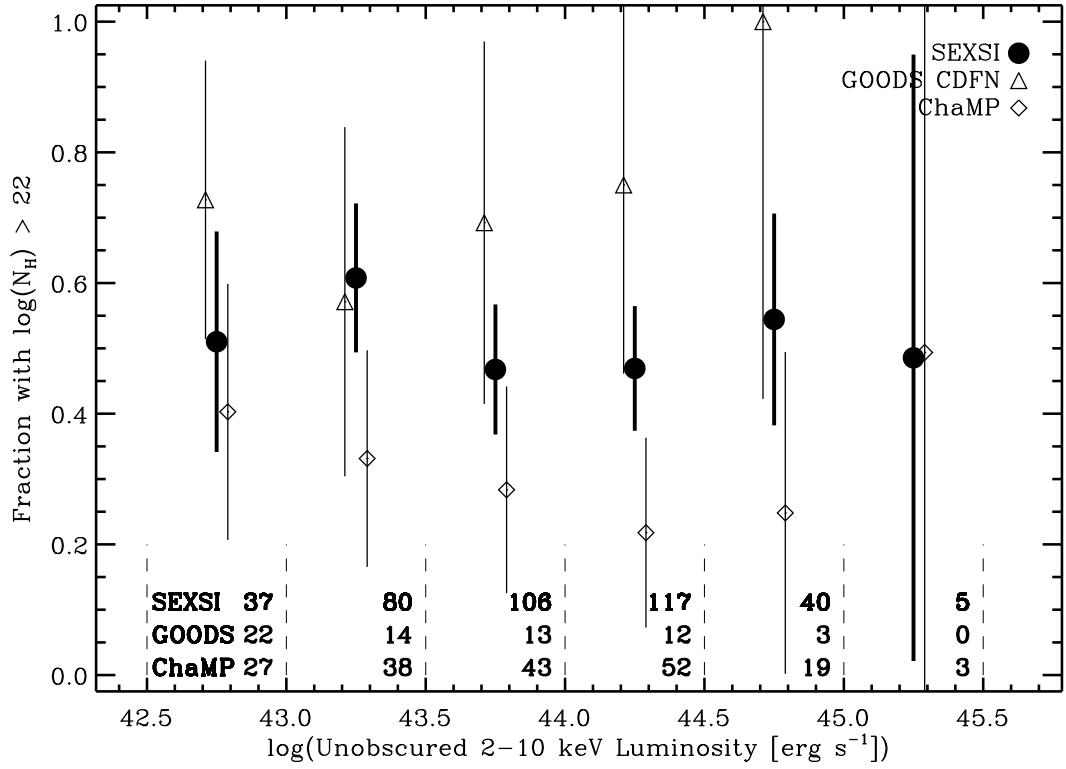}
\caption{Fraction of obscured sources as a function of {\em unobscured} luminosity for 
spectroscopically identified sources. The luminosities
are calculated based on the observed luminosity corrected for intrinsic \nh. Figure 
\ref{fig:frac_obsc_flux}~gives references to the catalogs used to calculate the non-SEXSI points.  
The luminosity bins are shown by the vertical dashed lines 
at the bottom of the plot; the errors
are 1 $\sigma$, calculated using Poisson counting statistics and the \nh\ errors when available
(for ChaMP and SEXSI).  The numbers printed near
the bottom of each bin show the number of sources in each bin for 
each survey.
Each datapoint is offset slightly along the x-axis for clarity.}
\label{fig:frac_obsc_lum}
\end{figure}

\begin{figure}
\epsscale{1.}
\plotone{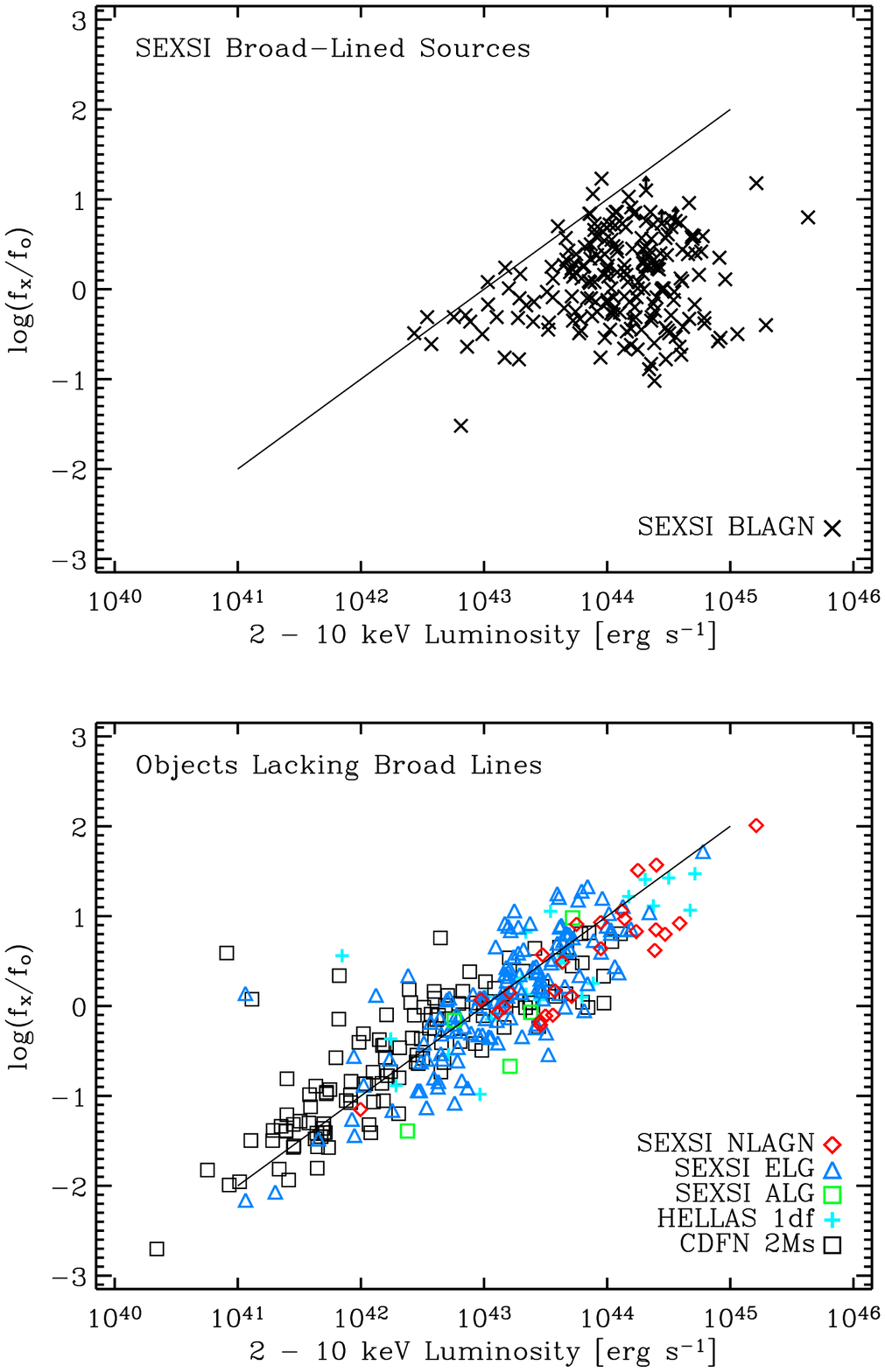}
\caption{These plots show the relationship between $\fxfo$ versus $\log{(L_{\rm 2-10~keV})}$ for SEXSI BLAGN
(top panel) and sources that lack broad emission lines in their optical spectrum including 
SEXSI NLAGN, ELG, and ALG as well as the HELLAS2XMM 1 degree field \citep{Fiore:03}~and 
the CDFN 2Ms sample \citep{Barger:03}. The objects that lack broad lines show a correlation 
between $\fxfo$ and  $\log{(L_{\rm 2-10~keV})}$, which would be expected were the optical 
photometry dominated by galactic light (see \S \ref{sec:elg}) as opposed to emission from 
the AGN. The BLAGN do not show
the correlation. The line shown in both panels is a linear regression to the data of \citet{Fiore:03}.}
\label{fig:fxfo_lum2}
\end{figure}

\begin{figure}
\epsscale{1.}
\plotone{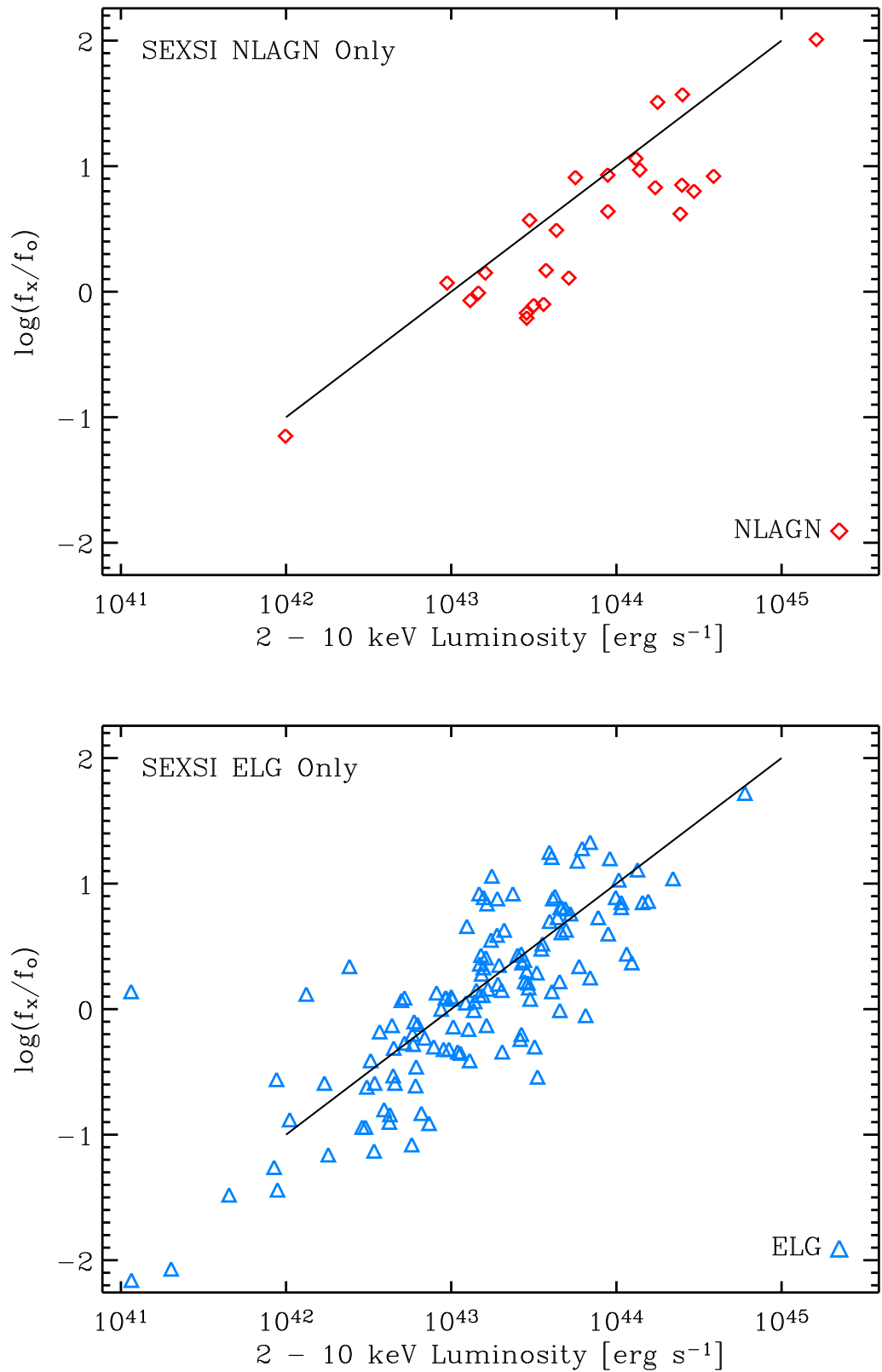}
\caption{Here we again plot the relationship between $\fxfo$ and $\log{(L_{\rm 2-10~keV})}$ for the NLAGN and ELG separately. The identified ELG may well be fit by the line \citep[again, from][]{Fiore:03}, but the NLAGN tend to fall below the line.}
\label{fig:fxfo_lum2_nlagn}
\end{figure}

\begin{figure}
\epsscale{1.0} 
\plotone{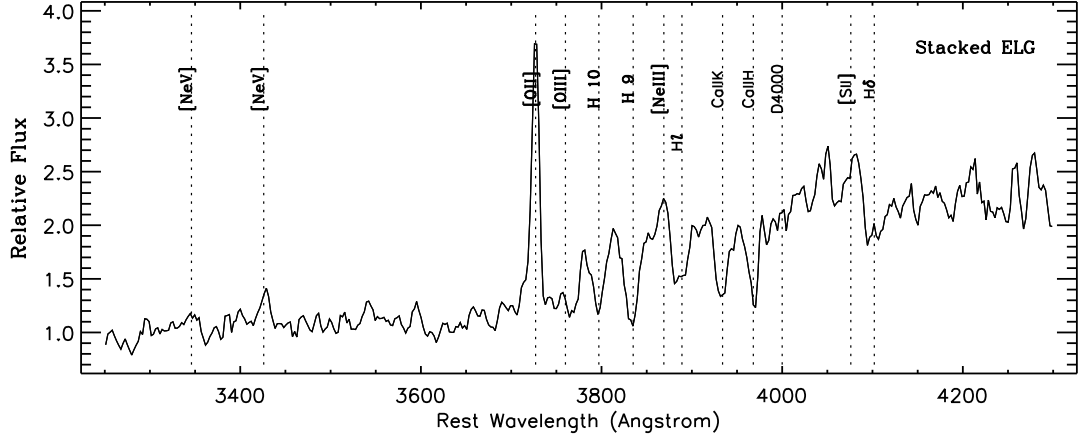} 
\caption{Stacked spectrum of 21 ELG 
all obtained with LRIS on the Keck I Telescope (individual ELG example 
spectra are shown in Figure \ref{fig:elg}, for reference). See \S~\ref{sec:faint_lines} for 
further details of the stacking procedure. 
Note that the individual ELG spectra do not show evidence of the 
\nevsingle~high-ionization emission line indicative of underlying 
AGN activity, while the increased S/N of the stacked spectrum does 
show the \nevsingle~emission. In addition, \neiii~emission
and several absorption features from $\sim 3800 - 4000$ \AA\ are 
well detected.}
\label{fig:stacked_elg}
\end{figure}

\end{document}